\documentclass[%
 prd,
 reprint,
superscriptaddress,
nofootinbib,
 amsmath,amssymb,
 aps,
]{revtex4-2}

\usepackage{graphicx}% Include figure files
\usepackage{dcolumn}% Align table columns on decimal point
\usepackage{bm}% bold math
\usepackage{hyperref}% add hypertext capabilities
\usepackage{url}
\usepackage{newtxtext,newtxmath}
\usepackage{bm}
\usepackage[T1]{fontenc}
\usepackage{xcolor}
\usepackage{tcolorbox}
\usepackage{orcidlink}

\newcommand{\mnras}{Mon.~Not.~Roy.~Astron.~Soc.}
\newcommand{\physrep}{Phys.~Rep.}
\newcommand{\jcap}{JCAP}
\newcommand{\pasj}{Publ.~Astron.~Soc.~Japan}
\newcommand{\apjl}{\apj~Lett.}
\newcommand{\apjs}{\apj~Suppl.}
\newcommand{\aap}{Astronomy~\&~Astrophysics}
\newcommand{\aj}{The~Astronomical~J.}
\newcommand{\pasp}{Publ.~of~the ASP}
\newcommand{\ssr}{Space~Science~Rev.}
\newcommand{\araa}{Ann.~Rev.~Astro.~Astrophys.}

\newcommand{\Cl}{C_{\ell}}

\begin{document}

\title{Hyper Suprime-Cam Year 3 Results: Cosmology from Cosmic Shear Power Spectra} 

\author{Roohi~Dalal\orcidlink{0000-0002-7998-9899}}
 \thanks{rdalal@princeton.edu}
\affiliation{Department of Astrophysical Sciences, Princeton University, Princeton, NJ 08544, USA}
 
\author{Xiangchong~Li\orcidlink{0000-0003-2880-5102}}
\affiliation{McWilliams Center for Cosmology, Department of Physics, Carnegie
Mellon University, 5000 Forbes Ave, Pittsburgh, PA 15213, USA}
\affiliation{Kavli Institute for the Physics and Mathematics of the Universe (WPI), The University of Tokyo Institutes for Advanced Study (UTIAS), The University of Tokyo, Chiba 277-8583, Japan}

\author{Andrina~Nicola\orcidlink{0000-0003-2792-6252}}
\affiliation{Argelander Institut f\"ur Astronomie, Universit\"at Bonn, Auf dem H\"ugel 71, 53121 Bonn, Germany}
\affiliation{Department of Astrophysical Sciences, Princeton University, Princeton, NJ 08544, USA}

\author{Joe~Zuntz\orcidlink{0000-0001-9789-9646}}
\affiliation{Institute for Astronomy, Royal Observatory Edinburgh, University of Edinburgh, EH9 3HJ, United Kingdom}

\author{Michael~A.~Strauss\orcidlink{0000-0002-0106-7755}}
\affiliation{Department of Astrophysical Sciences, Princeton University, Princeton, NJ 08544, USA}
 
 \author{Sunao~Sugiyama\orcidlink{0000-0003-1153-6735}}
\affiliation{Kavli Institute for the Physics and Mathematics of the Universe (WPI), The University of Tokyo Institutes for Advanced Study (UTIAS), The University of Tokyo, Chiba 277-8583, Japan}
\affiliation{Department of Physics, The University of Tokyo, Bunkyo, Tokyo 113-0031, Japan}

\author{Tianqing~Zhang\orcidlink{0000-0002-5596-198X}}
\affiliation{McWilliams Center for Cosmology, Department of Physics, Carnegie
Mellon University, 5000 Forbes Ave, Pittsburgh, PA 15213, USA}

\author{Markus~M.~Rau\orcidlink{0000-0003-3709-1324}}
\affiliation{McWilliams Center for Cosmology, Department of Physics, Carnegie
Mellon University, 5000 Forbes Ave, Pittsburgh, PA 15213, USA}
\affiliation{High Energy Physics Division, Argonne National Laboratory, Lemont,
IL 60439, USA}

\author{Rachel~Mandelbaum\orcidlink{0000-0003-2271-1527}}
\affiliation{McWilliams Center for Cosmology, Department of Physics, Carnegie
Mellon University, 5000 Forbes Ave, Pittsburgh, PA 15213, USA}

\author{Masahiro~Takada\orcidlink{0000-0002-5578-6472}}
\affiliation{Kavli Institute for the Physics and Mathematics of the Universe (WPI), The University of Tokyo Institutes for Advanced Study (UTIAS), The University of Tokyo, Chiba 277-8583, Japan}

\author{Surhud~More\orcidlink{0000-0002-2986-2371}}
\affiliation{The Inter-University Centre for Astronomy and Astrophysics, Post
bag 4, Ganeshkhind, Pune 411007, India}
\affiliation{Kavli Institute for the Physics and Mathematics of the Universe (WPI), The University of Tokyo Institutes for Advanced Study (UTIAS), The University of Tokyo, Chiba 277-8583, Japan}

\author{Hironao~Miyatake\orcidlink{0000-0001-7964-9766}}
\affiliation{Kobayashi-Maskawa Institute for the Origin of Particles and the
Universe (KMI), Nagoya University, Nagoya, 464-8602, Japan}
\affiliation{Institute for Advanced Research, Nagoya University, Nagoya
464-8601, Japan}
\affiliation{Kavli Institute for the Physics and Mathematics of the Universe (WPI), The University of Tokyo Institutes for Advanced Study (UTIAS), The University of Tokyo, Chiba 277-8583, Japan}

\author{Arun~Kannawadi\orcidlink{0000-0001-8783-6529}}
\affiliation{Department of Astrophysical Sciences, Princeton University, Princeton, NJ 08544, USA}

 \author{Masato~Shirasaki\orcidlink{0000-0002-1706-5797}}
\affiliation{National Astronomical Observatory of Japan, National Institutes of
Natural Sciences, Mitaka, Tokyo 181-8588, Japan}
\affiliation{The Institute of Statistical Mathematics, Tachikawa, Tokyo
190-8562, Japan}

\author{Takanori~Taniguchi}
\affiliation{Kavli Institute for the Physics and Mathematics of the Universe (WPI), The University of Tokyo Institutes for Advanced Study (UTIAS), The University of Tokyo, Chiba 277-8583, Japan}
\affiliation{Department of Physics, The University of Tokyo, Bunkyo, Tokyo 113-0031, Japan}

\author{Ryuichi~Takahashi}
\affiliation{Faculty of Science and Technology, Hirosaki University, 3 Bunkyo-cho, Hirosaki, Aomori 036-8561, Japan}

\author{Ken~Osato\orcidlink{0000-0002-7934-2569}}
\affiliation{Center for Frontier Science, Chiba University, Chiba 263-8522, Japan}
\affiliation{Department of Physics, Graduate School of Science, Chiba University, Chiba 263-8522, Japan}

\author{Takashi~Hamana}
\affiliation{National Astronomical Observatory of Japan, National Institutes of
Natural Sciences, Mitaka, Tokyo 181-8588, Japan}

\author{Masamune~Oguri\orcidlink{0000-0003-3484-399X}}
\affiliation{Center for Frontier Science, Chiba University, Chiba 263-8522, Japan}
\affiliation{Department of Physics, Graduate School of Science, Chiba University, Chiba 263-8522, Japan}
\affiliation{Kavli Institute for the Physics and Mathematics of the Universe (WPI), The University of Tokyo Institutes for Advanced Study (UTIAS), The University of Tokyo, Chiba 277-8583, Japan}

\author{Atsushi~J.~Nishizawa\orcidlink{0000-0002-6109-2397}}
\affiliation{Gifu Shotoku Gakuen University, 1-1 Takakuwanishi, Yanaizu, Gifu, 501-6194, Japan}
\affiliation{Kobayashi-Maskawa Institute for the Origin of Particles and the Universe (KMI), Nagoya University, Nagoya, 464-8602, Japan}

\author{Andr\'es~A.~Plazas Malag\'on\orcidlink{0000-0002-2598-0514}}
\affiliation{Kavli Institute for Particle Astrophysics and Cosmology, P.O. Box 20450, MS29, Stanford, CA 94309, USA}
\affiliation{SLAC National Accelerator Laboratory, 2575 Sand Hill Road, MS29, Menlo Park, CA  94025, USA}
\affiliation{Department of Astrophysical Sciences, Princeton University, Princeton, NJ 08544, USA}

\author{Tomomi~Sunayama\orcidlink{0009-0004-6387-5784}}
\affiliation{Department of Astronomy and Steward Observatory, University of Arizona, 933 N Cherry Ave, Tucson, AZ 85719, USA}
\affiliation{Kobayashi-Maskawa Institute for the Origin of Particles and the Universe (KMI), Nagoya University, Nagoya, 464-8602, Japan}

\author{David~Alonso\orcidlink{0000-0002-4598-9719}} 
\affiliation{Department of Physics, University of Oxford, Denys Wilkinson Building, Keble Road, Oxford OX1 3RH, United Kingdom}

\author{An\v{z}e~Slosar\orcidlink{0000-0002-8713-3695}}
\affiliation{Brookhaven National Laboratory, Upton, NY 11973, USA}

 \author{Robert~Armstrong}
\affiliation{Lawrence Livermore National Laboratory, Livermore, CA 94551, USA}

 \author{James~Bosch\orcidlink{0000-0003-2759-5764}}
\affiliation{Department of Astrophysical Sciences, Princeton University, Princeton, NJ 08544, USA}

\author{Yutaka~Komiyama\orcidlink{0000-0002-3852-6329}}
\affiliation{Department of Advanced Sciences, Faculty of Science and Engineering, Hosei University, 3-7-2 Kajino-cho, Koganei-shi, Tokyo 184-8584, Japan}

\author{Robert~H.~Lupton\orcidlink{0000-0003-1666-0962}}
\affiliation{Department of Astrophysical Sciences, Princeton University, Princeton, NJ 08544, USA}

\author{Nate~B.~Lust\orcidlink{0000-0002-4122-9384}}
\affiliation{Department of Astrophysical Sciences, Princeton University, Princeton, NJ 08544, USA}

\author{Lauren~A.~MacArthur}
\affiliation{Department of Astrophysical Sciences, Princeton University, Princeton, NJ 08544, USA}

\author{Satoshi Miyazaki\orcidlink{0000-0002-1962-904X}}
\affiliation{Subaru Telescope,  National Astronomical Observatory of Japan, 650 N Aohoku Place Hilo HI 96720 USA}

\author{Hitoshi~Murayama\orcidlink{0000-0001-5769-9471}}
\affiliation{Berkeley Center for Theoretical Physics, University of California, Berkeley, CA 94720, USA}
\affiliation{Theory Group, Lawrence Berkeley National Laboratory, Berkeley, CA 94720, USA}
\affiliation{Kavli Institute for the Physics and Mathematics of the Universe (WPI), The University of Tokyo Institutes for Advanced Study (UTIAS), The University of Tokyo, Chiba 277-8583, Japan}

 \author{Takahiro~Nishimichi\orcidlink{0000-0002-9664-0760}}
 \affiliation{Center for Gravitational Physics, Yukawa Institute for Theoretical
Physics, Kyoto University, Kyoto 606-8502, Japan}
\affiliation{Kavli Institute for the Physics and Mathematics of the Universe (WPI), The University of Tokyo Institutes for Advanced Study (UTIAS), The University of Tokyo, Chiba 277-8583, Japan}
\affiliation{Department of Astrophysics and Atmospheric Sciences, Faculty of Science, Kyoto Sangyo University, Motoyama, Kamigamo, Kita-ku, Kyoto 603-8555, Japan}

\author{Yuki~Okura\orcidlink{0000-0001-6623-4190}}
\affiliation{National Astronomical Observatory of Japan, National Institutes of Natural Sciences, Mitaka, Tokyo 181-8588, Japan}

\author{Paul~A.~Price\orcidlink{0000-0003-0511-0228}}
\affiliation{Department of Astrophysical Sciences, Princeton University, Princeton, NJ 08544, USA}

\author{Philip~J.~Tait}
\affiliation{Subaru Telescope,  National Astronomical Observatory of Japan, 650 N Aohoku Place Hilo HI 96720 USA}

\author{Masayuki~Tanaka}
\affiliation{National Astronomical Observatory of Japan, National Institutes of Natural Sciences, Mitaka, Tokyo 181-8588, Japan}

\author{Shiang-Yu~Wang}
\affiliation{Institute of Astronomy and Astrophysics, Academia Sinica, Taipei 10617, Taiwan}

\date{\today}

\begin{abstract}
We measure weak lensing cosmic shear power spectra from the three-year galaxy shear catalog of the Hyper Suprime-Cam (HSC) Subaru Strategic Program imaging survey. The shear catalog covers $416 \ \mathrm{deg}^2$ of the northern sky, with a mean $i$-band seeing of 0.59 arcsec and an effective galaxy number density of 15 $\mathrm{arcmin}^{-2}$ within our adopted redshift range. With an $i$-band magnitude limit of 24.5 mag, and four tomographic redshift bins spanning $0.3 \leq z_{\mathrm{ph}} \leq 1.5$ based on photometric redshifts, we obtain a high-significance measurement of the cosmic shear power spectra, with a signal-to-noise ratio of approximately 26.4 in the multipole range $300<\ell<1800$. The accuracy of our power spectrum measurement is tested against realistic mock shear catalogs, and we use these catalogs to get a reliable measurement of the covariance of the power spectrum measurements. We use a robust blinding procedure to avoid confirmation bias, and model various uncertainties  and sources of bias in our analysis, including point spread function systematics, redshift distribution uncertainties, the intrinsic alignment of galaxies and the modeling of the matter power spectrum. For a flat $\Lambda$CDM model, we find $S_8 \equiv \sigma_8 (\Omega_m/0.3)^{0.5} =0.776^{+0.032}_{-0.033}$, which is in excellent agreement with the constraints from the other HSC Year 3 cosmology analyses, as well as those from a number of other cosmic shear experiments. This result implies a $\sim$$2\sigma$-level tension with the \textit{Planck} 2018 cosmology \cite{Planck2018Cosmology}. We study the effect that various systematic errors and modeling choices could have on this value, and find that they can shift the best-fit value of $S_8$ by no more than $\sim$$0.5\sigma$, indicating that our result is robust to such systematics.
% \begin{description}
% Secondary publications and information retrieval purposes.
% You may use the \texttt{description} environment to structure your abstract;
% use the optional argument of the \verb+\item+ command to give the category of each item. 
% \end{description}
\end{abstract}

\maketitle

\section{\label{sec:introduction} Introduction}

The $\Lambda$ Cold Dark Matter ($\Lambda$CDM) model is considered the standard model for describing both the expansion history and the growth of large-scale structure of the universe. The parameters of the $\Lambda$CDM model have been measured to percent-level precision using a number of cosmological probes \cite{Weinberg2013}, including the cosmic microwave background (CMB) \cite{WMAP9, Planck2018Cosmology, ACTDR4Cosmology, SPT2022} and Type Ia supernovae \cite{Suzuki2012, Betoule2014, Scolnic2018, DES_Supernovae2019}, as well as the distribution of galaxies and other tracers of large-scale structure (LSS) \cite{Anderson2014, Alam2017, Alam2021, DESY3_3x2_2022, Hikage2019, Hamana2020, Amon2022, Secco2022, Doux2021, Asgari2021, Loureiro2022, Heymans2021}.

With these measurements, $\Lambda$CDM has been shown to be consistent across a wide range of cosmological experiments. However, as the precision of these measurements has grown, tensions have begun to emerge. In particular, a $4-5 \sigma$  tension has been seen between measurements of the Hubble constant \cite{DiValentino2021}, $\mathrm{H}_0$, from the local universe cosmic distance ladder \cite{Riess2022} and the early-universe CMB \cite{Planck2018Cosmology}. 
Cosmological surveys have also observed a $2-3 \sigma$ tension in measurements of the parameter $S_8 \equiv \sigma_8 (\Omega_m/0.3)^{0.5}$ \cite{Planck2018Cosmology, Hikage2019, Hamana2020, Amon2022, Secco2022, Doux2021, Asgari2021, Loureiro2022, GarciaGarcia2021}, where $\sigma_8$ is the root mean square variation in the mass in spheres of radius 8 $\mathrm{Mpc}/h$ (where $h\equiv \mathrm{H}_0/100$), and $\Omega_m$ is the matter density of the universe. Larger, improved data sets as well as more accurate and precise measurement and modeling methods will help us better understand these tensions, namely whether they suggest a need for new physics, or whether they are caused by unaccounted for, or as-yet unknown, systematic effects. 

Some of the strongest LSS constraints on $S_8$ come from the study of cosmic shear, i.e. the weak gravitational lensing of distant galaxies by the LSS along the line of sight \cite{Kilbinger2015, Mandelbaum2018_review}. These small, coherent distortions of galaxy shapes are sensitive to both the strength of matter density fluctuations and the growth of these fluctuations over cosmic time. This makes cosmic shear a powerful cosmological probe. Cosmology constraints from cosmic shear are known to be degenerate in the $\Omega_m$-$\sigma_8$ plane, but provide strong constraints on $S_8$, which is defined perpendicular to this degeneracy direction.

Measurements of cosmic shear, in practice, use two-point statistics, which describe the correlation between the shear of a galaxy at a given position with the shear of galaxies some distance away. Two methods by which this information can be summarized are the two-point correlation function (2PCF or $\xi_{\pm}$) and its Fourier transform, the angular power spectrum ($\Cl$). Other summary statistics include  Complete Orthogonal Sets of E/B-Integrals (COSEBIs, \cite{Schneider2010}) and bandpower estimates derived from correlation functions \cite{Schneider2002a, Becker2016, vanUitert2018}. This paper uses cosmic shear angular power spectra, while a companion paper \cite{Li2023} uses the two-point correlation function. These two statistics contain the same information in principle, but in practice, differ in terms of effective scale cuts, and sensitivity to different scales and systematic effects. For this reason, it is important that surveys carry out measurements of the cosmic shear signal and subsequent cosmological analyses using both sets of summary statistics.

Measurements of the cosmic shear signal require deep optical or near-infrared imaging with good seeing over large solid angles in multiple bands. The first detections of cosmic shear were made just over two decades ago \cite{Bacon2000, VanWaerbeke2000,Wittman2000, Kaiser2000}, and a number of studies over the past years have carried out analyses with $\sim$$10\%$ level precision (e.g. \cite{Massey2007, Lin2012, Heymans2012, Jee2016}). Today, cosmic shear analyses are being carried out by surveys covering larger sky areas, collectively known as ``Stage III" surveys \cite{Albrecht2006}. These include the Kilo-Degree Survey (KiDS; \cite{Kuijken2015}), the Dark Energy Survey (DES; \cite{DESAbbott2016, DESBecker2016}) and the Hyper Suprime-Cam (HSC) survey \cite{Aihara2018a, Aihara2018b}. As mentioned above, the measurements of $S_8$ from these surveys appear to be in $2-3 \sigma$ tension with those from the CMB. In the flat $\Lambda$CDM model, the \textit{Planck} temperature and polarization power spectra (excluding CMB lensing) constrain $S_8 = 0.834 \pm 0.016$ \cite{Planck2018Cosmology}. The HSC Year 1 analysis using angular power spectra constrains  $S_8 = 0.780^{+0.030}_{-0.033}$ \cite{Hikage2019}, while the two point correlation function analysis finds $S_8 = 0.823^{+0.032}_{-0.028}$ \cite{Hamana2020}. The most recent DES Year 3 analyses measure $S_8 = 0.759^{+0.025}_{-0.023}$ using 2PCFs \cite{Amon2022, Secco2022} and $S_8 = 0.793^{+0.038}_{-0.025}$ using $\Cl$s \cite{Doux2022}. The KiDS-1000 analysis with 2PCFs constrains $S_8 = 0.759^{+0.024}_{-0.021}$ \cite{Asgari2021} while the analysis with $\Cl$s finds $S_8 = 0.754^{+0.027}_{-0.029}$ \cite{Loureiro2022}. In this paper, we present a new measurement of $S_8$ using weak lensing data from the Year 3 data release of the HSC survey, along with careful modeling of systematics, as a step forward in understanding the $S_8$ tension. 

The Hyper Suprime-Cam Subaru Strategic Program (HSC-SSP; hereafter the HSC survey) is an imaging survey using the 8.2 m Subaru telescope \cite{Aihara2018b}. The Hyper Suprime-Cam is a wide-field camera with 870 Megapixels covering a 1.5 deg diameter field of view \cite{Miyazaki2012, Miyazaki2018, Komiyama2018, Furusawa2018}.  In this paper, we use data from the wide layer of the survey (although it also has deep and ultradeep components), which is observed in the \textit{grizy} broad-band filters \cite{Kawanomoto2018}. HSC is a very deep survey (a $5\sigma$ point-source depth in the wide layer of $i \sim 26 \ \mathrm{mag}$), with excellent seeing (a median $i$-band seeing of $\sim$$0.59''$). This allows the measurement of cosmic shear signals up to higher redshifts with higher accuracy than other Stage III surveys. Moreover, the HSC depth is just $\sim$$1$ magnitude shallower than the 10-year Wide, Fast, Deep survey depth of the Vera C. Rubin Observatory Legacy Survey of Space and Time (LSST; \cite{Ivezic2019}). This makes HSC the Stage III survey that, in terms of depth and galaxy number density, most closely resembles Stage IV cosmological surveys, including the Rubin LSST \cite{LSST2012}, \textit{Euclid} \cite{Euclid2011} and the \textit{Nancy Grace Roman Space Telescope} \cite{Roman2015, Roman2019}. 

There have been weak lensing studies with earlier incarnations of the HSC data, using the S16A (Year 1) internal data release, which covered $137 \ \mathrm{deg}^2$ of the sky \cite{Hikage2019, Hamana2020, Sugiyama2022, Miyatake2022, Hamana2022, Oguri2018}). The survey has continued, adding additional area and observations in each filter. In this work, we use data from the S19A internal data release, hereafter referred to as the Year 3, or Y3, data. This includes data collected between March 2014 and April 2019, and is part of the third public data release from the survey \cite{Aihara2022}. The Year 3 shear catalog for weak lensing science, presented in \cite{Li2022}, is based on this data release, and covers $416 \ \mathrm{deg}^2$ of the northern sky, with an effective galaxy number density of 15 $\mathrm{arcmin}^{-2}$ in the redshift range used in this analysis. By calibrating the measured galaxy shapes with image simulations, the galaxy property-dependent shear estimation bias is removed to the level of $|\delta m| < 9 \times 10^{-3}$ \cite{Li2022}. \cite{Li2022} further carry out a number of null tests for systematics related to Point Spread Function (PSF) modeling and shear estimation, and demonstrate that the shear catalog meets the requirements set out for using these data for cosmological analyses. 

In this paper, we present results from a tomographic cosmic shear analysis in harmonic space using the HSC-Y3 shear catalog. The use of tomography, i.e. adding the redshift information of source galaxies to the measurement, allows us to improve cosmological constraints by breaking parameter degeneracies \cite{Hu1999, Takada2004}. Our cosmic shear power spectrum measurement shows no significant detection of a $B$-mode signal, indicating that our measurement is not impacted by systematics such as those related to the PSF. In the process of our cosmological analysis, we have been careful to account for various sources of systematic error, including residual biases in our measurements, and uncertainties in our theoretical modeling. We show that our results are robust to these modeling choices. We also conduct a number of internal consistency checks to show that our results remain consistent across various splits of the data.  

This paper is accompanied by several other HSC-Y3 cosmology analysis papers, including  3$\times$2pt analyses combining cosmic shear, galaxy clustering, and galaxy-galaxy lensing \cite{Sugiyama2023, Miyatake2023, More2023}, as well as a cosmic shear analysis using two point correlation functions (2PCFs) \cite{Li2023}. In particular, we are careful to unify, as much as possible, the analysis choices between this analysis and the 2PCF analysis, facilitating the comparison of results between the two methods.

This paper is organized as follows. Section~\ref{sec:data} describes the data used in this paper, including the shear catalog, as well as our blinding procedure used to prevent confirmation bias from affecting our results. In Section~\ref{sec:measurement_methods}, we describe the measurement of our data vector using the Pseudo-$\Cl$ method to correct for biases due to incomplete sky coverage. We show the resulting measurement, as well as other ingredients in our analysis (the covariance matrix, source redshift distribution, and PSF systematics measurements) in Section~\ref{sec:measurement_results}. We describe our modeling and analysis choices in Section~\ref{sec:modeling}. Our cosmological constraints and their robustness to different systematics are presented in Section~\ref{sec:results}. Finally, we give our conclusions in Section~\ref{sec:conclusion}.

Throughout this paper, we quote the mode of the posterior distribution, along with 68\% credible intervals for parameter values and uncertainties, unless otherwise stated. We also report ``MAP'' values of parameters, where for notational convenience, we use ``MAP'' to refer to both the maximum a posteriori point (the maximum posterior of the parameter space, as determined using an optimization algorithm) as well as the maximum posterior (a noisy estimate of the maximum a posteriori based on the point in a sampling chain with the maximum posterior value). We assume a standard $\Lambda$CDM cosmological model, with no curvature ($\Omega_k = 0$), adiabatic Gaussian initial conditions, and a dark energy equation of state parameter $w=-1$. When discussing galaxy photometry, we quote \texttt{cmodel} magnitudes \cite{Bosch2018} on an AB system, with the extinction correction from \cite{Schlegel1998}. The figures presented throughout the paper, unless otherwise noted, were made prior to unblinding the analysis (see Section~\ref{sec:blinding}), with only the axes updated to show true parameter values after unblinding. 

\section{\label{sec:data} HSC Year 3 Data}

The Hyper Suprime-Cam Subaru Strategic Program \cite{Aihara2018b} is an imaging survey which uses the 8.2 m Subaru Telescope, along with the Hyper Suprime-Cam wide-field camera. The camera has 870 Megapixels covering a 1.5 deg diameter field of view \cite{Miyazaki2012, Miyazaki2018, Komiyama2018, Furusawa2018}. The survey consists of a wide, a deep and an ultradeep layer, each observed in the \textit{grizy} broad-band filters \cite{Kawanomoto2018}, along with narrowband filters in the deep and ultradeep layers. The median seeing for the wide layer in the \textit{i} band is 0.59 arcsec, allowing for excellent image quality for galaxy shape measurements. In this paper, we use the wide survey, which, for this data release, covers an area of $\sim$450 $\mathrm{deg}^2$ with a point source $5\sigma$ depth of $r \sim 26$ mag. The survey is spread out over six distinct fields, five of which are equatorial, which will eventually be combined into three fields with the added sky coverage from the final data release.

The survey data is reduced by a pipeline, presented in \cite{Bosch2018}, that has been developed in parallel with the pipeline for the Vera C. Rubin Observatory Legacy Survey of Space and Time (the LSST science pipelines) \cite{Ivezic2019, Juric2017, Bosch2019}. Astrometric and photometric calibrations are carried out by comparison with data from the Pan-STARRS1 survey \cite{Chambers2016}. The photometric calibration is accurate to $\sim$$1\%$ \cite{Aihara2019}, while the astrometric calibration accuracy is $\sim$$40 \ \mathrm{mas}$ \cite{Aihara2018a}. There have been three public data releases from the survey, presented in \cite{Aihara2018a, Aihara2019, Aihara2022}. 

Here, we describe the different data components that enter this analysis, including the shear catalog (Section~\ref{sec:shape_catalog}) and the blinding strategy (Section~\ref{sec:blinding}), as well as the photometric redshift catalogs, which are used for the source redshift distribution inference (Section~\ref{sec:photoz}), the star catalogs for quantifying PSF systematics (Section~\ref{sec:star_cat}), and the mock catalogs from which the covariance matrix is measured (Section~\ref{sec:mocks}). The analysis in this paper is done in parallel with an analysis using cosmic shear two point correlation functions \cite{Li2023}. As these analyses share the same data sets, much of the discussion in the sections below can also be found in \cite{Li2023}.

\subsection{\label{sec:shape_catalog} HSC-Y3 Shear Catalog}

The HSC-Y3 shear catalog is based on data from the S19A internal data release of the HSC survey, consisting of data taken between March 2014 and April 2019. The catalog is described in detail in \cite{Li2022}; we provide a brief summary below. 

In constructing the shear catalog, a number of cuts were applied to the S19A data in order to obtain a catalog which satisfied the requirements (defined in \cite{Li2022}) for carrying out cosmological weak lensing analyses. This includes considering only Full Depth Full Color regions of the sky, i.e. those reaching the approximate full depth of the survey in all five broadband filters, and thus allowing the measurement of accurate 
shapes and photometric redshifts for the galaxies. We use a magnitude-limited sample in the $i$ band, with a \texttt{cmodel} magnitude $i<24.5 \ \mathrm{mag}$ (see \cite{Bosch2018} for the definition of \texttt{cmodel} magnitudes in the context of HSC). Given the $5\sigma$ point source depth of $i = 26.2 \ \mathrm{mag}$, this is a relatively conservative magnitude cut \cite{Aihara2022}. We use additional cuts, detailed in Table 2 of \cite{Li2022}, to ensure that the sample is not contaminated by galaxies whose shapes are difficult to measure, including a signal-to-noise ratio cut in the $i$ band, requirements on the blendedness, shape measurement error, and the extent to which the galaxy is resolved compared to the PSF, as well as cuts on bad pixels and masks on regions of the sky around bright stars. The catalog covers $433.48 \ \mathrm{deg}^2$ of the sky, split into six disconnected fields: XMM, GAMA15H, HECTOMAP, GAMA09H, VVDS and WIDE12H, in order of increasing area. Five of these regions are on the celestial equator, and one (HECTOMAP) is at a declination of $\delta \sim +43^{\circ}$ (see Figure 2 of \cite{Li2022}). 

After the shear catalog was presented in \cite{Li2022}, we introduced a few additional cuts to improve the quality of the data for weak lensing science. In particular, we apply a cut to remove potential contamination from unresolved binary stars, following \cite{Hildebrandt2017}: we remove objects with an extremely large $i$-band ellipticity, $|e|>0.8$ and an $i$-band determinant radius $r_{\mathrm{det}} < 10^{-0.1r+1.8}$ arcsec (where $r$ is the $r$ band magnitude), amounting to 0.46\% of the sample. Additionally, as described in detail in Section~\ref{sec:photoz}, we remove 12\% of the galaxies in the shear catalog due to difficulties in calibrating their redshifts. Finally, we find that a small ($\sim$$20 \ \mathrm{deg}^2$) region of sky in the GAMA09H field has large PSF model residuals when looking at the fourth moments of the PSF (see Figure B1 of \cite{Zhang2022}). This is due to exceptionally good seeing in this region,  $\sim$$0.4"$. We remove this region due to our inability to accurately model the PSF in this limit. After these cuts, the final shear catalog contains $\sim$$25$ million galaxies over $416 \ \mathrm{deg}^2$ of the sky.

Given this galaxy sample, the galaxy shapes are estimated on the $i$-band coadded images using the re-Gaussianization (\texttt{reGauss}) shape measurement method \cite{Hirata2003}. This method measures the two components of galaxy ellipticity:
\begin{equation}
    (e_1, e_2) = \frac{1-(r_b/r_a)^2}{1+(r_b/r_a)^2} (\cos 2\phi, \sin 2\phi),
\end{equation}
where $r_b/r_a$ is the observed minor-to-major axis ratio and $\phi$ is the position angle of the major axis with respect to the equatorial coordinate system. The shear of each galaxy can then be estimated from the measured ellipticity:
\begin{equation}
    \check{\gamma}_{\alpha} = \frac{1}{1+\left<m\right>}\left(\frac{e_{\alpha}}{2\mathcal{R}} - ae^\text{psf}_{\alpha}\right),
    \label{eq:shear_ellip}
\end{equation}
where $\alpha = 1,2$. Here, $\left<m\right>$ and $ae^\text{psf}_{\alpha}$ are the multiplicative and additive biases in shear estimation (where $a$ is the fractional additive bias), and $\mathcal{R}$ is the responsivity, i.e. the response of the
average galaxy ellipticity to a small shear distortion \cite{Kaiser1995, Bernstein2002}, and is given by:
\begin{equation}
    \mathcal{R} = 1 - \frac{\sum_i w_i e^2_{\mathrm{rms}, i}}{\sum_i w_i},
\end{equation}
where $e_{\mathrm{rms}, i}$ is the intrinsic root mean square ellipticity per component for galaxy $i$, and the weight $w_i$ is the inverse variance of the shape noise:
\begin{equation}
    w_i = \left(\sigma_{\mathrm{e}, i}^2 + e^2_{\mathrm{rms}, i}\right)^{-1}.
    \label{eq:gal_weight}
\end{equation}
Here $\sigma_{\mathrm{e}, i}$ is the shape measurement error for each galaxy. 

The multiplicative and additive shear estimation biases introduced in Equation~\ref{eq:shear_ellip} are estimated for each object using image simulations which downgrade Hubble Space Telescope images from the COSMOS region to HSC survey quality. The image simulations are described in detail in \cite{Mandelbaum2018_image, Li2022}. The additive bias is corrected for each object, while the multiplicative bias is corrected using a weighted average, $\left<m\right>$, over the ensemble of galaxies in each field, in each tomographic bin. 

A bias may also be introduced by selection cuts that correlate with the true lensing shear and/or anisotropic PSF systematics (see \cite{Li2022} for more details). The bias caused by the former is multiplicative ($m^{\mathrm{sel}}$), while that caused by the latter is additive ($a^{\mathrm{sel}}$). 

These biases are also measured from the image simulations described in \cite{Li2022}, and are used to correct the estimated shear, using a weighted average over the ensemble of galaxies in each tomographic bin:
\begin{equation}
    \hat{\gamma}_\alpha =
    \frac{\check{\gamma}_\alpha - \hat{c}^\text{sel}_\alpha}{1+m^\text{sel}},
\end{equation}
where
\begin{equation}
    \hat{c}^\text{sel}_\alpha =
    \frac{ a^\text{sel} \sum_{i}
    w_i e^\text{psf}_{\alpha,i}}{\sum_i w_i}.
\end{equation}
Here, $\bm{e}^{\mathrm{psf}}_i$ is the PSF ellipticity for galaxy $i$. The estimated galaxy shear of a given object is then:
\begin{equation}
    \gamma_{\alpha} = \frac{1}{1+m^{\mathrm{sel}}}\left(\frac{1}{1+\left<m\right>}\left(\frac{e_{\alpha}}{2\mathcal{R}} - ae^\text{psf}_{\alpha}\right) - \hat{c}^{\mathrm{sel}}_{\alpha}\right).
\end{equation}

\subsection{\label{sec:blinding} Blinding Strategy}

In the era of precision cosmology, with a persistent tension in certain measurements between different probes, it is essential to protect one's analysis against experimenters' biases, particularly confirmation bias. To avoid confirmation bias in the HSC cosmology analyses, we proceed with the analyses in a blinded manner. This is done by adding a random multiplicative bias to the values in the shear catalog. We use a two-level blinding scheme, such that:
\begin{equation}
    \bm{m}^i_{\mathrm{cat}} = \bm{m}_{\mathrm{true}} + \mathrm{d}\bm{m}_1^i + \mathrm{d}\bm{m}_2^i, 
\end{equation}
where $\bm{m}_{\mathrm{true}}$ is the actual multiplicative bias estimated from image simulations, and the index $i$ runs from 0 to 2, resulting in three different shear catalogs for each analysis. Each analysis uses catalogs with different values of $\mathrm{d}\bm{m}_1$ and $\mathrm{d}\bm{m}_2$.

The first blinding term, $\mathrm{d}\bm{m}_1$ is a user-level blinding, to prevent accidental unblinding via comparison of blinded catalogs across different analysis teams (where ``analysis team'' refers to each of the cosmology analyses simultaneously underway, including those described in \cite{Li2023, Sugiyama2023, Miyatake2023, More2023}. The values of $\mathrm{d}\bm{m}_1$ are encrypted with the public keys from the analysis lead of each team. These values are decrypted and subtracted from the multiplicative bias values for each catalog entry before the analysis begins. 

The collaboration-level blinding is done through $\mathrm{d}\bm{m}_2$. The values for $\mathrm{d}\bm{m}_2$ for the three blinded catalogs are randomly selected from the following three different choices ($\mathrm{d}\bm{m}_2^0$, $\mathrm{d}\bm{m}_2^1$, $\mathrm{d}\bm{m}_2^2$): $(-0.1, -0.05, 0.)$, $(-0.05, 0., 0.05)$, $(0., 0.05, 0.1)$. Thus, the true catalog (with $\mathrm{d}\bm{m}_2 = 0$) can be any of the three blinded catalogs. The values of $\mathrm{d}\bm{m}_2$ are encrypted by a public key from a person not involved in the cosmology analyses.

We carry out the same analysis on all three catalogs after subtracting $\mathrm{d}\bm{m}_1$ from each catalog. In addition to the catalog-level blinding, we also employ analysis-level blinding. We do not compare the cosmic shear power spectra obtained from any blinded catalog to theoretical predictions from known cosmological parameters. Additionally, prior to unblinding, any plots of cosmological constraints from the data are centered at zero, by subtracting the mean values from the chains. 

Prior to starting the analysis, we define a number of tests and conditions that must be passed prior to unblinding. These include code reviews by members of the collaboration, tests of the code with mock data sets (Appendix~\ref{app:cosmosis_validation}), as well as a number of tests of the cosmological constraints to ensure the goodness of fit, internal consistency and robustness to modeling choices (Sections~\ref{sec:results_fiducial}, \ref{sec:results_robustness} and \ref{sec:results_internal_consistency}). 

Once the collaboration agrees to unblind the analysis, the analysis-level unblinding is first removed by the analysis team. The final catalog-level unblinding happens a few hours later, and the analysis setup and fiducial results are not changed after unblinding. The figures shown in this paper, unless otherwise noted, were made prior to unblinding, with only the axes changed after unblinding to show true values.

\subsection{\label{sec:photoz} Photometric Redshift Catalogs}

The HSC S19A data include photometric redshift (photo-$z$) estimates based on three different photo-$z$ codes. These are \texttt{DNNz} \cite{Nishizawa_inprep} and \texttt{DEmPz} \cite{Hsieh2014, Tanaka2018}, which are both neural network-based conditional density estimation algorithms, as well as \texttt{mizuki}, which uses a Spectral Energy Distribution (SED) fitting technique \citep{Tanaka2015}. All three codes were trained with available spectroscopic redshifts as well as 30 band photometric redshifts from COSMOS2015 \cite{Laigle2016}. We refer the reader to \cite{Nishizawa2020} for further details.

As described in \cite{Rau2022}, the shear catalog galaxies are divided into four tomographic redshift bins, in the intervals $(0.3, 0.6]$, $(0.6, 0.9]$, $(0.9, 1.2]$ and $(1.2, 1.5]$, using the ``best'' photo-$z$ estimation by the \texttt{DNNz} algorithm.  Here, the ``best'' estimate refers to the point estimate where a given risk function is minimized (see Section~4.2 of \cite{Tanaka2018}). 

It was shown in \cite{Rau2022} that a number of galaxies have secondary solutions at $z \gtrsim 3.0$ in their redshift distributions from \texttt{DNNz} and \texttt{mizuki}. Since these secondary solutions lie outside of the redshift coverage of the Luminous Red Galaxy (LRG) sample used to calibrate our source redshift distribution (see Section~\ref{sec:n(z)}), we remove them from our sample. Galaxies with such secondary solutions are identified by the distance between the 2.5 and 97.5 percentiles of the \texttt{mizuki} and \texttt{DNNz} photo-$z$ PDF estimations:
\begin{equation}
    \left(z_{\text{0.975}; i}^\text{mizuki} - z_{\text{0.025}; i}^\text{mizuki}\right) < 2.7
    \quad \text{AND} \quad
    \left(z_{\text{0.975}; i}^\text{dnnz} - z_{\text{0.025}; i}^\text{dnnz}\right) < 2.7 \,,
\end{equation}
where $z_{\text{0.975}; i}^\text{mizuki (dnnz)}$ and $z_{\text{0.025};
i}^\text{mizuki (dnnz)}$ denote the $97.5$ and $2.5$ percentiles for galaxy
$i$ derived using the \texttt{mizuki} (\texttt{DNNz}) estimates of the photo-$z$ PDF.  This cut reduces the number of galaxies in the first redshift bin by 31\%, and the number of galaxies in the second redshift bin by 8\%. No galaxies are removed from the third and fourth bins.

After this cut, the number of galaxies in each tomographic bin, in order of increasing redshift, is 5,889,826, 8,445,233, 7,023,314, and 3,902,504 galaxies respectively.

\subsection{\label{sec:star_cat} Star Catalog}

We use a star catalog from Y3 data to estimate PSF systematics and their impact on our cosmological analysis, particularly PSF leakage due to imperfect shear estimation and PSF modeling error due to an incorrect PSF model (see Section~\ref{sec:psf_systematics}). We refer readers to \cite{Zhang2022} for the details of the star catalog selection.

The star catalog contains stars that were used for the PSF modeling of the co-added images in the HSC-Y3 data release (``PSF stars''), as well as stars that were not used in the process (``non-PSF stars''). We refer the reader to \cite{Bosch2018} for details regarding the selection of PSF stars. In the processing of single exposures, $\sim 20\%$ of stars are randomly selected to not be used in the PSF modeling, and are instead reserved for validation of the PSF model. Each $i$-band co-added image is made using at least four exposures. Different exposures will not necessarily use the same sets of PSF and non-PSF stars, so at the level of a co-added image, stars that were used in the PSF model in at least $20\%$ of the input visits are labeled as ``PSF stars'', while the others are labeled as ``non-PSF stars''.

We consider both PSF and non-PSF stars when estimating the impact of PSF systematics on our cosmic shear power spectrum measurement. As shown in \cite{Zhang2022}, we find that the additive bias to the power spectrum from PSF systematics is consistent for PSF and non-PSF stars. As the PSF star sample is much larger than the non-PSF star sample, the estimation of the PSF systematics from the former has a higher signal-to-noise ratio. For this reason, we use the PSF star sample in our PSF systematics estimation and modeling (described in Sections~\ref{sec:psf_systematics} and \ref{sec:psf_model}).

\subsection{\label{sec:mocks} Mock Catalogs}

We use a suite of 1404 mock shape catalogs to  measure the covariance matrix of our cosmic shear power spectra. These mock catalogs are generated following the method described in \cite{Shirasaki2019}, with updates to account for the new survey footprint, galaxy shape noise, shape measurement error, and photometric redshift error of the Y3 shear catalog, compared to the Year 1 version. 

In short, the mock catalogs use actual galaxy positions and shapes from the Y3 shear catalog, but erase the true lensing signal and impose a simulated lensing signal following the method described in \cite{Shirasaki2014,Shirasaki2017} and summarized here. The galaxies are populated in the mock catalog following their measured angular positions and redshifts (from \texttt{DNNz}) in the Y3 shape catalog. The galaxies are then rotated at random to erase any lensing or intrinsic alignment signal. Finally, the lensing distortion on each galaxy is simulated by adding the lensing contribution at each foreground lens plane, following the full-sky lensing simulations of \cite{Takahashi2017}. The full-sky lensing simulation is a ray-tracing simulation based on 108 realizations of $N$-body simulations using the WMAP9 cosmology \cite{WMAP9}. The light-ray deflection on the celestial sphere is calculated using the projected matter density field in $38$ spherical shells. Each shell has a radial thickness of $150~h^{-1}\mathrm{Mpc}$. The angular resolution of the shear map is $0.43~\mathrm{arcmin}$. To get a larger number of mock catalogs, each full-sky map is divided into 13 regions with the HSC-Y3 survey geometry, resulting in $108 \times 13 = 1404$ mock catalogs. After each galaxy's intrinsic shape is distorted based on the shear map, a shape measurement error is also added, which is generated from a zero-mean Gaussian distribution with the standard deviation measured in the HSC-Y3 shear catalog. These mock catalogs are then used to measure the covariance matrix of the cosmic shear power spectra, as described in Section~\ref{sec:covariance}. We note that unlike in the real universe, the galaxy positions in these mock catalogs are not correlated with the matter density in the lensing simulations, however we expect the impact of this to be negligible.

\section{\label{sec:measurement_methods} Cosmic Shear Power Spectrum Measurement Method}

The shear catalog allows us to make a map of the shear field, which is a spin-2 field, and therefore can be decomposed into two scalar fields on the basis of spherical harmonics as:
\begin{equation}
    (\gamma_1 \pm i\gamma_2)(\bm{\theta}) = - \sum_{\ell m}[E_{\ell m} \pm i B_{\ell m}] {}_{\pm 2}Y_{\ell m}(\bm{\theta}),
\end{equation}
where $_{s}Y_{\ell m}$ are the spin-weighted spherical harmonics \cite{Hikage2011}. This is a Fourier transform on a sphere. The $E$ mode is the curl-free component of the field, and the $B$ mode is the divergence-free component.

In order to fit a cosmological model to this shear field, we need to compress our data into a summary statistic. The statistical properties of the shear field can be measured through the cosmic shear angular power spectrum. The power spectrum is defined as the expectation value of the product of the spherical harmonic coefficients $\psi_{\ell m}$ and $\phi_{\ell m}$ (where $\psi$ and $\phi$ are either $E$ or $B$):
\begin{equation}
    \left<\psi_{\ell m}^{*} \phi_{\ell' m'}\right> = \delta_{\ell \ell'}\delta_{m m'} C_{\ell}^{\psi\phi}.
\end{equation}
For a full-sky map of the shear field, the power spectrum estimator averages over the $m$s for each $\ell$. An optimal estimator for the different power spectra on the full sky is:
\begin{equation}
    \hat{C}_{\ell}^{EE}= \frac{1}{2\ell +1} \sum_m E_{\ell m} E^{*}_{\ell m},
\end{equation}
\begin{equation}
    \hat{C}_{\ell}^{EB} = \frac{1}{2\ell +1} \sum_m E_{\ell m} B^{*}_{\ell m},
\end{equation}
\begin{equation}
    \hat{C}_{\ell}^{BB} = \frac{1}{2\ell +1} \sum_m B_{\ell m} B^{*}_{\ell m}.
\end{equation}

However, this method does not account for incomplete sky coverage of the map, due to the finite survey area as well as pixels that are masked due to bright stars or other reasons. In reality, we apply a survey mask, or weight map, $W(\bm{\theta})$ to the map, giving a observed, masked map of:
\begin{equation}
    \hat{\gamma}(\bm{\theta}) = W(\bm{\theta})\gamma(\bm{\theta}).
\end{equation}
The power spectrum obtained from a spherical harmonic transform of this pixelized shear field is biased due to the convolution with the survey window function $W$. The mask also leads to coupling between $E$ and $B$ modes. To correct for these effects, we estimate our angular power spectra using the pseudo-$\Cl$ formalism, developed for shear fields in \cite{Hikage2011} and implemented in \texttt{NaMaster} \cite{Alonso2019}. We briefly summarize the formalism here and refer the reader to \cite{Nicola2021} for a more detailed description.

For two shear fields $\hat{\bm{\gamma}}^{i} (\bm{\theta})$
 and $\hat{\bm{\gamma}}^{j} (\bm{\theta})$ in redshift bins $i$ and $j$, with survey window functions $W^i(\bm{\theta})$ and $W^j(\bm{\theta})$, the cross-power spectrum of the two fields, i.e. the pseudo-spectrum of the fields, has an expectation value:
 \begin{equation}
    \left<\tilde{\mathbf{C}}_{\ell}^{ij} \right> = \sum_{a,b,\ell’} \mathbf{M}^{(ij),(ab)}_{\ell\ell’} \mathbf{C}^{(ij),(ab)}_{\ell’}
    \label{eq:pseudo_cl}
 \end{equation}
where $a$ and $b$ refer to $E$ or $B$ modes, and $\mathbf{M}_{\ell\ell'}^{ij}$ is the mode-coupling matrix of the window functions, which describes how the window function correlates different multipoles, as well as the leakage between $E$ and $B$ modes. Here $\mathbf{C}_{\ell}^{ij}$ describes the cross-power spectrum of the two fields in the ideal of a full-sky survey. The mode-coupling matrix can be computed analytically from the spherical harmonic coefficients of the window functions (see \cite{Nicola2021} for details). We can estimate the power spectrum by inverting the mode coupling matrix, which requires binning in $\ell$. The estimator for the binned power spectrum is then:
 \begin{equation}
     \left<\hat{\mathbf{C}}_{L}^{ij}\right> = \sum_{a,b,L'} \left(\mathbf{M}^{(ij), (ab)}_{LL'}\right)^{-1} \tilde{\mathbf{C}}_{L'}^{(ij), (ab)},
 \end{equation}
where the binning is done by:
 \begin{equation}
     \mathbf{C}_{L'}^{(ij), (ab)} \equiv \sum_{\ell \in L} \omega_L^{\ell} \mathbf{C}_{\ell}^{(ij), (ab)},
 \end{equation} 
and
  \begin{equation}
     \mathbf{M}^{(ij),(ab)}_{LL'} \equiv \sum_{\ell \in L} \sum_{\ell' \in L'} \omega_L^{\ell} \omega_{L'}^{\ell'} \mathbf{M}^{(ij),(ab)}_{\ell \ell'}.
 \end{equation}
Here, $\omega_L^{\ell}$ is a set of weights defined for multipoles $\ell$ in bandpower $L$, normalized such that $\sum_{\ell \in L} \omega_L^{\ell} = 1$. In this work, we use an equal-weight binning scheme, such that $\omega_L^{\ell} = 1/|L|$ if $\ell \in L$, and 0 otherwise. The mean multipole of each bin is defined as $L' \equiv \sum_{\ell \in L} \omega_L^{\ell} \ell$.

Additionally, one must also subtract from the power spectrum the additive noise bias, which arises from shape noise (due to the intrinsic ellipticities of galaxies). We analytically estimate the constant binned noise pseudo-power spectrum following \cite{Nicola2021}:
 \begin{equation}
     \mathbf{N}_L = \Omega_{\mathrm{pix}} \left< \sum_{i \in \mathrm{pix}} w_i^2 \frac{e^2_{1, i} + e^2_{2, i}}{2} \right>_{\mathrm{pix}},
 \end{equation}
where $i$ now represents each galaxy in a given pixel, and $w_i$ is the weight of the galaxy, defined in Eq.~(\ref{eq:gal_weight}). One could also estimate this term empirically by creating many realizations of randomized galaxy shapes and taking the mean of the power spectra over the realizations. \cite{Nicola2021} show that the analytical and empirical approach agree to within $\sim$$3\%$ for the range of multipoles considered here, and that any disagreement is due to stochasticity in the empirical approach, rather than any biases in the analytic estimate.

In this work, we create a pixelized map of the shear field, $\gamma(\theta)$, for each of the six fields on the sky. We make use of the flat-sky approximation, as most fields are equatorial and cover a small area of the sky (between $33 \ \mathrm{deg}^2$ and $120 \ \mathrm{deg}^2$). \cite{Taniguchi_inprep} demonstrates that using this approximation is appropriate for the HSC-Y3 data. In making the maps, we follow \cite{Nicola2020} and \cite{Nicola2021}, and use a rectangular pixelization scheme which uses the Plate Carrée projection \cite{WCS}. In this scheme, pixels are defined in equal intervals of co-latitude $\theta$ and azimuth $\phi$. For this analysis, we use pixels of size 1 arcmin $\times$ 1 arcmin. In order to minimize the distortions caused by the flat-sky approximation, we place the projection reference point, i.e. a point on the equator ($\theta = \pi/2$), at the center of each field. To compute the power spectra, the galaxy weights, $w_i$, which are used as the mask $W(\bm{\theta})$, are taken from the Y3 shear catalog. We compute the auto- and cross-correlation cosmic shear power spectra for each tomographic redshift bin using an implementation of the pseudo-$\Cl$ method in the \texttt{NaMaster} code \cite{Alonso2019}. We measure the power spectra in 17 bins between $\ell_{\mathrm{min}} = 100$ and $\ell_{\mathrm{max}} = 15800$. These bins are approximately logarithmically spaced, with linear spacing at low multipoles to avoid bins that would be too narrow at large scales.

We measure $EE$, $EB$, and $BB$ power spectra for each of the ten combinations of our four redshift bins. Although we do not expect a cosmological $B$ mode signal (i.e. $\Cl^{EB}$ and $\Cl^{BB}$ should be zero), $B$ modes can be used to test for potential systematic effects in the data, including contamination by the PSF. For this reason, we preserve both components of the field, and use the $BB$ and $EB$ power spectra as a null test (see Section~\ref{sec:null_tests}). The incomplete sky coverage could cause a contamination from $B$ modes in the $E$ mode component of a given map, and vice versa \cite{Lewis2001, Bunn2003, Smith2006, Zhao2010, Kim2010, Bunn2011}. This mixing of $E$ and $B$ modes is fully described by the Pseudo-$\Cl$ method, and corrected for by the mode-coupling matrix. 

The power spectrum is not the only summary statistic that is commonly used for cosmic shear data. A companion paper \cite{Li2023} conducts a similar analysis using the cosmic shear angular two point correlation function (2PCF, or $\xi_{\pm}$), which is the Fourier transform of the power spectrum. While these two statistics, in principle, contain the same information, in practice, we employ scale cuts due to the finite survey area and theoretical uncertainties in the modeling of baryonic effects and intrinsic alignments. The scale cuts in 2PCFs and $\Cl$s do not directly translate, since the two statistics are related to each other through an integral of a Bessel function.
\begin{equation}
    \xi_{\pm} (\theta) = \frac{1}{2\pi} \int d\ell \ \ell \Cl J_{0/4}(\ell \theta).
\end{equation}
This means that a hard scale cut in real space corresponds to an oscillatory scale cut in Fourier space, and vice-versa (see \cite{Doux2021} for a detailed study of the consistency between the two statistics).

$\Cl$s and 2PCFs are complementary approaches to the cosmic shear analysis. Not only are the two statistics sensitive to different scales, they also use different treatments of observational effects, such as discrete galaxy sampling and the correction of irregular survey geometries. Moreover, the two approaches have different noise properties and different sensitivities to systematic effects. For these reasons, the HSC collaboration performs both sets of analyses in a coordinated manner. A comparison of the results of this power spectrum based analysis to the 2PCF analysis of \cite{Li2023} is presented in Section~\ref{sec:results_fiducial}.

\section{\label{sec:measurement_results} Measurements}

Here, we present in Section~\ref{sec:power_spectra} the measurement of tomographic cosmic shear power spectra from the HSC-Y3 shear catalog, following the method described in Section~\ref{sec:measurement_methods}. We additionally describe the measurement of the covariance matrix of these spectra from realistic mock catalogs (Section~\ref{sec:covariance}), as well as the results of the null tests carried out with the spectra and covariance (Section~\ref{sec:null_tests}). Finally, in Section~\ref{sec:psf_systematics}, we estimate the impact of PSF systematics on our cosmic shear measurement (as described in \cite{Zhang2022}), and in Section~\ref{sec:n(z)}, describe our source redshift distribution inference (presented in \cite{Rau2022}). 

\subsection{\label{sec:power_spectra} Cosmic Shear Power Spectra}

\begin{figure*}
\includegraphics[width=0.95\textwidth]{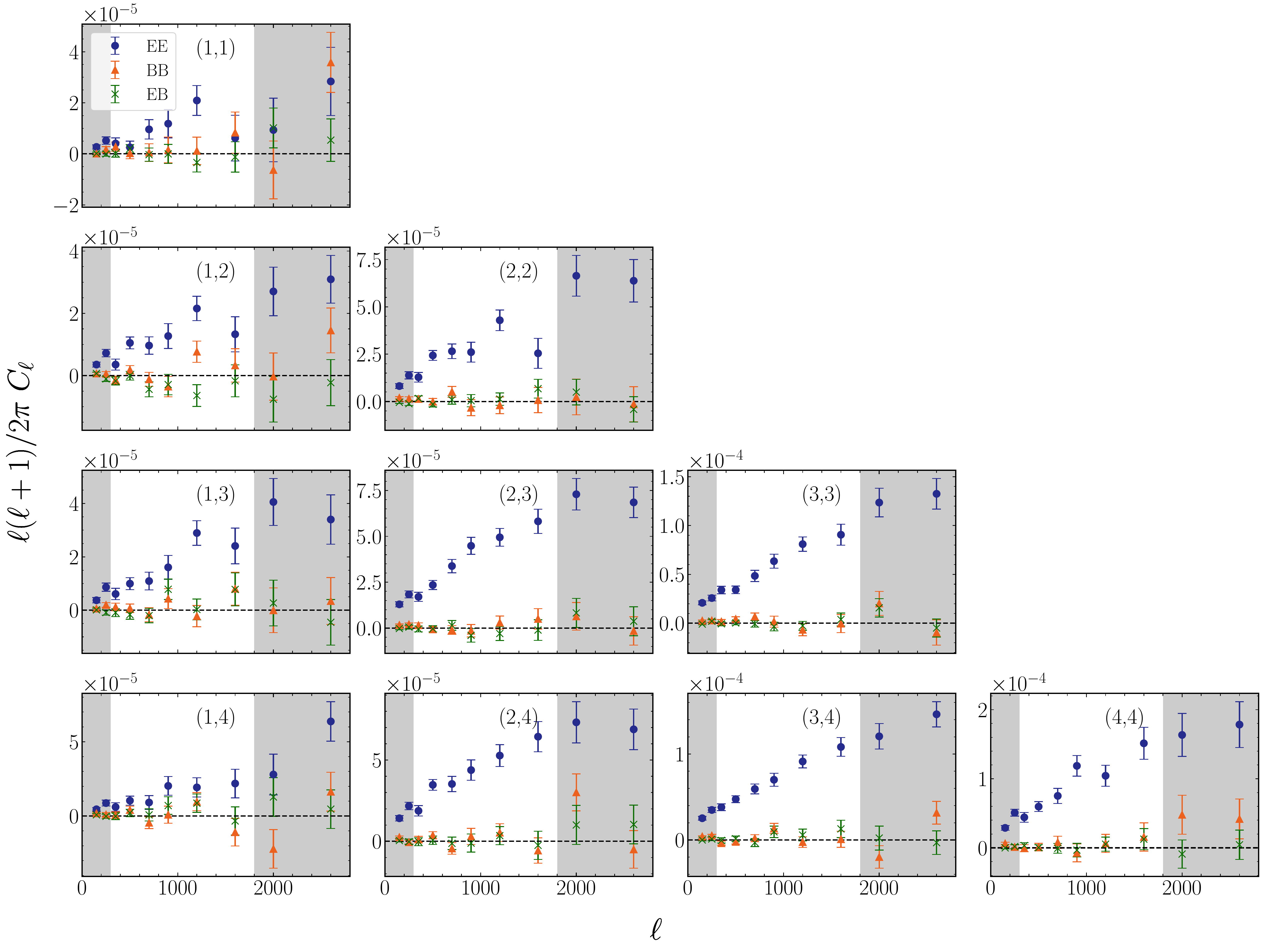}
\caption{\label{fig:power_spectra} Tomographic cosmic shear power spectra of $EE$ (blue circle), $BB$ (orange triangle), and $EB$ (green cross) modes. The galaxy sample is divided into four tomographic bins, with redshift ranges $(0.3, 0.6]$, $(0.6, 0.9]$, $(0.9, 1.2]$ and $(1.2, 1.5]$, using the ``best'' photo-$z$ estimation by the \texttt{DNNz} algorithm. These bins are referred to as bin numbers 1 to 4 respectively. The scales $\ell<300$ and $\ell>1800$ (shaded regions) are excluded in the cosmological analysis, based on the scale cuts described in Sections~\ref{sec:null_tests}, \ref{sec:model_sufficiency}, and \ref{sec:ia}. The combined total signal-to-noise ratio of the $EE$ spectra is 26.4 in the range of our fiducial scale cuts. Both the $BB$ and $EB$ spectra are consistent with zero.}
\end{figure*}

We use the pseudo-$\Cl$ method described in Section~\ref{sec:measurement_methods} to measure the tomographic cosmic shear power spectra from the HSC-Y3 shear catalog, including the $EE$, $BB$, and $EB$ spectra in 17 multipole bins between $\ell_{\mathrm{min}} = 100$ and $\ell_{\mathrm{max}} = 15800$. While we present, in Figure~\ref{fig:power_spectra}, the power spectra up to only $\ell_{\mathrm{max}}=3000$, we measure the power spectra and mode-coupling matrices to larger $\ell$ to correctly account for mode coupling between smaller scale modes and those we use in our analysis.

We first measure the power spectra, using the flat-sky approximation, in each of the six individual fields of the survey. We then co-add the spectra from individual fields using inverse variance weighting with the covariance described in Section~\ref{sec:covariance}. 

We use the fiducial scale cuts of $300 < \ell < 1800$ for our analysis. We cut large scales with $\ell < 300$ due to excess $B$ mode signals at these scales (see Section~\ref{sec:null_tests}). We also cut small scales with $\ell > 1800$ due to theoretical uncertainties in our modeling of these nonlinear scales (based on tests described in Sections~\ref{sec:model_sufficiency} and \ref{sec:ia}). This gives us six multipole bins that we use in our fiducial analysis, from the 17 that we originally measure.

We are able to obtain a high-significance measurement of the power spectrum, shown in Figure~\ref{fig:power_spectra}, with a total signal-to-noise ratio within our fiducial scale cuts of 26.4, as measured from the $\chi^2$ value of the data points. 

\subsection{\label{sec:covariance} Covariance Matrix}

We derive the covariance matrix of our cosmic shear power spectra using the mock catalogs described in Section~\ref{sec:mocks}. We do so by measuring the cosmic shear power spectra, following the same method as our measurement from data, from all 1404 realizations of the mocks, and then using these 1404 power spectra to compute the covariance matrix.

The cosmic shear signal in the mock catalogs is based on full-sky ray-tracing simulations that account for nonlinear structure formation \cite{Takahashi2017}, so the covariance estimated from mock catalogs includes both Gaussian and non-Gaussian components. Moreover, the mock catalog survey geometry is identical to that of the data, so the derived covariance also includes super-sample covariance \cite{Hamilton2006}. 

\begin{figure}
\includegraphics[width=0.45\textwidth]{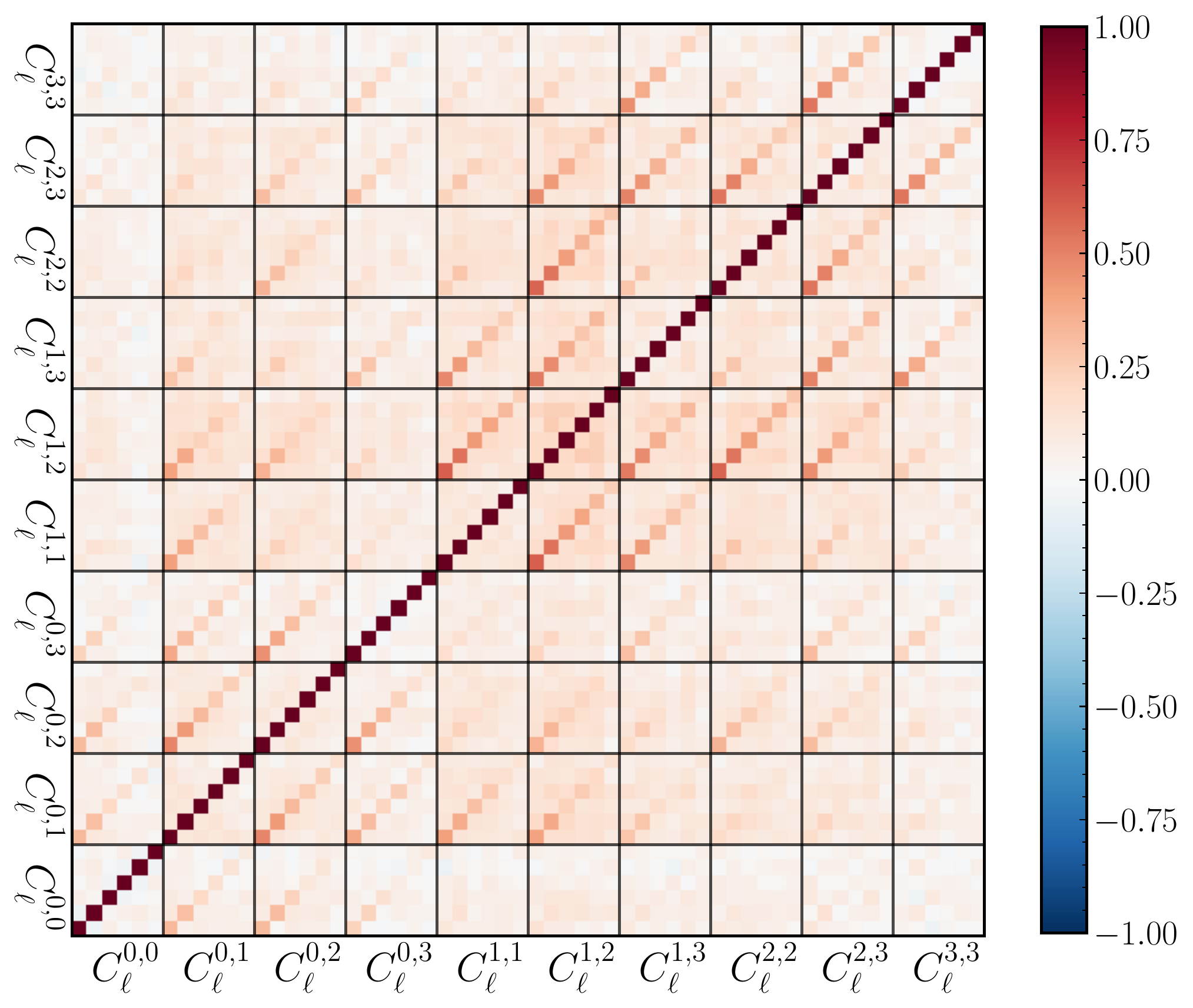}
\caption{\label{fig:covariance} Normalized covariance matrix (correlation coefficients), measured as described in Section~\ref{sec:covariance}, for the fiducial scale cuts $300 < \ell < 1800$, and the auto- and cross-correlations of four tomographic redshift bins (10 spectra in total, with 6 bins each in $\ell$).}
\end{figure}

We treat these mock catalogs exactly like our real data. We first measure the power spectra from each mock catalog on a field-by-field basis. We analytically compute the additive noise bias from shape noise, and subtract it from the measured power spectra. Finally, we co-add the spectra from each field for each mock realization using inverse variance weighting, with the individual covariances measured from the mock power spectra for each of the fields. We then use the 1404 co-added spectra to compute the covariance matrix. Figure~\ref{fig:covariance} shows the correlation matrix from this covariance matrix.

We do not account for the dependence of our covariance on cosmological parameters, since our mock catalogs are generated based on a set of lensing simulations \cite{Takahashi2017} that all adopt the WMAP9 cosmology \cite{WMAP9}. \cite{Kodwani2019} have shown that not accounting for this dependence will bias cosmological parameter constraints for current weak lensing surveys by at most $\sim 10\%$ of the statistical uncertainties. For this reason, our covariance estimation should be appropriate for this analysis. 

In addition, we correct for the effect of finite angular resolution, finite redshift resolution, and finite shell thickness in the lensing simulations (described in detail in \cite{Shirasaki2019}) by applying a correction factor to the measured $\Cl$s from the mock catalogs based on the ratio of the measured spectra to the theory prediction. This correction factor is computed for each redshift bin (1.14, 0.98, 0.95 and 0.99 in order of increasing redshift), and it is approximately constant across the range of scales considered. \cite{Shirasaki2019} show that such a correction, for the case of $\xi_+$, is reliable for scales larger than $\theta \sim 1 \ \mathrm{arcmin}$, or approximately $\ell \sim 10,000$. This limit is much smaller than the scales considered in this analysis.

The likelihood, described in Section~\ref{sec:sampler}, uses the inverse of the covariance matrix. To approximately account for bias in the inversion of the covariance matrix, due to the fact that a finite number of simulations were used to estimate the covariance, we multiply the inverse covariance by the Hartlap factor \cite{Hartlap2007}: $(1404-60-2)/(1404-1) = 0.96$, where 1404 is the number of mock realizations and 60 is the number of data points used in the analysis. This correction assumes Gaussian noise and statistically independent data vectors. 

\subsection{\label{sec:null_tests} \texorpdfstring{$B$}{B} Mode Null Test}

To first order, we do not expect to see any cosmological $B$ mode signal, as lensing is caused by a scalar gravitational potential in the lens, and should therefore be curl-free. The presence of $B$ modes may be an indicator of potential systematic effects in the data, such as contamination by the PSF. $B$ modes may also be generated by second-order lensing effects \cite{Krause2010}, intrinsic alignments \cite{Blazek2019}, and the clustering of source galaxies \cite{Schneider2002}. We check the statistical significance of the $B$ mode signal to assess whether any of these effects may be impacting our measurements.

As shown in Figure~\ref{fig:power_spectra}, we find that the $B$ mode signal for multipoles larger than $\ell = 300$, is consistent with zero, as expected. The overall $p$-value, i.e. the probability that these data could have occurred under the hypothesis that the $B$ modes should be zero, across all ten auto- and cross-correlation spectra is 0.50, for the $BB$ spectra within our fiducial scale cuts ($300 < \ell < 1800)$. Each of the ten auto- and cross-correlation $BB$ spectra also do not show significant $B$ modes, with $p$-values ranging from 0.14 to 0.91. The $EB$ spectra have an overall $p$-value of 0.95, with the values for individual spectra ranging from 0.12 to 0.99, further confirming that the $B$ modes are not significant. We note that prior to removing the $\sim$20 $\mathrm{deg}^2$ region in GAMA09H with extremely good seeing and large fourth moment PSF residuals (see the discussion in Section~\ref{sec:shape_catalog}), we had a significant detection of $B$ modes, motivating the removal of this region. 

We do see excess $B$ modes at multipoles smaller than $\ell = 300$ as shown in Figure~\ref{fig:bb_eb_spectra} in Appendix~\ref{app:bmodes}. This is consistent with the findings of the HSC Year 1 analysis \cite{Hikage2019}. $\ell = 300 \sim 0.6 \ \mathrm{deg}$ corresponds roughly to the scale of the field of view of the HSC camera, and could indicate systematic effects due to variations with pointings on the sky. To ensure that such systematic effects do not contaminate our cosmological analysis, we cut multipoles below $\ell_{\mathrm{min}} = 300$. However, \cite{Asgari2018} showed that there may be residual systematic effects on $E$ modes from the source of the $B$ modes, even after removing scales with significant $B$ modes. With this in mind, our modeling of PSF systematics in our cosmological analysis (as described in Section~\ref{sec:psf_systematics}) should further mitigate systematic effects that could be causing $B$ modes.

\subsection{\label{sec:psf_systematics} PSF Systematics}

Systematics tests of the HSC-Y3 shear catalog in \cite{Li2022} showed evidence of PSF model shape residual correlations and additive systematics from star-galaxy correlations. These effects could produce artificial two-point correlations and hence bias our cosmic shear results. The PSF systematics model and measurement methods for the HSC-Y3 cosmology analyses are described in detail in \cite{Zhang2022} (where Appendix~F describes the model, map making, and measurements in Fourier space), so we provide only a brief summary here. 

The PSF can contaminate cosmic shear measurements in two ways. First, if the PSF model inaccurately describes the actual PSF shape (``PSF modeling error''), then the inferred shear can get an additive systematics term. In addition, even if the PSF model is perfect, but the PSF deconvolution is not, then the PSF will coherently contaminate the inferred shear (``PSF leakage''). In the past, cosmic shear analyses have only accounted for these PSF systematics due to the second moments of the PSF. However, \cite{Zhang2022} showed that the fourth moment terms can be significant, potentially leading to $\sim 0.3\sigma$ biases in cosmological parameters if not accounted for in the PSF model. 

In our PSF model, we assume that the measured galaxy shears have an additive bias due to PSF systematics, given by
\begin{equation}
\label{eq:psf_sys_full_fourier}
g_{\rm sys} = \alpha^{\rm (2)} e_{\rm PSF} + \beta^{\rm (2)} \Delta e_{\rm PSF}
+ \alpha^{\rm (4)} M^{\rm (4)}_{\rm PSF} + \beta^{\rm (4)} \Delta M^{\rm (4)}_{\rm PSF}.
\end{equation}
The terms, from left to right, represent second-moment PSF leakage, second-moment PSF modeling error, fourth-moment PSF leakage, and fourth-moment PSF modeling error.

Upon adding $g_{\rm sys}$ to the observed galaxy ellipticity, the measured cosmic shear power spectrum becomes:
\begin{equation}
\label{eq:delta_cl_fourier}
C_{\ell} \rightarrow C_{\ell} +
\sum_{i=1}^4 \sum_{j=1}^4 p_i p_j C_{\ell}^{S_i S_j},
\end{equation}
where the parameter vector is defined as $p = [\alpha^{\rm (2)},
\beta^{\rm (2)},\alpha^{\rm (4)}, \beta^{\rm (4)}]$, and the PSF moment vectors are $S = [e_{\rm PSF}, \Delta e_{\rm PSF},
M^{\rm (4)}_{\rm PSF}, \Delta M^{\rm (4)}_{\rm PSF}]$. We refer to the additive term in Eq.~\eqref{eq:delta_cl_fourier} as $\Delta C_{\ell}$. The correlations $C_{\ell}^{S_i S_j}$ are estimated from maps of the measured PSF and PSF modeling error using \texttt{NaMaster}.

\begin{figure}
\includegraphics[width=0.45\textwidth]{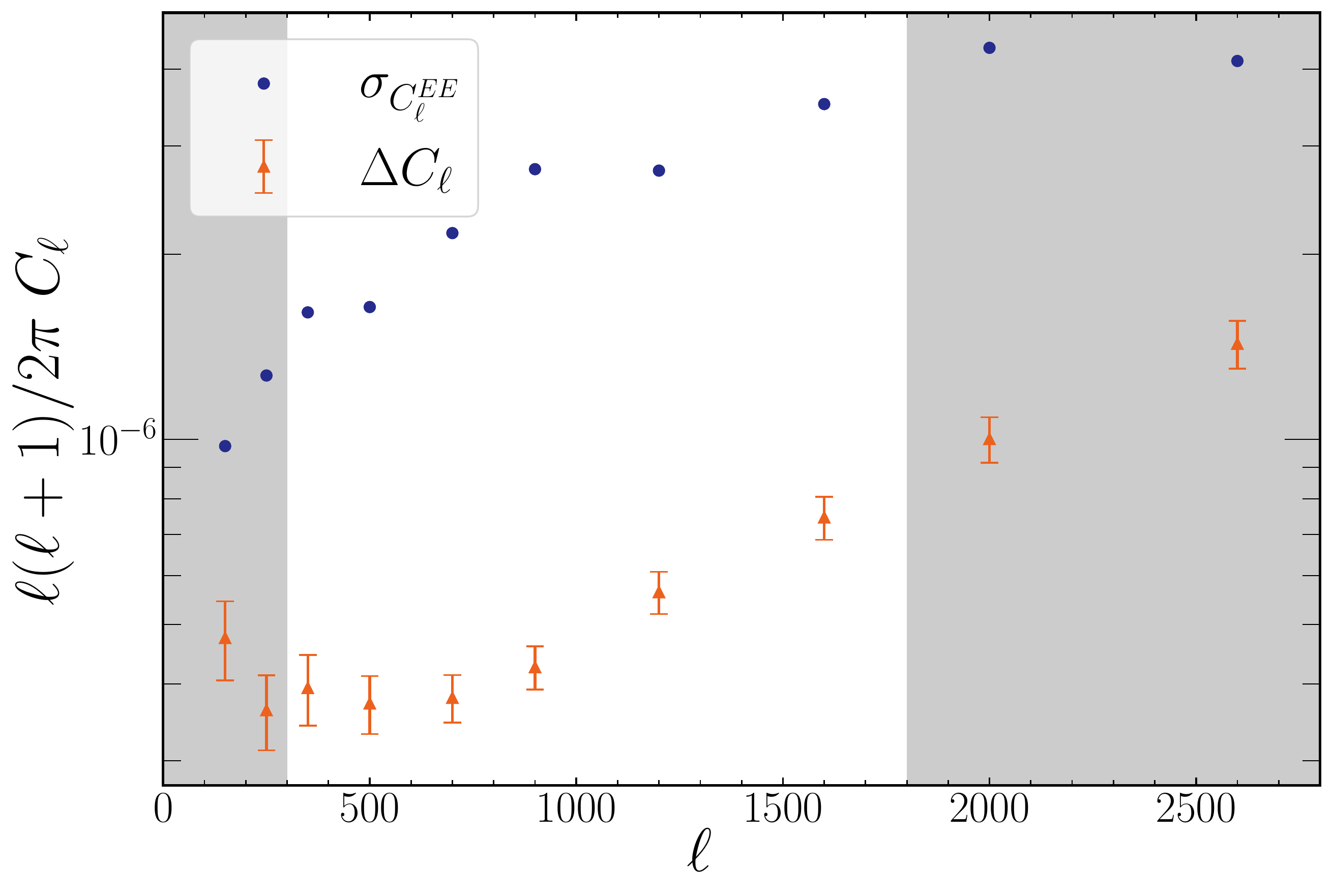}
\caption{\label{fig:delta_cl} Additive bias to the cosmic shear power spectrum from PSF systematics, $\Delta \Cl$ (orange triangles; see Section~\ref{sec:psf_systematics} for details), compared to the uncertainty of the non-tomographic cosmic shear
power spectrum, $\sigma_{\Cl^{EE}}$ (filled circles). The scales $\ell < 300$ and $\ell > 1800$ (shaded regions) are excluded in the cosmological analysis. We find that the contribution of PSF systematics is about $30\%$ of the uncertainty in the power spectra, so we marginalize over these systematics parameters in our cosmological analysis.}
\end{figure}

We include $\Delta C_{\ell}$ in the $\Cl$ model for our cosmological analysis, and jointly fit the PSF systematics parameters $p$ as nuisance parameters in the analysis. To get priors on these parameters, we estimate them by measuring the shear-star correlations for the HSC-Y3 shear catalog and star catalog, and comparing them to our model prediction, which assumes no redshift dependence of these correlations:
\begin{widetext}
\begin{align}
    \label{eq:null1_fourier}C_{\ell}^{\hat{g}_{\rm gal} e_{\text{PSF}}} &= \alpha^{\rm (2)} C_{\ell}^{e_{\text{PSF}} e_{\text{PSF}}}  + \beta^{\rm (2)}C_{\ell}^{\Delta e_{\text{PSF}}  e_{\text{PSF}}}  + \alpha^{\rm (4)} C_{\ell}^{ M^{\rm (4)}_{\text{PSF}}  e_{\text{PSF}}} + \beta^{\rm (4)} C_{\ell}^{\Delta M^{\rm (4)}_{\text{PSF}}  e_{\text{PSF}}},  \\ 
    \label{eq:null2_fourier} C_{\ell}^{\hat{g}_{\rm gal} \Delta e_{\text{PSF}}} &= \alpha^{\rm (2)} C_{\ell}^{e_{\text{PSF}} \Delta e_{\text{PSF}}}  + \beta^{\rm (2)} C_{\ell}^{\Delta e_{\text{PSF}} \Delta e_{\text{PSF}}} + \alpha^{\rm (4)} C_{\ell}^{M^{\rm (4)}_{\text{PSF}} \Delta e_{\text{PSF}}}  + \beta^{\rm (4)} C_{\ell}^{\Delta M^{\rm (4)}_{\text{PSF}} \Delta e_{\text{PSF}}}, \\
    \label{eq:null3_fourier} C_{\ell}^{\hat{g}_{\rm gal}  M^{\rm (4)}_{\text{PSF}}} &= \alpha^{\rm (2)} C_{\ell}^{e_{\text{PSF}}  M^{\rm (4)}_{\text{PSF}}}  + \beta^{\rm (2)} C_{\ell}^{\Delta e_{\text{PSF}} M^{\rm (4)}_{\text{PSF}}} + \alpha^{\rm (4)} C_{\ell}^{M^{\rm (4)}_{\text{PSF}}  M^{\rm (4)}_{\text{PSF}}}  + \beta^{\rm (4)} C_{\ell}^{\Delta M^{\rm (4)}_{\text{PSF}}  M^{\rm (4)}_{\text{PSF}}},\\
    \label{eq:null4_fourier}C_{\ell}^{\hat{g}_{\rm gal} \Delta M^{\rm (4)}_{\text{PSF}}} &= \alpha^{\rm (2)} C_{\ell}^{e_{\text{PSF}} \Delta M^{\rm (4)}_{\text{PSF}}}  + \beta^{\rm (2)} C_{\ell}^{\Delta e_{\text{PSF}} \Delta M^{\rm (4)}_{\text{PSF}}}  + \alpha^{\rm (4)} C_{\ell}^{M^{\rm (4)}_{\text{PSF}} \Delta M^{\rm (4)}_{\text{PSF}}}  + \beta^{\rm (4)} C_{\ell}^{\Delta M^{\rm (4)}_{\text{PSF}} \Delta M^{\rm (4)}_{\text{PSF}}}.
\end{align}
\end{widetext}
We obtain the following constraints on the parameters $p$, which are then used as priors in the cosmological analysis: $\alpha^{(2)} = 0.027 \pm 0.004$, $\beta^{(2)} = -0.39 \pm 0.04$, $\alpha^{(4)} = 0.17 \pm 0.02$, $\beta^{(4)} = -0.19 \pm 0.08$. As described in Section~\ref{sec:psf_model}, in practice, we sample four uncorrelated, normally distributed parameters, which are then transformed into the parameters $p$. Figure~\ref{fig:delta_cl} shows the measured $\Delta \Cl$ based on these best fit values of the PSF systematics parameters, compared to the uncertainty in the cosmic shear power spectrum. We find that the contribution of PSF systematics is approximately 30\% of the uncertainty in the cosmic shear power spectra, so we marginalize over the PSF systematics in our cosmological analysis (see Section~\ref{sec:psf_model} for more details). For a complete description of the results of the PSF systematics measurements, we refer the reader to Appendix~F of \cite{Zhang2022}.

\subsection{\label{sec:n(z)} Source Redshift Distribution Inference}

The source redshift distribution for the HSC-Y3 shear catalog galaxies was presented in \cite{Rau2022}. We summarize the method and results here, and refer the reader to \cite{Rau2022} for further details. The posterior source redshift distributions were constructed through a combined Bayesian Hierarchical Inference using the galaxies' photometry (specifically the \texttt{DNNz} redshifts), and spatial cross-correlations with a catalog of well-measured redshifts. The inference accounts for cosmic variance due to the limited survey area. 

The spatial cross-correlations, or clustering redshifts, utilize the photometric Luminous Red Galaxy (LRG) sample \cite{Oguri2018} from the same data release of HSC (S19A). The photometric LRGs are selected using the stellar population synthesis based red-sequence technique developed for the CAMIRA (Cluster-finding Algorithm based on Multi-band Identification of Red-sequence gAlaxies) optical cluster finding algorithm \cite{Oguri2014, Oguri2018_camira}. The photometric LRG sample has good redshift quality out to $z\sim1.2$ (with a bias of $\Delta z/(1+z)$ $\lesssim 0.005$ and a scatter of $\sim$0.02) and well understood clustering properties, making it an ideal sample to use for clustering redshifts, i.e. redshift inference for a photometric sample (the HSC-Y3 shear catalog) using spatial cross-correlations with a sample with known redshifts \cite{Schneider2006, Newman2008, Schmidt2013, Menard2013, Morrison2017}. 

\begin{figure}
\includegraphics[width=0.45\textwidth]{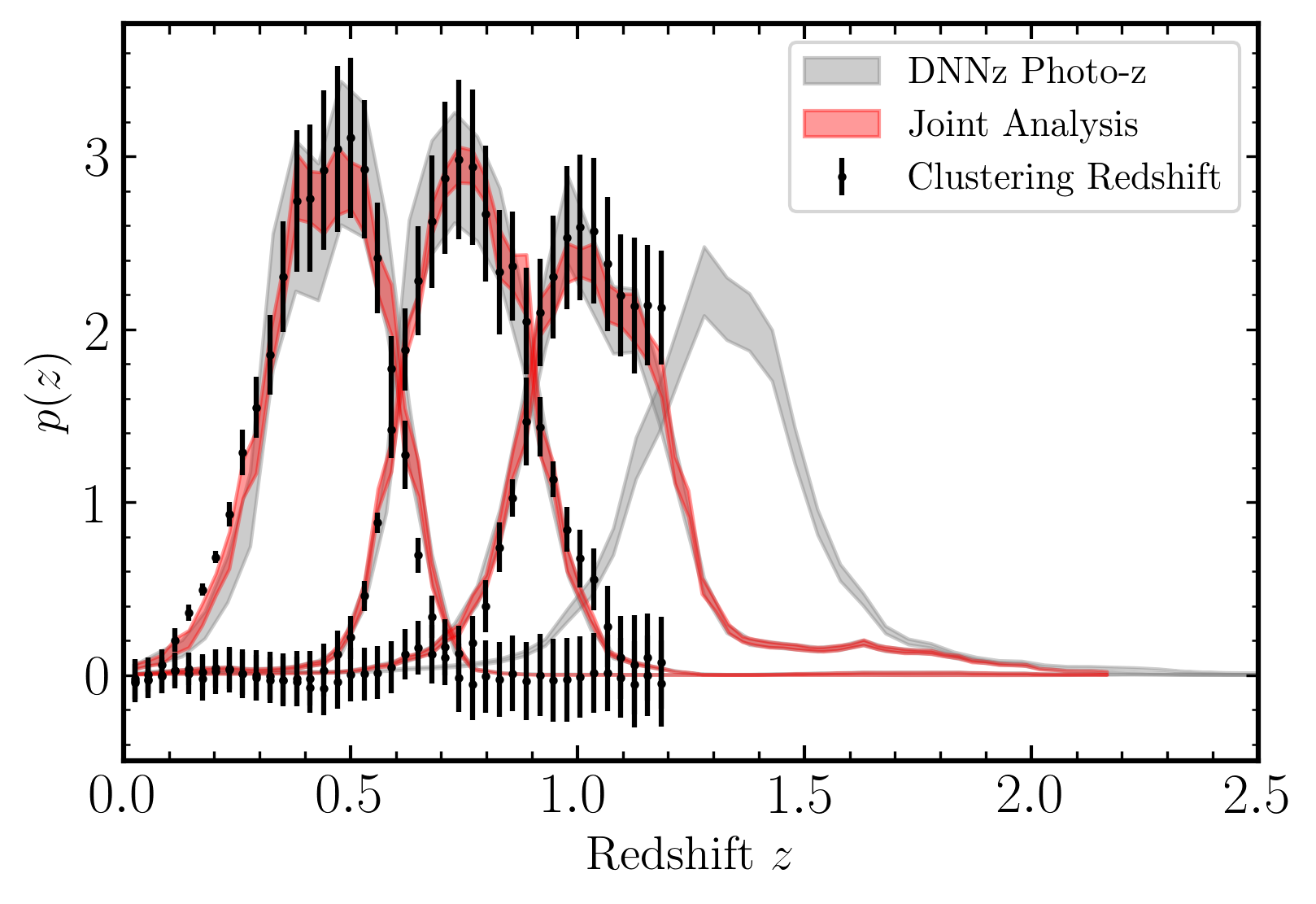}
\caption{\label{fig:nz_dist} Source redshift distribution inferred from the stacked \texttt{DNNz} photometric redshifts, with a cosmic variance correction (grey contour), the clustering redshifts using the CAMIRA LRG sample (black points), and the joint inference combining the two methods (red contour). The mean of the joint inference contour is used as our fiducial source redshift distribution model. This figure has been adapted from Figure 8 of \cite{Rau2022}.}
\end{figure}

As shown in Figure~\ref{fig:nz_dist}, we find that the redshift distributions inferred from photometry and clustering redshifts alone are consistent with the joint inference, and that the joint inference gives tighter constraints on the redshift distribution compared to the individual methods. However, the LRG sample from CAMIRA only extends up to a redshift of $z\sim 1.2$, and therefore we are unable to calibrate the last redshift bin in our analysis with this method. For this bin, we use the redshift distribution inferred from photometry alone, i.e. the stacked individual galaxy redshift posteriors, with a correction for cosmic variance. 

As described in Section~\ref{sec:nz_uncertainty}, we model the uncertainty in this redshift distribution in our cosmological analysis, by allowing the shift in the mean redshift of each bin to be a free parameter. 

\section{\label{sec:modeling} Modeling and Analysis Choices}

In this section, we describe the methods by which we fit a cosmological model to the power spectrum measurement presented in Section~\ref{sec:measurement_results}. The following sections will motivate the choice of a fiducial model comprised of a total of 23 free parameters, including cosmological, astrophysical and observational systematics (nuisance) parameters. Table~\ref{tab:fid_parameters} shows the full set of parameters used. First, in Section~\ref{sec:sampler} we describe the likelihood inference method, including the computation of the likelihood as well as the nested sampling implementation. We then describe in detail our model of the power spectrum. This includes our linear and nonlinear cosmic shear power spectrum model (Section~\ref{sec:power_spectrum_model}), our intrinsic alignment model (Section~\ref{sec:ia}), and our modeling of systematics from the Point Spread Function (Section~\ref{sec:psf_model}), redshift distribution uncertainties (Section~\ref{sec:nz_uncertainty}), and shear calibration biases (Section~\ref{sec:shear_calib}).  We discuss our choice of cosmological parameters to sample, and their prior ranges in Section~\ref{sec:priors}.

We are careful to coordinate our analysis choices with the parallel analysis using two point correlation functions \cite{Li2023}. The analysis tests and choices described below have also been also adopted and described by \cite{Li2023}.

\begin{table}%[b]
\caption{\label{tab:fid_parameters}%
Fiducial parameters and priors for our cosmological analysis. $\cal{U}$ represents a flat (uniform) prior, while $\cal{N}$ represents a Gaussian prior.}
\begin{ruledtabular}
\begin{tabular}{ll} 
Parameter & Prior \\ \hline
\multicolumn{2}{l}{\bf{Cosmological parameters} (Sections~\ref{sec:power_spectrum_model}, \ref{sec:priors})}\\
$\Omega_\mathrm{m}$                 & ${\cal U}(0.1, 0.7)$\\
$A_\mathrm{s} \,(\times 10^{-9})$   & ${\cal U}(0.5, 10)$\\
$h_0$                               & ${\cal U}(0.62, 0.80)$\\
$\omega_\mathrm{b} \equiv \Omega_b h^2$                 & ${\cal U}(0.02, 0.025)$\\
$n_\mathrm{s}$                      & ${\cal U}(0.87, 1.07)$\\
\hline
\multicolumn{2}{l}{\bf{Baryonic feedback parameters} (Section~\ref{sec:power_spectrum_model})}\\
$A_\mathrm{b}$                      & ${\cal U}(2, 3.13)$ \\
\multicolumn{2}{l}{\bf{Intrinsic alignment parameters} (Section~\ref{sec:ia})}\\
$A_1$                               & ${\cal U}(-6, 6)$ \\
$\eta_1$                            & ${\cal U}(-6, 6)$ \\
$A_2$                               & ${\cal U}(-6, 6)$ \\
$\eta_2$                            & ${\cal U}(-6, 6)$ \\
$b_\mathrm{ta}$                     & ${\cal U}(0, 2)$ \\
\hline
\multicolumn{2}{l}{\bf{PSF systematics} (Section~\ref{sec:psf_model})}\\
$\tilde{\alpha}_2$                          & ${\cal N}(0, 1)$\\
$\tilde{\beta}_2$                           & ${\cal N}(0, 1)$\\
$\tilde{\alpha}_4$                          & ${\cal N}(0, 1)$\\
$\tilde{\beta}_4$                           & ${\cal N}(0, 1)$\\
\multicolumn{2}{l}{\bf{Redshift distribution uncertainties} (Section~\ref{sec:nz_uncertainty})}\\
$\Delta z^{(1)}$                    & ${\cal N}(0, 0.024)$ \\
$\Delta z^{(2)}$                    & ${\cal N}(0, 0.022)$ \\
$\Delta z^{(3)}$                    & ${\cal U}(-1, 1)$ \\
$\Delta z^{(4)}$                    & ${\cal U}(-1, 1)$ \\
\multicolumn{2}{l}{\bf{Shear calibration biases} (Section~\ref{sec:shear_calib})}\\
$\Delta m^{(1)}$                    & ${\cal N}(0,0.01)$ \\
$\Delta m^{(2)}$                    & ${\cal N}(0,0.01)$ \\
$\Delta m^{(3)}$                    & ${\cal N}(0,0.01)$ \\
$\Delta m^{(4)}$                    & ${\cal N}(0,0.01)$ \\
\hline
\multicolumn{2}{l}{\bf Fixed Parameters}\\
$\sum m_{\nu} (\mathrm{eV})$ & $0.06$\\
$w$ & $-1$\\
$w_a$ & $0$\\
$\Omega_k$ & $0$\\
$\tau$ & $0.0851$\\
\end{tabular}
\end{ruledtabular}
\end{table}

\subsection{\label{sec:sampler} Likelihood Inference}

To constrain cosmological, astrophysical and observational systematics parameters from our data, we use a Monte Carlo Bayesian analysis to evaluate the posterior in the twenty-three dimensional space. We use a Gaussian likelihood $\mathcal{L}$: 
\begin{equation}
    -2 \ln \mathcal{L} (\hat{C_{\ell}} \,|\, \Theta ) = \left(\hat{C_{\ell}} - C_{\ell}(\Theta)\right)^\mathrm{T} \mathbf{C}^{-1} \left( \hat{C_{\ell}} - C_{\ell}(\Theta)\right),
\end{equation}
where $\Theta$ is the set of 23 parameters, sampled based on the priors in Table~\ref{tab:fid_parameters}, $C_{\ell}(\Theta)$ is the model prediction for a given set of parameters (described in detail in the following sections), $\hat{C_{\ell}}$ are the measured power spectra, and $\mathbf{C}^{-1}$ is the inverse of the covariance matrix estimated from mock catalogs. As discussed in Section~\ref{sec:covariance},  we multiply the inverse covariance by the Hartlap factor \cite{Hartlap2007}, in this case 0.96, to account for biases in the inversion, and we do not account for the cosmological parameter dependence of $\mathbf{C}$.

The theoretical power spectrum predictions are convolved with the bandpower windows when comparing to the measured spectra. The bandpower windows are given by:
\begin{equation}
    \mathcal{F}^{ij}_{L\ell} = \sum_{L'} (\mathbf{M}^{ij})^{-1}_{LL'}\sum_{\ell' \in L'} \omega_{L'}^{\ell'} \mathbf{M}_{\ell\ell'}^{ij},
\end{equation}
where $\mathbf{M}$ is the mode coupling matrix introduced in Equation~\ref{eq:pseudo_cl} and the convolution is:
\begin{equation}
    \left<\tilde{C}_L^{ij}\right> = \sum_{\ell} \mathcal{F}^{ij}_{L\ell} C_{\ell}^{ij}.
\end{equation}

We use nested sampling to sample the posterior in this high-dimensional parameter space. Nested sampling algorithms start with a large number of points (called ``live'' points), and then repeatedly eliminate the live point with the smallest value of the posterior density. This now ``dead" point is replaced with a live point that has a posterior value larger than that of the point that was eliminated. The set of both live and dead points can then be used to calculate the evidence while also serving as a (weighted) sample of the posterior. We use two nested sampling algorithms in this work: \texttt{PolyChord} \cite{Handley2015} and \texttt{MultiNest} \cite{Feroz2009}. These algorithms have different methods of finding new live points (see \cite{Lemos2022} for a detailed explanation). We use the implementations of these algorithms in the \texttt{CosmoSIS} package \cite{Zuntz2015}.

We report our fiducial constraints using the \texttt{PolyChord} nested sampling algorithm, as this has been shown to provide a more consistent and accurate estimate of the Bayesian evidence compared to \texttt{MultiNest}, and more conservative, i.e. wider, parameter credible intervals \cite{Lemos2022}. However, \texttt{MultiNest} is a much faster nested sampling implementation, so we use it for the consistency checks described in Sections~\ref{sec:results_robustness} and \ref{sec:results_internal_consistency}, where we do not need the Bayesian evidence or highly accurate parameter credible intervals. We also check in Section~\ref{sec:sampler_consistency} that our $S_8$ constraints are not significantly changed when using other samplers, specifically \texttt{emcee} \cite{Foreman-Mackey2013} and \texttt{zeus} \cite{Karamanis2021}.

For our \texttt{MultiNest} runs, we use 500 live points, with a sampling efficiency = 0.3, and a tolerance of 0.1 (see \cite{Feroz2009} for a description of these parameters). For the \texttt{Polychord} runs, we again use 500 live points, with the number of repeats set to 20, and a tolerance of 0.01 (see \cite{Handley2015} for a description of these parameters).

We assess the convergence of our chains by checking that the normalized sample weight (the weight at each point, divided by the maximum weight) has stopped increasing and is close to zero. Additionally, we also use \texttt{nestcheck} \cite{Higson2018, Higson2018_nestcheck, Higson2019} to check that the posterior mass, i.e. the total weight assigned to all samples in that region, has peaked out. This indicates that most of the posterior mass contribution is well sampled. 

For our fiducial constraints, we report the 1D marginalized mode and its asymmetric ±34\% confidence intervals, along with the point with the maximum posterior (MAP, or maximum a posteriori) in the chain returned by the nested sampling algorithm:
\begin{equation}
    \text{marginalized mode}^{+34\%\,\text{CL}}_{-34\%\,\text{CL}}
    ~(\text{MAP})\,.
\end{equation}
We report the marginalized mode, rather than the marginalized mean, as the former is less sensitive to the tails of the projected 1D posterior. Additionally, the marginalized mode is more stable than the MAP with a mathematically well-defined uncertainty on the estimation. However, statistics reported from marginalizing the full posterior over the rest of the multi-dimensional space are subject to projection effects, i.e. biases due to significant non-Gaussianities in the posterior (e.g. see Section~IV of \cite{Pandey2022} and Section~VA of \cite{Sugiyama2020}). In tests with mock data vectors, we find that projection effects can cause significant biases in the marginalized mode. For example, when simulating a data vector with our fiducial model, and analyzing it with the same model, we find a bias of $-0.77\sigma$ in the marginalized mode of $\Omega_m$, and a bias of $-0.48 \sigma$ in the marginalized mode of $S_8$. Upon conducting the same test using a minimizer to get the MAP, rather than using the marginalized mode from nested sampling, we recover these parameters with biases of $-0.01 \sigma$ and $-0.02 \sigma$ respectively. This indicates that our likelihood is unbiased, but projection effects can cause biases in the reported parameter values. We recommend that the reader keep this in mind when interpreting reported constraints. The MAP estimate should be robust to projection effects, however the maximum posterior point in the chain is a noisy estimate of the true maximum of the parameter space. More detailed descriptions of our investigation into projection effects and our validation of the likelihood inference setup can be found in Appendix~\ref{app:cosmosis_validation}.

\subsection{\label{sec:power_spectrum_model} Cosmic Shear Power Spectrum Model}

As described above, we constrain cosmological parameters by comparing the observed cosmic shear power spectra measured in Section~\ref{sec:power_spectra} to model-predicted power spectra. We compute the latter using the Limber approximation \cite{Limber1953}, which is valid as we are not considering very large scales (small multipoles) in this analysis \cite{Kitching2017, Kilbinger2017}. Under this assumption, the power spectra can be computed as:
\begin{equation}
    C_{\ell}^{ij} = \int_0^{\chi_H} d\chi \frac{q^i(\chi)q^j(\chi)}{\chi^2} P_{\mathrm{M}} \left[ k = \frac{\ell + 1/2}{\chi}, z(\chi)\right],
\end{equation}
where $i$ and $j$ are tomographic bins, $\chi$ is the comoving distance, $\chi_H$ is the comoving horizon distance, and $P_{\mathrm{M}}$ is the nonlinear matter power spectrum. The lensing efficiency, $q^i(\chi)$ is given by:
\begin{equation}
    q^i(\chi) \equiv \frac{3}{2} \Omega_m \frac{\mathrm{H}_0^2}{c^2} \frac{\chi}{a(\chi)}\int_{\chi}^{\chi_H} d \chi' n^i(\chi')\frac{\chi-\chi'}{\chi'},
\end{equation}
where $\Omega_m$ is the matter density parameter, $\mathrm{H}_0$ is the Hubble constant today, $a = 1/(1+z)$ is the scale factor, and $n^i(\chi)$ is the redshift distribution of source galaxies in tomographic bin $i$. In practice, $n^i(\chi)$ is taken as the mean of the joint inference contour described in Section~\ref{sec:n(z)} and shown in Figure~\ref{fig:nz_dist}. 

In practice, the linear part of the power spectrum is commonly computed using publicly available codes, such as \texttt{CAMB} \cite{Lewis2000} and \texttt{CLASS} \cite{Blas2011, Lesgourgues2011}. However this can be a computationally expensive calculation to implement in a cosmological analysis which has to recompute the power spectrum at each of the thousands of steps in the parameter sampling. For our analysis, we speed up this process by obtaining the linear matter power spectrum from the neural network-based \texttt{BACCO} emulator \cite{Arico2021}, which is able to provide the linear power spectrum with 0.5\% accuracy for redshifts $z\leq9$ and scales $10^{-4}\leq k \leq50 \ h \mathrm{Mpc}^{-1}$. Our linear power spectrum model has five cosmological parameters, $\Omega_m$, $\mathrm{H}_0$ (both described above), as well as the amplitude, $A_s$, and tilt, $n_s$, of the primordial power spectrum, and the baryon density, $\omega_b \equiv \Omega_b h^2$.

However, the linear matter power spectrum is not a complete theoretical description of matter clustering. At small scales, the growth of structure is nonlinear. Moreover, at small scales, the matter power spectrum is also affected by baryonic feedback from supernovae and active galactic nuclei \cite{Chisari2018}. We model the nonlinear matter power spectrum, including baryonic effects, using \texttt{HMCode 2016} \cite{Mead2016}, implemented in \texttt{pyhmcode} \cite{Troster2022}. \texttt{HMCode} is a nonlinear matter power spectrum prediction scheme based on the halo model \cite{Ma2000, Peacock2000, Seljak2000, Cooray2002}, which uses physically motivated parameters, fit to both N-body simulations and hydrodynamical simulations. \texttt{HMCode} parameterizes baryonic effects using the halo bloating parameter, $\eta_b$ and the amplitude of the halo mass concentration, $A_{\mathrm{bary}}$. The value $A_{\mathrm{bary}}=3.13$ corresponds to an absence of baryonic feedback. Following \cite{Joachimi2021}, we sample $A_{\mathrm{bary}}$, and define $\eta_b$ in terms of it:
\begin{equation}
    \eta_b = 0.98 - 0.12 A_{\mathrm{bary}}.
\end{equation}

\subsubsection{\label{sec:model_sufficiency} Nonlinear Model Sufficiency}

We investigate the sufficiency of \texttt{HMCode} for our nonlinear modeling by investigating the bias in $S_8$ and $\Omega_m$ when using \texttt{HMCode} to analyze a mock data vector with realistic baryonic feedback and nonlinear dark matter clustering, within a given range of scales. We investigate two possible scale cuts, $\ell_{\mathrm{max}}=1800$ and $\ell_{\mathrm{max}}=2200$, keeping the large scale cut fixed ($\ell_{\mathrm{min}}=300$). We consider our nonlinear model to be sufficient for describing realistic nonlinear physics if the 2D bias in the $S_8$-$\Omega_m$ plane is smaller than $0.3 \sigma_{\mathrm{2D}}$, following \cite{Krause2021}. Here, $\sigma_{\mathrm{2D}}$ is the error in the 2D $S_8$-$\Omega_m$ plane (see Equation~\ref{eq:map_bias}).

In practice, we simulate noiseless data vectors in \texttt{CosmoSIS} with baryonic effects based on the $P_{\mathrm{M}}(k)$ suppression observed in the AGN mode of the cosmological hydrodynamical simulations, OverWhelming Large Simulations (owlsAGN, \cite{Schaye2010, vanDaalen2011}), and nonlinear dark matter clustering based on CosmicEmu 2022 \cite{Heitmann2016, Moran2022} (an emulator for the matter power spectrum based on gravity-only dark matter simulations). Following \cite{Krause2021} and \cite{Huang2021}, we assume that the owlsAGN data vector represents a realistic level of baryonic feedback. Similarly, CosmicEmu has been shown to agree well with other nonlinear dark matter-only power spectrum estimates \cite{EuclidEmu2019}. We isolate the two nonlinear effects by introducing each type of ``contamination'' individually, i.e. we separately study the bias in parameter recovery due to different models of baryonic feedback and due to different models of nonlinear dark matter clustering. We analyze these ``contaminated'' data vectors with the fiducial model described above, and a given $\ell_{\mathrm{max}}$, and determine the bias on $S_8$ and $\Omega_m$ relative to the true values used in generating the data vectors.

Bearing in mind that the statistics we report, such as the marginalized mode, are subject to projection effects, we investigate the bias due to model mis-specification in terms of the maximum a posteriori (MAP) point. This allows us to separate any bias due to model mis-specification from biases due to projection effects, as the impact of projection effects will vary from model to model. However, the MAP estimation is not straightforward, as this is a high-dimensional parameter space with a non-trivial structure, which can make it difficult to converge on the global maximum posterior point, and the MAP estimate can be very dependent on the starting point. To alleviate these effects, we run the MAP estimation with 50 different starting points and take the final MAP to be the end point with the largest posterior, following \cite{Joachimi2021}. The starting points are varied in the parameters of interest in relation to the model choice, namely $\Omega_m$, $A_s$, $A_{\mathrm{bary}}$ and four intrinsic alignment parameters, $A_1$, $A_2$, $\eta_1$ and $\eta_2$ (see Section~\ref{sec:ia}). With these starting points, we then use \texttt{CosmoSIS}\footnote{https://cosmosis.readthedocs.io/en/latest/} \cite{Zuntz2015} to run the \texttt{scipy} \cite{Scipy2020} implementation of the Powell minimizer \cite{Powell1964} to find the MAP estimate for each point. We note that all 23 model parameters are varied during the minimization, even though the starting points are sampled in only seven dimensions. We find that our MAP estimation generally converges with $\sim$$20$ starting points, i.e. the maximum posterior point does not change upon adding additional points.

We require that our fiducial model, when used to analyze these ``contaminated'' data vectors, recover $S_8$ and $\Omega_m$ with a bias smaller than $0.3 \sigma_{\mathrm{2D}}$ in the $S_8$-$\Omega_m$ plane. In practice, this requirement is checked by computing the distance $\bm{d} = \left[S_{8, \mathrm{MAP}} - S_{8, \mathrm{true}}, \Omega_{m, \mathrm{MAP}} - \Omega_{m, \mathrm{true}}\right]$. We use the covariance matrix of our fiducial run on actual data, $\Sigma$ (based on the blinded catalog with the most constraining power, and therefore the most conservative choice), to check that:
\begin{equation}
    \sqrt{\bm{d} \, \Sigma^{-1} \, \bm{d}^T} < 0.3.
    \label{eq:map_bias}
\end{equation}

The results of this test are summarized in Table~\ref{tab:nonlinear_map_bias}. We find that our fiducial model passes the test for contamination from both baryonic effects and nonlinear dark matter clustering, for both sets of scale cuts, $300 < \ell < 1800$ and $300 < \ell < 2200$. Somewhat surprisingly, we find that the bias in the case of the CosmicEmu data vector, is larger when using the more conservative scale cut of $\ell_{\mathrm{max}}=1800$, rather than $\ell_{\mathrm{max}}=2200$. We believe that this is due to differences between \texttt{HMCode} (and other nonlinear dark matter models) and CosmicEmu at large scales, as has been noted in the literature (see, for example, Figure 9 of \cite{EuclidEmu2019}). 

\begin{table*}%[b]
\caption{\label{tab:nonlinear_map_bias}%
Results of model sufficiency tests using the MAP estimation, in terms of the bias in the $S_8$-$\Omega_m$ plane from analyzing a ``contaminated'' data vector with the fiducial model, relative to the parameter uncertainty from the fiducial run on data.
}
\begin{ruledtabular}
\begin{tabular}{lccl}
\textrm{Mock Data}&
\textrm{2D bias (Equation~\ref{eq:map_bias}) ($\sigma_{\mathrm{2D}}$)}&
\textrm{$S_8$1D bias ($\sigma$)}&
\textrm{$\Omega_m$1D bias ($\sigma$)}\\
\colrule
fiducial ($\ell_{\mathrm{max}}=1800$) & 0.03 & -0.01 & -0.02\\
owlsAGN ($\ell_{\mathrm{max}}=1800$) & 0.05 & 0.05 & -0.03\\
CosmicEmu ($\ell_{\mathrm{max}}=1800$) & 0.28 & 0.08 & -0.27\\
owlsAGN ($\ell_{\mathrm{max}}=2200$) & 0.11 & 0.07 & -0.08\\
CosmicEmu ($\ell_{\mathrm{max}}=2200$) & 0.19 & 0.02 & -0.19\\
\end{tabular}
\end{ruledtabular}
\end{table*}

Since both sets of scale cuts, $\ell_{\mathrm{max}}=1800$ and $\ell_{\mathrm{max}}=2200$ meet the requirements with the MAP estimation described above, we look at intrinsic alignment modeling to determine our fiducial scale cuts, as described in Section~\ref{sec:ia}. 

We also check the sensitivity of our nonlinear $P(k)$ model to other baryonic feedback and nonlinear dark matter clustering prescriptions, from other hydrodynamical simulations and nonlinear $P(k)$ emulators. In particular, we repeat the tests of MAP bias in the $\Omega_m$-$S_8$ plane described above with simulated data vectors based on the Illustris \cite{Nelson2015}, Horizon-AGN \cite{Chisari2018}, EAGLE \cite{Schaye2015, Crain2015} and cosmo-OWLS \cite{LeBrun2014} cosmological hydrodymanical simulations, as well as the Euclid Emulator \cite{EuclidEmu2019} for the nonlinear dark matter-only power spectrum. The results of these tests are presented in Table~\ref{tab:nonlinear_map_bias_other_tests}. We find that only one of these tests shows a significant bias (larger that $0.3\sigma_{\mathrm{2D}}$), namely the analysis of the data vector based on cosmo-OWLS. This may be due to the rather extreme baryonic physics prescription of this simulation (a minimum heating temperature for AGN feedback of $\log T_{\mathrm{AGN}} = 8.5$). From the small bias in the inferred cosmological parameters for the other hydrodynamic simulations and for Euclid Emulator, we infer that our modeling is robust to fairly large variations in the nonlinear physics model.

\begin{table*}%[b]
\caption{\label{tab:nonlinear_map_bias_other_tests}%
Results of additional tests of model sensitivity to different nonlinear physics using the MAP estimation, in terms of the bias in the $S_8$-$\Omega_m$ plane. These biases were determined from analyzing a ``contaminated'' data vector with the fiducial model, and are presented relative to the parameter uncertainty from the fiducial run on data.
}
\begin{ruledtabular}
\begin{tabular}{lccl}
\textrm{Mock Data}&
\textrm{2D bias (Equation~\ref{eq:map_bias}) ($\sigma_{\mathrm{2D}}$)}&
\textrm{$S_8$1D bias ($\sigma$)}&
\textrm{$\Omega_m$1D bias ($\sigma$)}\\
\colrule
fiducial ($\ell_{\mathrm{max}}=1800$) & 0.03 & -0.01 & -0.02\\
Illustris ($\ell_{\mathrm{max}}=1800$) & 0.26 & -0.25 & 0.06\\
Horizon-AGN ($\ell_{\mathrm{max}}=1800$) & 0.20 & 0.11 & -0.17\\
EAGLE ($\ell_{\mathrm{max}}=1800$) & 0.15 & 0.02 & -0.15\\
cosmo-OWLS ($\ell_{\mathrm{max}}=1800$) & 0.57 & -0.49 & 0.31\\
Euclid Emulator ($\ell_{\mathrm{max}}=1800$) & 0.21 & 0.09 & -0.19\\
\end{tabular}
\end{ruledtabular}
\end{table*}

\subsection{\label{sec:ia} Intrinsic Alignments}

In addition to baryonic effects, cosmic shear analyses also suffer from astrophysical systematic effects due to the intrinsic alignment (IA) of galaxy shapes (for recent reviews, see \cite{Kiessling2015, Kirk2015, Joachimi2015}). Since galaxies are extended objects, they are subject to tidal forces. As a result, their intrinsic shapes tend to align with the tidal field of the gravitational potential, and therefore with each other, rather than being completely random \cite{Hirata2004}. The observed shear power spectrum thus has additional contributions from the correlation of intrinsic shapes, $C_{\ell, \mathrm{II}}$ (arising from galaxies being spatially close to one another), and the cross-correlation of intrinsic shapes with cosmological shear, $C_{\ell, \mathrm{GI}}$. The cross-correlation terms are caused by galaxies at different distances along the same line of sight being lensed by, or experiencing gravitational tidal interaction with the same large scale structure. The observed signal for tomographic bins $i$ and $j$ is then:
\begin{equation}
    C_{\ell}^{ij, \mathrm{obs}} = C_{\ell, \mathrm{GG}}^{ij} + C_{\ell, \mathrm{GI}}^{ij} + C_{\ell, \mathrm{GI}}^{ji} + C_{\ell, \mathrm{II}}^{ij},
\end{equation}
where the angular power spectra can be expressed in terms of the 3D power spectra, assuming the Limber approximation:
\begin{equation}
    C_{\ell, \mathrm{II}}^{ij} = \int_0^{\chi_H} d\chi \frac{n^i(\chi)n^j(\chi)}{\chi^2} P_{\mathrm{II}} \left[ k = \frac{\ell + 1/2}{\chi}, z(\chi)\right],
\end{equation}
\begin{equation}
    C_{\ell, \mathrm{GI}}^{ij} = \int_0^{\chi_H} d\chi \frac{q^i(\chi)n^j(\chi)}{\chi^2} P_{\mathrm{GI}} \left[ k = \frac{\ell + 1/2}{\chi}, z(\chi)\right].
\end{equation}
In order to compute the power spectra $P_{\mathrm{II}} (k)$ and $P_{\mathrm{GI}}(k)$, we consider two different intrinsic alignment models for this analysis, as described below (see \cite{Campos2022} for a more detailed summary).

\subsubsection{\label{sec:ia_tatt} Tidal Alignment and Tidal Torquing (TATT)}

The Tidal Alignment and Tidal Torquing (TATT) model \cite{Blazek2019} uses nonlinear perturbation theory to expand the field of intrinsic galaxy shapes $\gamma^{\mathrm{I}}$ in terms of the tidal field, $s$, and the matter overdensity, $\delta$. We consider terms up to quadratic order in the tidal field:
\begin{equation}
    \gamma^{\mathrm{I}}_{ij} = C_1 s_{ij} + C_2 \sum_k s_{ik} s_{kj} + b_{\mathrm{TA}}C_1 \delta s_{ij},
\end{equation}
where $C_1$, $C_2$ and $b_{\mathrm{TA}}$ are free parameters. This gives us the 3D power spectra:
\begin{equation}
\begin{split}
P^{EE}_{\mathrm{GI}}(k)
    &= C_1 P_\delta(k) + b_\mathrm{TA} C_1 P_{0|0E}(k) + C_2 P_{0|E2}(k)\,, \\
\end{split}
\end{equation}
\begin{equation}
\begin{split}
    P^{EE}_{\mathrm{II}}(k)
    &= C^2_1 P_\delta(k) + 2 b_\mathrm{TA} C_1^2 P_{0|0E}(k) \\
    &+ b^2_\mathrm{TA} C_1^2 P_{0E|0E}(k) + C_2^2 P_{E2|E2}(k) \\
    &+ 2 C_1 C_2 P_{0|E2}(k) + 2b_\mathrm{TA} C_1 C_2 P_{0E|E2}(k)\,,\\
\end{split}
\end{equation}
where the subscripts of the power spectra indicate correlations between different order terms in the expansion of $\gamma^{\mathrm{I}}$ (see \cite{Blazek2019} for the full definitions).

The redshift-dependent amplitudes, $C_1$ and $C_2$ are modeled as power laws in $1+z$:
\begin{equation}
    \label{eq:ia_amp}
    C_1(z) = -A_1  \frac{\bar{C_1} \rho_{\mathrm{c}}
    \Omega_\mathrm{M}}{D(z)} \left ( \frac{1+z}{1+z_0} \right )^{\eta_1},
\end{equation}
\begin{equation}
    C_2(z) = 5 A_2  \frac{\bar{C_1} \rho_{\mathrm{c}}
    \Omega_\mathrm{M}}{D^2(z)} \left ( \frac{1+z}{1+z_0} \right )^{\eta_2},
\end{equation}
where $D(z)$ is the growth function, and $\rho_\mathrm{c}$ is the critical
density. By convention, we set the constant $\bar{C_1} = 5\times10^{-14} M_\odot h^{-2} \mathrm{Mpc}^2$ and the redshift pivot $z_0 = 0.62$.

Thus, our implementation of the TATT model has five free parameters: $A_1$, $A_2$, $\eta_1$, $\eta_2$ and $b_{\mathrm{TA}}$. $A_1$ and $A_2$ capture the IA power spectra that scale linearly and quadratically with the tidal field, while  $\eta_1$ and $\eta_2$ model possible redshift evolution beyond what is already encoded in the model. $b_\mathrm{TA}$ is a bias parameter which models the fact that galaxies are over-sampled in the highly clustered regions. In this analysis, we adopt wide, flat priors for these parameters, where $A_1, A_2, \eta_1, \eta_2 \in [-6, 6]$, and $b_\mathrm{TA} \in [0,2]$. Since the IA signal is very sensitive to the properties of a galaxy sample, it is difficult to derive reliable Gaussian priors on these model parameters. The implementation of the TATT model in \texttt{CosmoSIS} is powered by the \texttt{FAST-PT} algorithm, which rapidly performs the mode-coupling integrals \cite{McEwen2016}.

\subsubsection{\label{sec:ia_nla} Non-Linear Alignments (NLA)}

A simpler, and more commonly used model is the Non-Linear Alignment (NLA) model \cite{Bridle2007}. This model assumes that galaxies align linearly with the tidal field. Under this model, the 3D power spectra are:
\begin{equation}
    P^{EE}_\mathrm{GI} = C_1(z) P_{\delta}, \qquad
    P^{EE}_\mathrm{II} = C^2_1(z) P_{\delta},
\end{equation}
where $C_1$ is parameterized as in Equation~\ref{eq:ia_amp}. Here, $P_{\delta}$ is the full nonlinear matter power spectrum, generated using \texttt{HMCode}. Our implementation of the NLA model has two free parameters: $A_1$ and $\eta_1$. As with TATT, we adopt wide, flat priors on these parameters, with $A_1, \eta_1, \in [-6, 6]$. We note that NLA is a subset of the TATT model, with $A_2 = 0$ and $b_{\mathrm{TA}} = 0$.

To choose an intrinsic alignment model, we follow the empirical approach suggested by \cite{Campos2022} which is based on the $\chi^2$ difference between fits to the two models based on the actual data. We take this approach rather than generating mock ``contaminated'' data vectors, as it can be difficult to understand what a realistic level of IA contamination would be, when the TATT model is not yet well constrained by data. 

\begin{figure}
\includegraphics[width=0.45\textwidth]{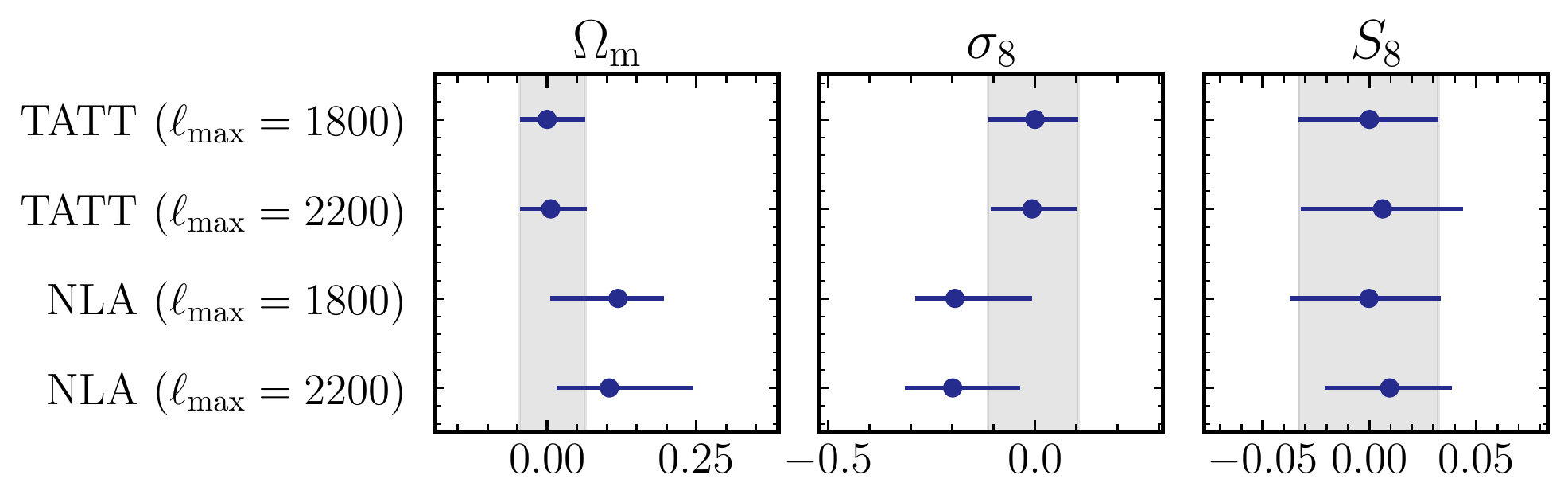}
\caption{\label{fig:ia_models_scale_cut} Blinded parameter constraints from different IA models (described in Section~\ref{sec:ia}) and different scale cuts, run on blinded catalog 0, which has the most constraining power. These results were used to choose the fiducial IA model and small scale cut. Since the constraints on our parameter of interest, $S_8$, are comparable for both NLA and TATT (with no shift in the central value, or appreciable increase in error bars), we choose to use the more complete model, TATT. The constraints on $S_8$ are also similar for the two sets of scale cuts, so we use the more conservative scale cut of  $\ell_{\mathrm{max}} = 1800$.}
\end{figure}

We analyze all three blinded catalogs with both TATT and NLA, using scale cuts of $\ell_{\mathrm{max}} = 1800$ and $\ell_{\mathrm{max}} = 2200$. These resulting parameter constraints are summarized in Figure~\ref{fig:ia_models_scale_cut}. We look at the goodness of fit from each setup, as well as the shifts and errors on our parameter of interest, $S_8$. We find that the constraints on $S_8$ are comparable for all four combinations of models and scale cuts\footnote{We note that in the case of blinded catalog 0, which was used for the model selection tests, the error on $S_8$ increases when extending the analysis to smaller scales ($\ell_{\mathrm{max}} = 2200$) using TATT, as shown in Figure~\ref{fig:ia_models_scale_cut}. This is counterintuitive, as adding data points should increase the constraining power of the analysis, and this implied to us that we did not fully understand how the TATT model was actually making use of data on smaller scales.  For this reason, we made the conservative choice to use the stricter scale cut. (For the true catalog, blinded catalog 2, the error on $S_8$ decreases when going to smaller scales, as expected, so the behavior observed with blinded catalog 0 is not universal and may couple with other parameters such as multiplicative bias).}. We do not see a significant shift in $S_8$ from any choice of model or scale cut. We use TATT for our fiducial model, as it is a more complete description of IA and doesn't appear to degrade our constraints. Given that our model sufficiency tests pass for both sets of possible scale cuts, $\ell_{\mathrm{max}} = 1800$ and $\ell_{\mathrm{max}} = 2200$, we choose to use the more conservative option of $\ell_{\mathrm{max}} = 1800$ as our fiducial scale cut. However, as shown in Section~\ref{sec:results_internal_consistency}, our results are unchanged by extending the analysis to $\ell_{\mathrm{max}} = 2200$.

\subsection{\label{sec:psf_model} PSF systematics}

As described in Section~\ref{sec:psf_systematics}, we model PSF systematics using four parameters: $p = $ [$\alpha^{\rm (2)}$, $\beta^{\rm (2)}$, $\alpha^{\rm (4)}$, $\beta^{\rm (4)}$], which describe the second-moment leakage, the second-moment modeling error, the fourth-moment leakage and the fourth-moment modeling error respectively. In practice, as shown by \cite{Zhang2022}, the parameters $\alpha^{\rm (2)}$ and $\alpha^{\rm (4)}$ are correlated. In order to account for this correlation in the sampling, we instead sample four uncorrelated parameters ($\tilde{p}$ = [$\tilde{\alpha}^{\rm (2)}$, $\tilde{\beta}^{\rm (2)}$, $\tilde{\alpha}^{\rm (4)}$, $\tilde{\beta}^{\rm (4)}$]), each from a standard normal distribution. We then transform these parameters into our original parameters by the following:
\begin{equation}
    p = \Lambda^{1/2} \mathrm{U} \tilde{p} + {\bar{p}},
\end{equation}
where $\bar{p}$ is the mean of each of the parameters $p$, determined from the measurements described in Section~\ref{sec:psf_systematics}, $\Lambda$ is a diagonal matrix with the eigenvalues of $p - \bar{p}$, and $\mathrm{U}$ is a matrix of the eigenvectors. 

We find that not accounting for this correlation has little impact on our parameter estimates, as described in Section~\ref{sec:results_robustness}. 

\subsection{\label{sec:nz_uncertainty} Redshift Distribution Uncertainties}

The source redshift distribution, whose inference is described in Section~\ref{sec:n(z)}, is subject to uncertainties. We marginalize over these uncertainities in our cosmological analysis with a shift model for the mean redshift of each tomographic bin $i$:
\begin{equation}
n^i(z) \rightarrow n^i(z + \Delta z_i).
\end{equation}
In this model, a negative value of $\Delta z_i$ corresponds to a shift of the $n^i(z)$ distribution to higher redshifts. \cite{Zhang2023_nz} demonstrated that this model is sufficient for capturing the redshift distribution uncertainty for HSC-Y3 data, in addition to being computationally inexpensive. 

\begin{figure}
\includegraphics[width=0.45\textwidth]{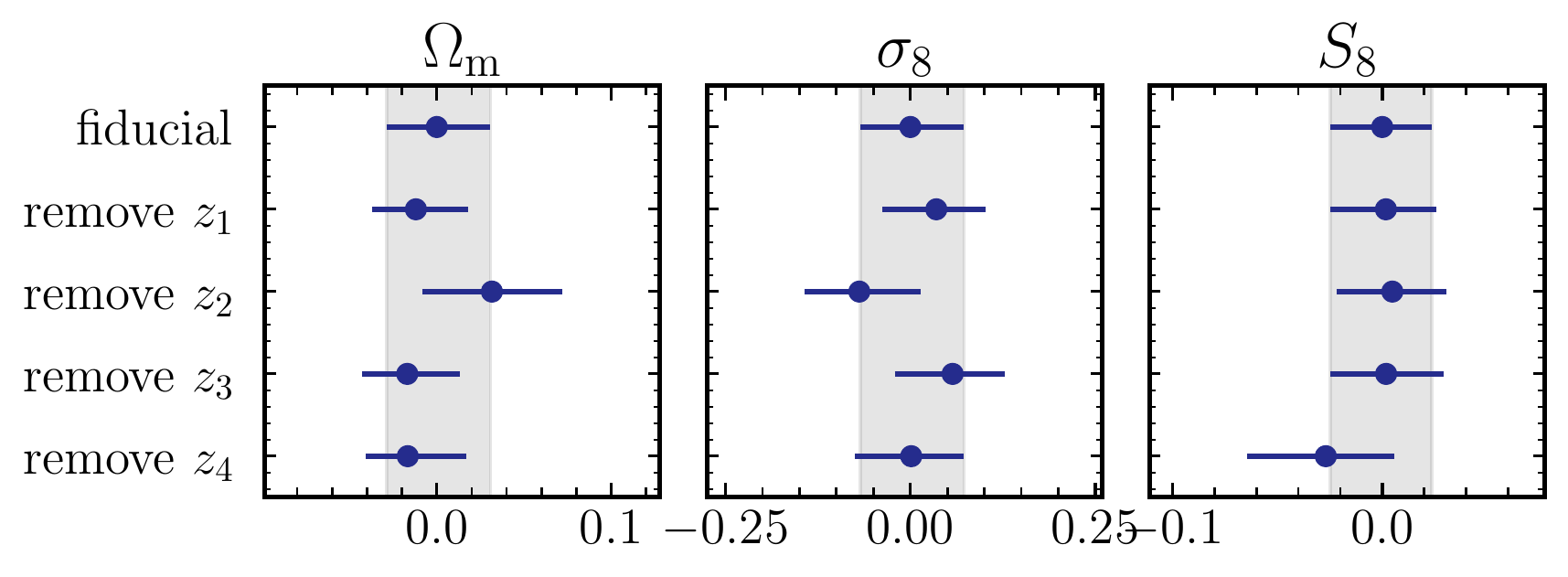}
\caption{\label{fig:remove_z_bins_old_prior} Blinded parameter constraints for an analysis using Gaussian priors for the redshift shift parameters, $\Delta z_i$, from \cite{Rau2022}, compared to the results when removing each of the four redshift bins from the analysis. The large shift in $S_8$ when removing the fourth bin is an indication that the calibration of $n(z)$ in this redshift bin is poor. Since the third redshift bin is also only partially calibrated by the LRG clustering redshifts, we choose to use conservative, flat priors on both $\Delta z_3$ and $\Delta z_4$.}
\end{figure}

This model adds a total of four parameters (one per redshift bin) to our analysis. The priors on these parameters are listed in Table~\ref{tab:fid_parameters}. The priors on $\Delta z_1$ and $\Delta z_2$ are Gaussian priors taken from \cite{Rau2022} (see Section~5.7), and are dominated by the differences in the inferred $n(z)$ from the three different HSC-Y3 photometric redshift codes. \cite{Rau2022} also contains a recommendation for Gaussian priors on $\Delta z_3$ and $\Delta z_4$: ${\cal N}(0, 0.031)$ and ${\cal N}(0, 0.034)$ respectively. However, the third redshift bin is only partially calibrated by the LRG clustering redshifts, and the fourth is not at all calibrated. Moreover, we suspect that the redshift distributions in these bins are systematically biased by all three photo-$z$ codes. This idea is supported by the fact that we see a large shift in $S_8$ when using the Gaussian priors from \cite{Rau2022} for all four bins and then excluding the fourth redshift bin from the analysis (Figure~\ref{fig:remove_z_bins_old_prior}). For this reason, we adopt a wide, flat prior for these parameters: $\Delta z_3, \Delta z_4 \in [-1, 1]$. We shall see that these parameters are constrained by the data to be well within this range. 

\subsection{\label{sec:shear_calib} Shear Calibration Bias}

As described in \cite{Li2022}, the HSC-Y3 shear catalog is calibrated using image simulations \cite{Mandelbaum2018_image}, where Hubble Space Telescope images are downgraded to HSC observing conditions. \cite{Li2022} use the image simulations to model and calibrate the shear estimation bias, including galaxy model bias \citep{Bernstein2010}, noise bias \citep{Refregier2012}, selection bias \citep{Kaiser2000b}, and detection bias \citep{Sheldon2020}. It is also shown that the shear estimation bias due to the blending of galaxies located at different redshifts is small for the HSC-Y3 weak lensing science.

We model and marginalize over any residual uncertainty in the multiplicative bias after this calibration procedure, following the model from \cite{Amon2022}. We introduce the nuisance parameters $\Delta m^{(i)}$ to represent the residual redshift-dependent multiplicative bias in each tomographic bin $i$. The cosmic shear power spectra then become:
\begin{equation}
    C_{\ell}^{ij} \longrightarrow (1+\Delta m^{(i)}) (1+\Delta m^{(j)}) C_{\ell}^{ij}.
\end{equation}

We use a Gaussian prior for each of the $\Delta m^{(i)}$ parameters, with a mean of 0 and a standard deviation of $0.01$, since the calibration in \cite{Li2022} confirmed that the multiplicative bias residual is smaller than $1\%$.

\subsection{\label{sec:priors} Prior Distributions}

The above sections motivate a fiducial model with 23 free parameters to describe cosmological, astrophysical and observational systematic effects. The fiducial prior distributions for these parameters are listed in Table~\ref{tab:fid_parameters}. In this section, we discuss the choice of cosmological parameters to sample, as well as the impact of the priors for these parameters on our analysis. 

Our fiducial analysis, as described in Section~\ref{sec:power_spectrum_model}, samples the cosmological parameters $\Omega_m$ and $A_s$. We have adopted conservative, wide priors for these two parameters, $\Omega_m \in [0.1, 0.7]$ and $A_s \in [0.5 \times 10^{-9}, 10^{-8}]$. A value outside this prior range would be unrealistic. The $A_s$ prior in particular, is tens of $\sigma$ wider than the posterior determined by \textit{Planck} \cite{Planck2018Cosmology}. Moreover, we find that our $S_8$ constraint is unaffected by increasing these prior ranges.

\begin{figure*}
     \centering
     \includegraphics[width=0.49\textwidth]{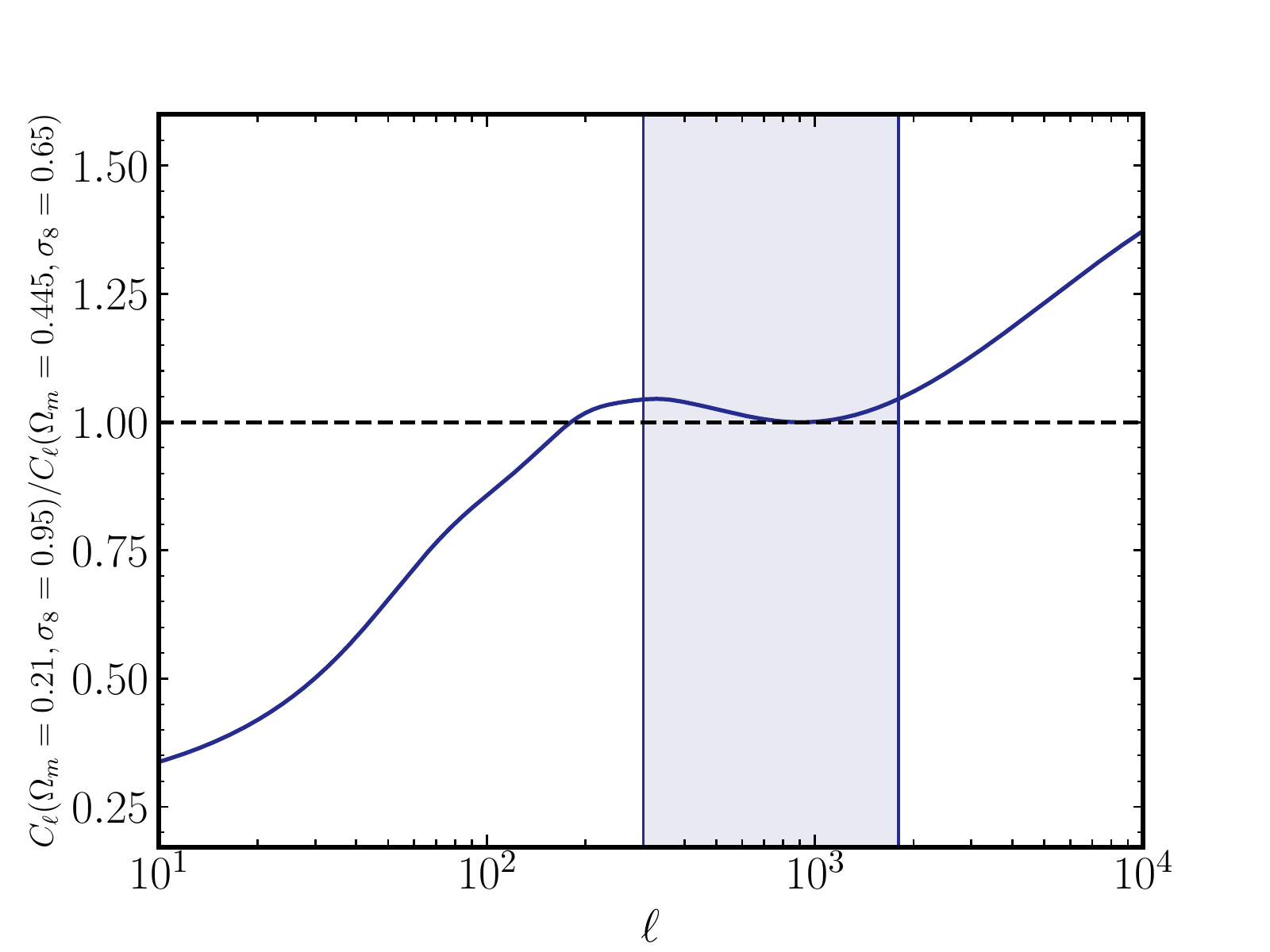}
     \includegraphics[width=0.49\textwidth]{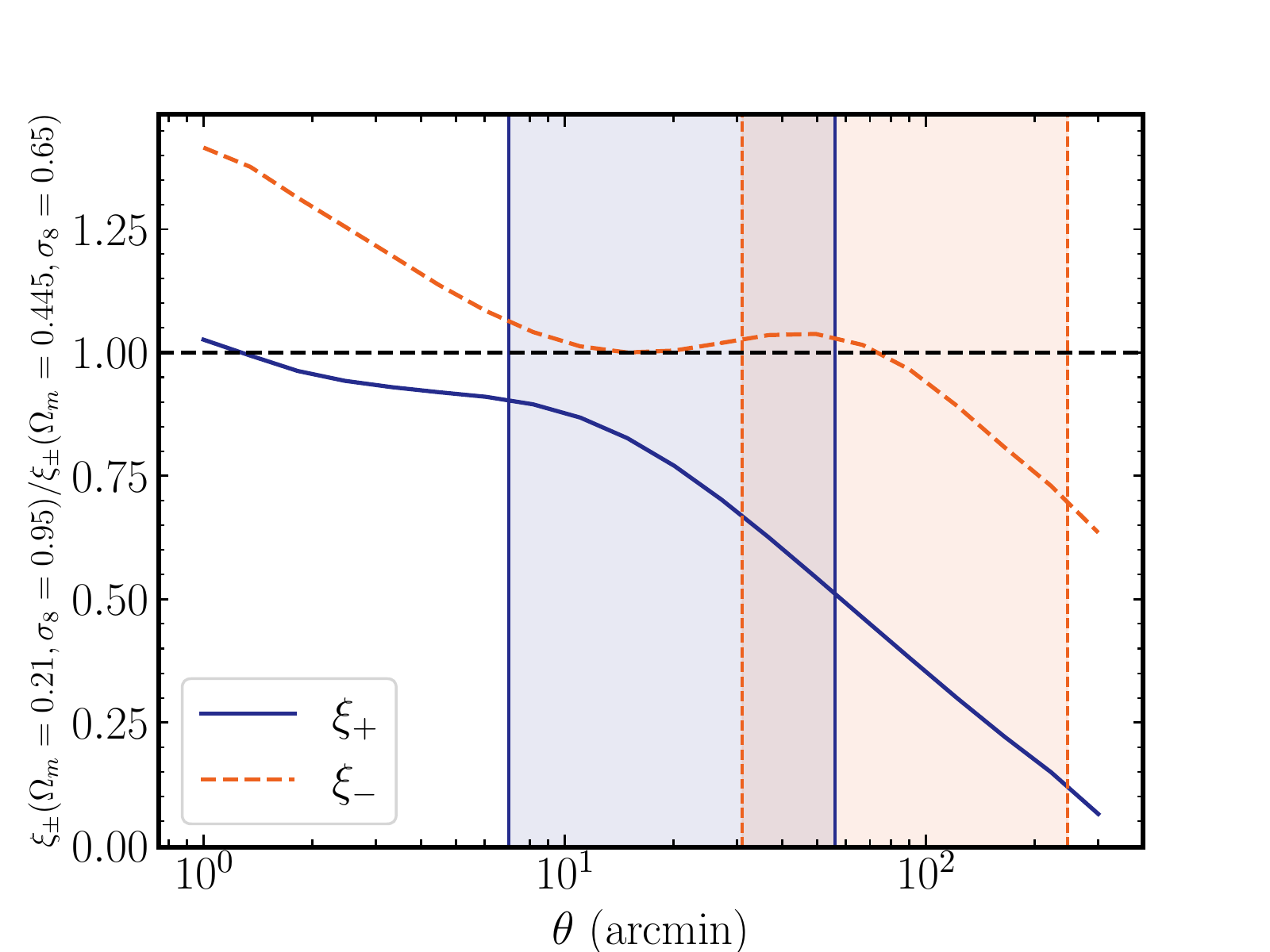}
     \caption{Ratio of different summary statistics - $\Cl$ on the left, $\xi_{+}$ (blue, solid) and $\xi_{-}$ (orange, dashed) on the right - for the fourth redshift bin autocorrelation, for two sets of cosmological parameters along the $\Omega_m$-$\sigma_8$ degeneracy with the same $S_8$ value: $(\Omega_m = 0.21, \sigma_8 = 0.95)$ and $(\Omega_m = 0.445, \sigma_8 = 0.65)$, computed using our fiducial model. The fiducial scale cuts for this analysis and for the 2PCF analysis \cite{Li2023} are the shaded regions. While the real space statistics, $\xi_{\pm}$, are able to distinguish between these extremely different parameter values, the ratio of $\Cl$s in our fiducial scale cut range is nearly unity, making it difficult to constrain parameters along this degeneracy direction.
     }
    \label{fig:cl_degeneracy}
\end{figure*}

We note, in particular, our choice to sample $A_s$, the amplitude of the primordial power spectrum, rather than $\sigma_8$ or $S_8$. This is motivated by our finding that for our fiducial scale cuts $300 < \ell < 1800$, there is a strong degeneracy between $\Omega_m$ and $\sigma_8$, much stronger than is seen in the 2PCF analysis \cite{Li2023}. We can illustrate the origin of this degeneracy by examining the change in our observables along it. Figure~\ref{fig:cl_degeneracy} shows ratios between theory predictions at extreme points along the degeneracy, for both Fourier and real-space measurements. While both statistics are sensitive to the degeneracy direction when we look at their complete possible ranges, the $\Cl$ ratio is close to unity within our scale cuts, unlike $\xi_{\pm}$ which varies more strongly. For this reason, $\Omega_m$ and $\sigma_8$ are poorly constrained by our analysis, and we do not focus on these quantities when reporting parameter values, or when conducting internal consistency and model robustness checks. 

\begin{figure}
     \centering
     \includegraphics[width=0.45\textwidth]{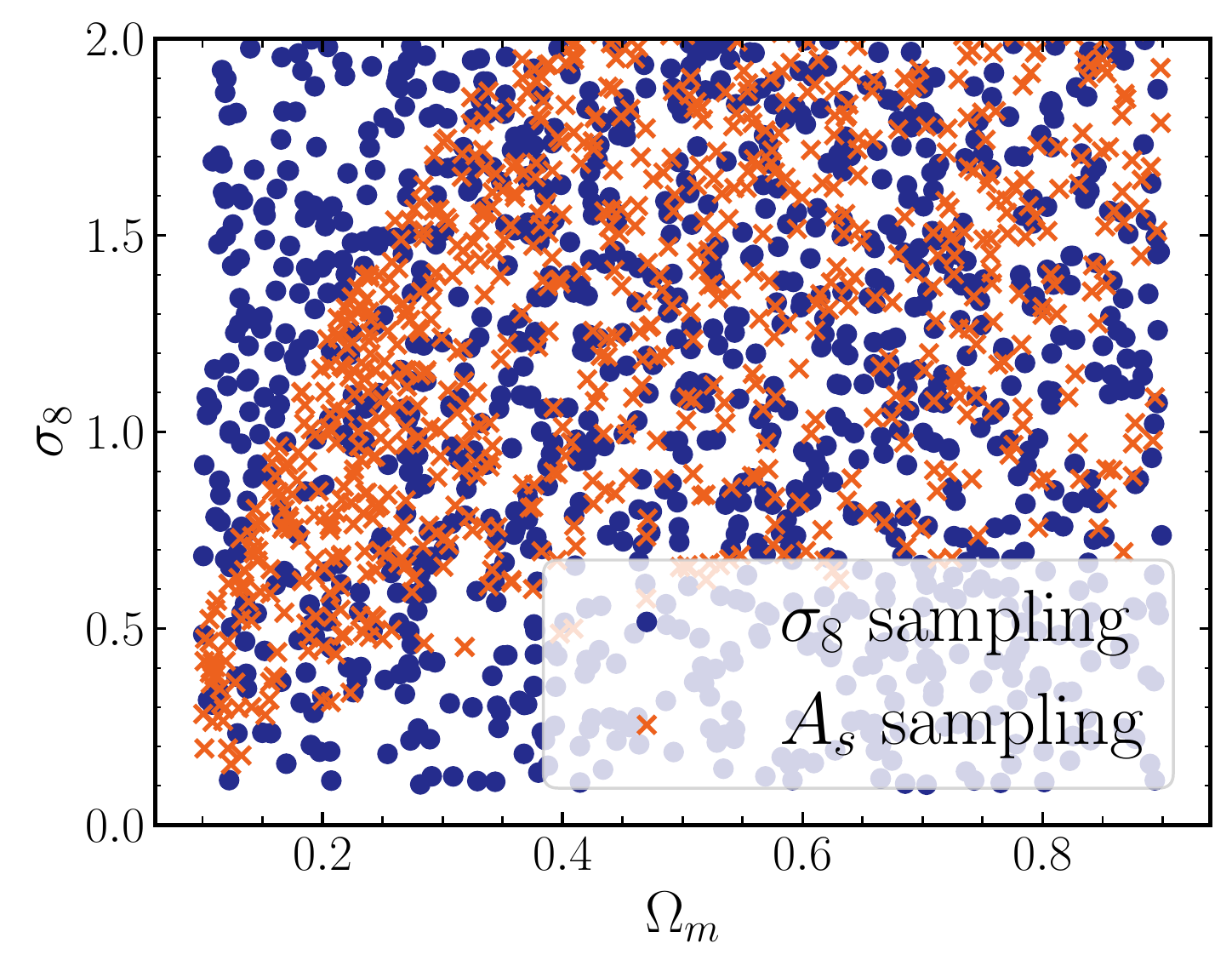}
     \caption{Sampling in the $\Omega_m$-$\sigma_8$ space when sampling uniformly in $\Omega_m$ and $\sigma_8$ (blue dots), and when sampling uniformly in $\Omega_m$ and $A_s$ (orange crosses). Given the inability of our summary statistic and scale cuts to distinguish between extreme points along the $\Omega_m$-$\sigma_8$ degeneracy (as shown in Figure~\ref{fig:cl_degeneracy}), sampling $A_s$ provides a more informative prior without affecting the central value of our $S_8$ constraint.}
    \label{fig:sigma8_vs_As_sampling}
\end{figure}

Sampling in $A_s$ rather than $\sigma_8$ helps to overcome this degeneracy. Sampling uniformly in $\Omega_m$ and $A_s$ is not equivalent to sampling uniformly in the $\Omega_m$-$\sigma_8$ parameter space, as seen in Figure~\ref{fig:sigma8_vs_As_sampling}. Instead, the wide, uniform prior on $A_s$ acts as an informative prior, which excludes the most extreme regions along the $\Omega_m$-$\sigma_8$ degeneracy. For this reason, we find from both mock analyses, and our data, that our constraints on $S_8$ are tighter when sampling $A_s$, rather than $S_8$ or $\sigma_8$, without affecting the inferred central value of $S_8$. (see Section~\ref{sec:results_robustness}).

We find that, for our fiducial analysis, the inferred posterior hits the prior boundary of $A_s$ and $\Omega_m$ (this is seen for $\Omega_m$ in Figure~\ref{fig:real_fourier_space_constraints}, due to the upper boundary on $A_s$). This is likely to happen because, as explained above, these two parameters are poorly constrained by our data, given our scale cuts. However, given our conservative priors on these parameters, we are not excluding realistic regions of parameter space. For these reasons, we believe our $S_8$ inference to be robust, but the constraints on $\Omega_m$, $A_s$, and $\sigma_8$ to be less so.

After obtaining a parameter chain from a converged sampling run, we reweight the chain in order to obtain flat priors for $\sigma_8$ and $\Omega_m$. This correction involves simply multiplying the weight of each sample by $\sigma_8/A_s$ (see \cite{Sugiyama2020} for a derivation of this correction). 

\section{\label{sec:results} Cosmological Constraints}

In this section, we present cosmological constraints from the tomographic cosmic shear power spectra measured with the HSC-Y3 shear catalog in the flat $\Lambda$CDM model (Section~\ref{sec:results_fiducial}). In Section~\ref{sec:results_robustness}, we test the robustness of our cosmological constraints against different modeling choices, including the modeling of baryons and intrinsic alignments, as well as redshift distribution uncertainties and PSF systematics. As described in Section~\ref{sec:results_internal_consistency}, we also perform internal consistency checks that ensure our constraints are not significantly affected by changing the scale cuts and redshift bins adopted for the analysis. Finally, in Section~\ref{sec:comparison}, we compare our results to other constraints from the literature. 

\subsection{\label{sec:results_fiducial} Fiducial Constraints}

We derive marginalized posterior contours in the $\Omega_m$-$\sigma_8$ plane from our fiducial model fit to the measured $\Cl$s. However, constraints from cosmic shear are known to be degenerate in this plane, and we have found that with our scale cuts, these parameters are poorly constrained (see Section~\ref{sec:priors}). Instead, we focus on a combination of these parameters, $S_8(\alpha) \equiv \sigma_8(\Omega_m/0.3)^{\alpha}$. While the Year 1 Fourier space analysis was able to get the tightest constraints with $\alpha=0.45$ \cite{Hikage2019}, we found that the standard value of $\alpha=0.5$ gives us the tightest constraints with the Y3 data.

\begin{figure*}
     \centering
     \includegraphics[width=0.95\textwidth]{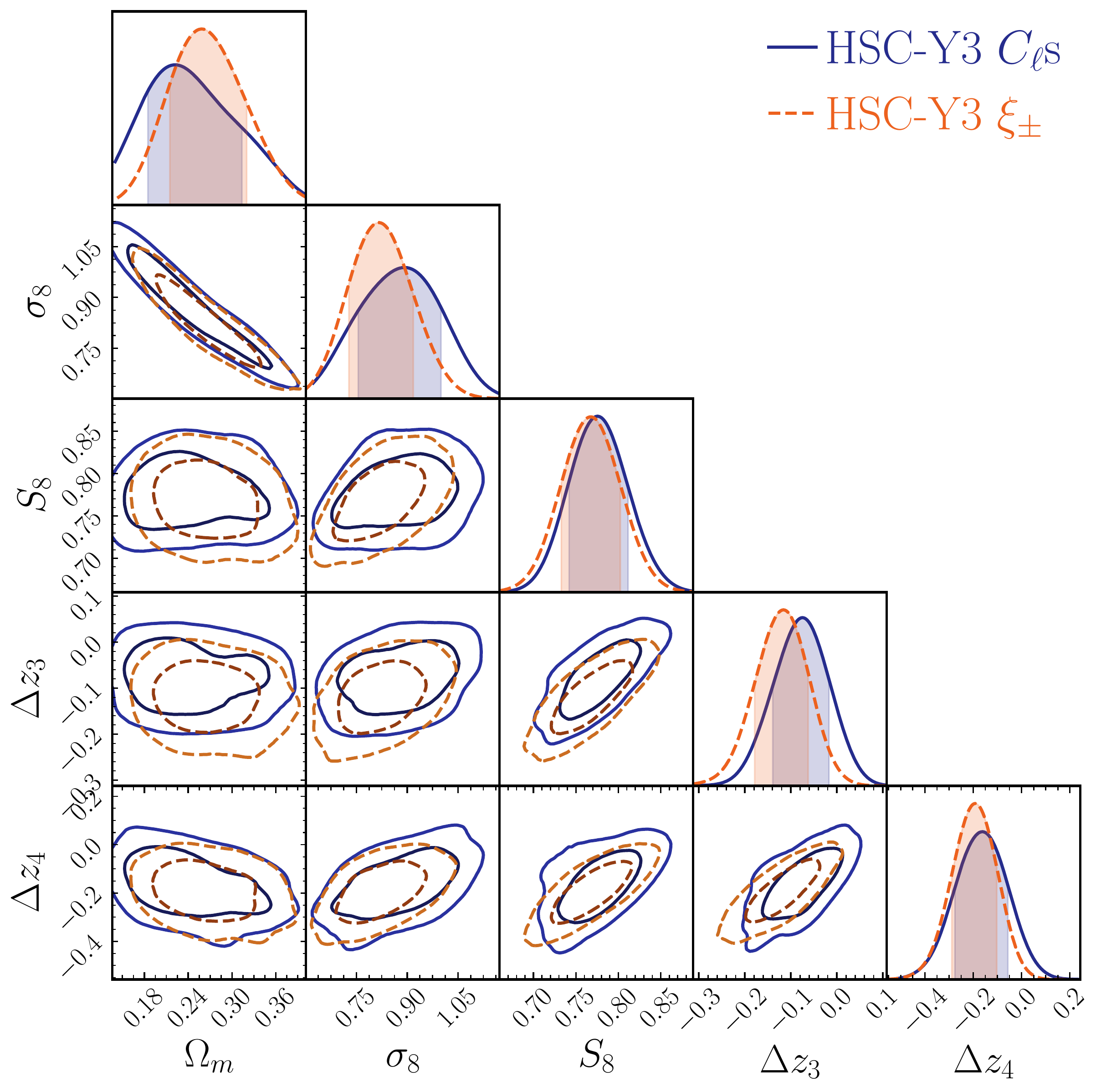}
     \caption{Cosmological constraints ($1\sigma$ and $2\sigma$ contours) from this analysis (blue, solid), which uses angular cosmic shear power spectra ($\Cl$s), and the constraints from a parallel cosmic shear analysis (\cite{Li2023}, orange, dashed), which uses two point correlation functions (2PCFs) along with the same data and coordinated analysis choices. The shading in the 1D posteriors represents the 68\% confidence interval for the constraints. We find excellent agreement between the two analyses.  We also include $\Delta z_3$ and $\Delta z_4$, the two ``nuisance'' parameters that have the strongest covariance with $S_8$.}
    \label{fig:real_fourier_space_constraints}
\end{figure*}

For $\alpha = 0.5$, we obtain the following constraints for the cosmological parameters of interest. We report the 1D marginalized mode and its asymmetric $\pm 34\%$ cconfidence intervals, with the maximum posterior point (MAP) of the nested sampling chain shown in parentheses:
\begin{itemize}
    \item $S_8 = 0.776^{+0.032}_{-0.033} (0.792)$,
    \item $\Omega_m = 0.219^{+0.075}_{-0.052} (0.226)$,
    \item $\sigma_8 = 0.900^{+0.100}_{-0.139} (0.913)$.
\end{itemize}
The 1D and 2D posteriors of these parameters are shown in Figure~\ref{fig:real_fourier_space_constraints}.

A parallel analysis used the same data, with coordinated analysis choices, to constrain $S_8$ using two point correlation functions (2PCFs) \cite{Li2023}. We also show the results of this analysis in Figure~\ref{fig:real_fourier_space_constraints}, and find excellent agreement between the inferred value of $S_8$ from the two analyses. 

\begin{figure*}
     \centering
     \includegraphics[width=0.95\textwidth]{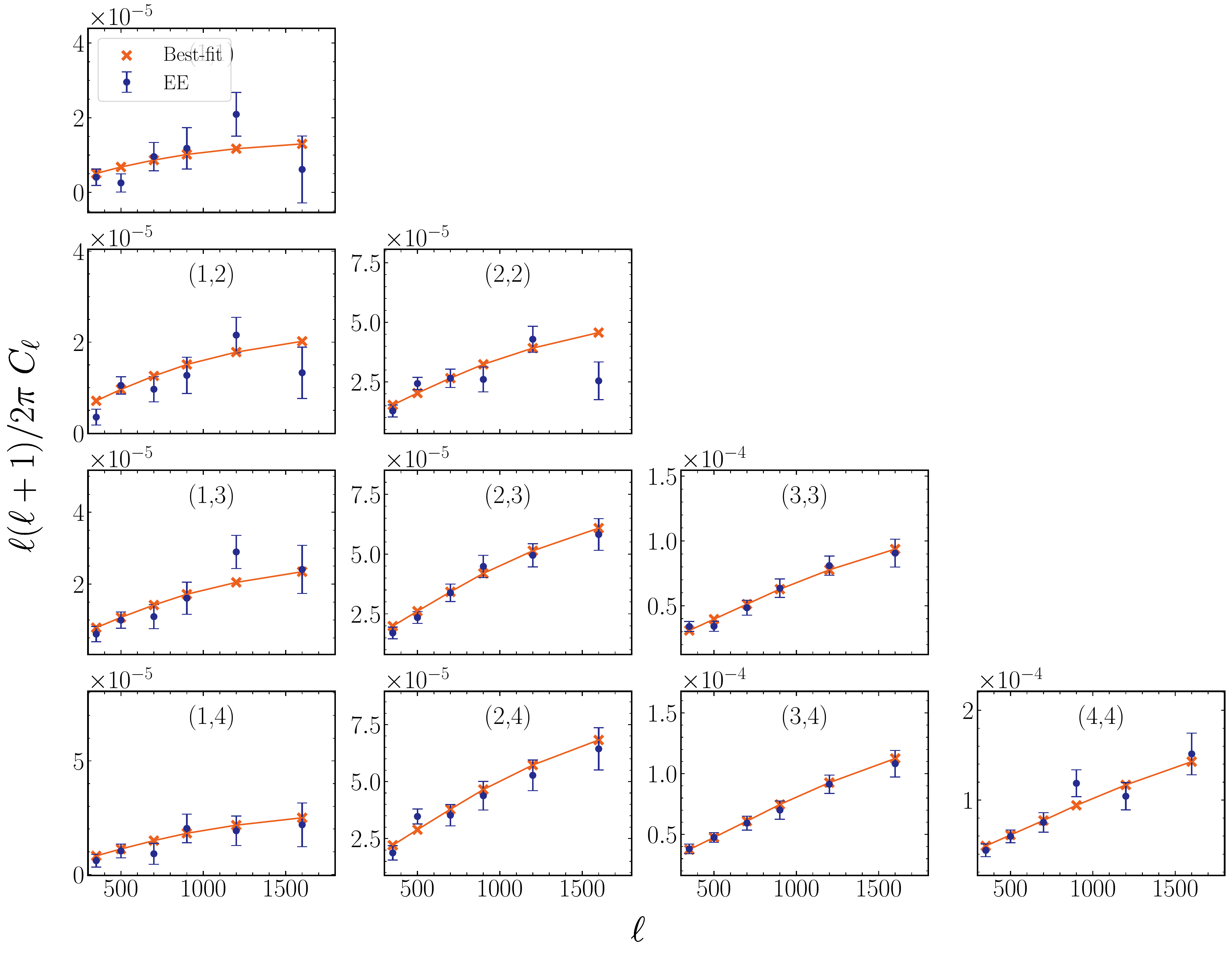}
     \caption{Comparison of the measured tomographic cosmic shear power spectra (blue dots, same as Figure~\ref{fig:power_spectra}) with our theoretical model using the best-fit values of our fiducial analysis (orange).}
    \label{fig:best_fit_spectra}
\end{figure*}

We compare our measured cosmic shear power spectra with our best-fitting model in Figure~\ref{fig:best_fit_spectra}. Before unblinding, we determine the goodness of fit of our data by taking the $\chi^2$ for the best-fit point, i.e. the MAP, or maximum posterior point, from the \texttt{Polychord} chain. Rather than attempting to estimate the effective number of degrees of freedom in our analysis (accounting for the correlations between parameters), we follow the method described in  \cite{Sugiyama2022}: we compare the $\chi^2$ of our data to a reference $\chi^2$ distribution obtained from simulating 50 noisy mock data vectors and applying the same analysis to them as the real data (using, however, \texttt{MultiNest} rather than \texttt{Polychord} due to its faster speed). By comparing our observed $\chi^2=55.38$ value to the reference distribution, we find a p-value of $p=0.42$ and conclude that the model provides a good fit to the data. 

We show the one- and two-dimensional posteriors of the other cosmological, astrophysical and observational systematics parameters in Appendix~\ref{app:other_params}.

\subsection{\label{sec:results_robustness} Robustness to Modeling and Analysis Choices}

To check the robustness of our results to our modeling and analysis choices, we change the setup of the analysis in various ways to test the impact on our cosmological constraints. This includes tests of different samplers, different sampled parameters, different astrophysical models for intrinsic alignments and baryonic feedback, as well as different models for observational systematics, including redshift distribution uncertainties and PSF systematics. Prior to unblinding, we check that the shift in our $S_8$ constraint is smaller than $1\sigma$ in each of these tests. Any shifts larger than $1\sigma$ are further tested and understood before unblinding.

\begin{figure}
\includegraphics[width=0.45\textwidth]{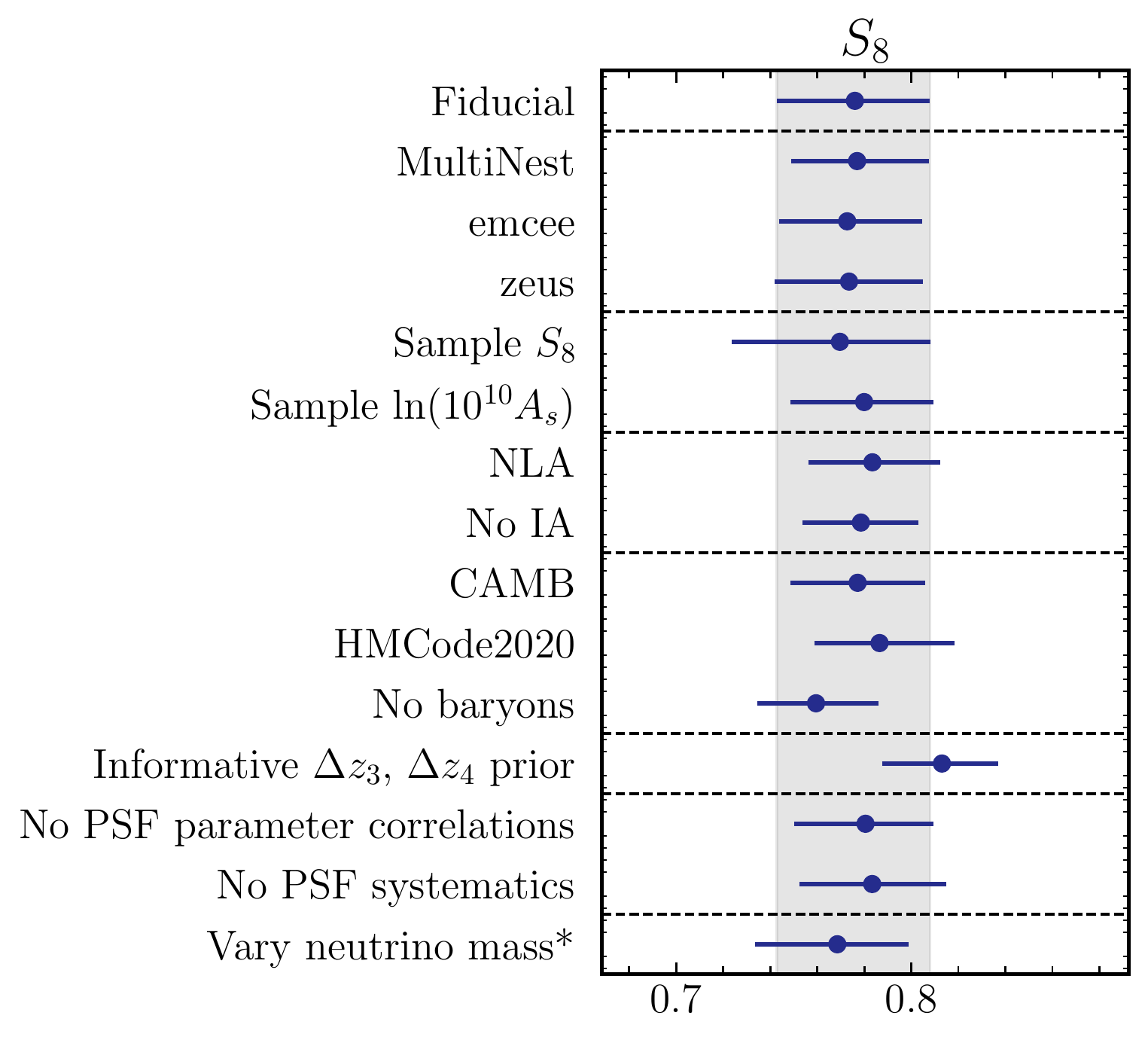}
\caption{\label{fig:consistency_checks} Constraints on $S_8$ for the tests of modeling and analysis choice robustness described in Section~\ref{sec:results_robustness}. The asterisk indicates that only the test with varying neutrino mass was performed after unblinding the analysis.}
\end{figure}

The results of these tests are summarized in Table~\ref{tab:consistency_checks} and Figure~\ref{fig:consistency_checks}. As mentioned in Section~\ref{sec:sampler}, we use the \texttt{MultiNest} sampler for these checks, as it is significantly faster to run than \texttt{PolyChord}. The resulting constraints are compared to our fiducial constraint, which uses \texttt{PolyChord}. As shown in Section~\ref{sec:sampler_consistency}, the central values of the constraints on $S_8$ from fiducial runs with \texttt{PolyChord} and \texttt{MultiNest} are consistent.  However, \texttt{MultiNest} underestimates the width of the posterior by $\sim$$10\%$, which causes the constraints from some of these consistency checks to appear to be stronger than those for our data. We describe each of the different setups used for these robustness checks in detail below. 
\begin{table}
\renewcommand*{\arraystretch}{1.8}
\caption{\label{tab:consistency_checks}% 
$S_8$ constraints for the tests of modeling and analysis choice robustness described in Section~\ref{sec:results_robustness}. We report the  posterior mode value and 68\% confidence bounds of $S_8$ for each of these tests, along with the maximum posterior point of the chain (MAP) in parentheses. We report the shift in each of these tests from the fiducial constraint as a fraction of the fiducial 68\% confidence bounds.}
\begin{ruledtabular}
\begin{tabular}{p{4cm} p{3.5cm}p{0.75cm}}
\textrm{Consistency test}&
\textrm{$S_8$ posterior mode, (MAP)}&
\textrm{Shift ($\sigma_{\mathrm{fid}}$)}\\
\colrule
Fiducial & $0.776^{+0.032}_{-0.033} (0.792)$ & N/A\\
\texttt{Multinest} & $0.777^{+0.031}_{-0.028} (0.775)$ & 0.03\\
\texttt{emcee} & $0.773^{+0.032}_{-0.029} (0.804)$ & -0.09\\
\texttt{zeus} & $0.774 \pm 0.032 (0.814)$ & -0.06\\
\hline
Sample $S_8$ & $0.770^{+0.039}_{-0.046} (0.770)$ & -0.18\\
Sample $\ln(10^{10}A_s)$ & $0.780^{+0.030}_{-0.031} (0.795)$ & 0.12\\
\hline
NLA & $0.783^{+0.029}_{-0.027} (0.799)$ & 0.22\\
No IA & $0.779 \pm 0.025 (0.798)$ & 0.09\\
\hline
\texttt{CAMB} & $0.777 \pm 0.029 (0.778)$ & 0.03\\
\texttt{HMCode2020} & $0.786^{+0.032}_{-0.028} (0.783)$ & 0.31\\
No baryons & $0.759^{+0.026}_{-0.025} (0.792)$ & -0.52\\
\hline
Informative $\Delta z_3$, $\Delta z_4$ prior & $0.813^{+0.025}_{-0.024} (0.788)$ & 1.14\\
\hline
No PSF parameter correlations & $0.781^{+0.029}_{-0.030} (0.771)$ & 0.15\\
No PSF systematics & $0.783 \pm 0.031 (0.774)$ & 0.22\\
\hline
Vary neutrino mass ($\sum m_{\nu}$)\footnote{We note that this test alone was performed after unblinding the analysis.} & $0.769^{+0.030}_{-0.035} (0.758)$ & -0.22\\
\end{tabular}
\end{ruledtabular}
\end{table}

\subsubsection{\label{sec:sampler_consistency} Sampler Choice}

While our fiducial constraints are from the \texttt{PolyChord} algorithm, we check that our $S_8$ constraint is not significantly affected by the choice of sampler. In addition to \texttt{MultiNest}, we also consider two other Markov Chain Monte Carlo (MCMC) samplers that use ensembles of ``walkers'' to explore the parameter space: \texttt{emcee} \cite{Foreman-Mackey2013} and \texttt{zeus} \cite{Karamanis2020, Karamanis2021}. In all of these cases, we run these samplers from within the \texttt{CosmoSIS} code \cite{Zuntz2015}.

As shown in Table~\ref{tab:consistency_checks} and Figure~\ref{fig:consistency_checks}, we find that the central value of $S_8$ is not affected by the choice of sampler, with shifts of at most $0.1\sigma$. However, \texttt{PolyChord}, \texttt{emcee} and \texttt{zeus} find a $\sim$$10\%$ wider posterior on $S_8$ compared to \texttt{MultiNest}. This is consistent with the findings of \cite{Lemos2022}, who have shown that \texttt{MultiNest} can underestimate the posterior width compared to \texttt{PolyChord}. We find that the constraints on $\Omega_m$ are shifted (by $0.36\sigma_{\mathrm{polychord}}$) with a severely underestimated posterior width (by a factor of 0.5) when using \texttt{MultiNest} compared to \texttt{PolyChord}, \texttt{emcee} and \texttt{zeus}. We interpret this as an inability of \texttt{MultiNest} to recover accurate posterior widths for parameters that are poorly constrained by the data.

\subsubsection{\label{sec:sampled_param_consistency} Sampled Parameters}

While our fiducial analysis samples $A_s$ (see Table~\ref{tab:fid_parameters}), we test the impact on our $S_8$ constraints if we instead sample $S_8$ or $\ln (10^{10}A_s)$. As shown in Table~\ref{tab:consistency_checks} and Figure~\ref{fig:consistency_checks}, we find that the shift in our central $S_8$ value from these different setups is not significant (smaller than $0.2\sigma$), although the $S_8$ constraint is degraded (by $\sim$$30\%$) when sampling $S_8$ as opposed to $A_s$ or $\ln (10^{10}A_s)$. This could be related to the informative prior that sampling $A_s$ provides in order to break the degeneracy between $\sigma_8$ and $\Omega_m$, which we are not able to do with our data and scale cuts alone (see Section~\ref{sec:priors}).

\subsubsection{\label{sec:ia_consistency} Intrinsic Alignment Model}

In our fiducial analysis, we use the Tidal Alignments and Tidal Torquing (TATT) intrinsic alignments model (see Section~\ref{sec:ia_tatt}). We test the robustness of our cosmological constraints to the intrinsic alignment model used by checking the $S_8$ constraints obtained using two different models. First, we use the Non-Linear Alignments model, described in Section~\ref{sec:ia_nla}. We also test the constraints with no intrinsic alignment modeling, i.e. assuming that there is no IA signal. The results of these tests are shown in Table~\ref{tab:consistency_checks} and Figure~\ref{fig:consistency_checks}. We find no significant shift in our $S_8$ constraint from changing the IA model (at most $\sim$$ 0.2\sigma$), although the constraints are stronger when using these simpler models, given that they have fewer parameters. As discussed in Section~\ref{sec:astro_syst}, we do not detect an IA signal in our fiducial analysis, further confirming that our analysis is not sensitive to the choice of IA model.

\subsubsection{\label{sec:baryons_consistency} Matter Power Spectrum/Baryonic Feedback Model}

Our fiducial analysis models the linear matter power spectrum with the \texttt{BACCO} emulator \cite{Arico2021}, and the nonlinear power spectrum using \texttt{HMCode 2016}  \cite{Mead2016}, implemented in \texttt{pyhmcode} \cite{Troster2022}. We explore the impact of using other power spectrum models on our $S_8$ constraint by changing the linear power spectrum modeling to \texttt{CAMB} \cite{Lewis2000}, and subsequently, the nonlinear modeling to \texttt{HMCode 2020} \cite{Mead2021}. We also test the impact of continuing to use the \texttt{BACCO} emulator and \texttt{HMCode 2016}, without the modeling of baryonic effects, by fixing the $A_{\mathrm{bary}}$ parameter to 3.13. As shown in Table~\ref{tab:consistency_checks} and Figure~\ref{fig:consistency_checks}, we find consistent $S_8$ constraints in the first two cases, and a $0.5\sigma$ shift to a lower $S_8$ value when not modeling baryonic effects. This shift is not surprising, as we do have a $1.5\sigma$ detection of baryonic effects from our data ($A_{\mathrm{bary}} = 2.43^{+0.46}_{-0.25}$, as described in Section~\ref{sec:astro_syst}). We have shown, in Section~\ref{sec:model_sufficiency}, that our model and choice of scale cuts should be robust to a range of different baryonic feedback scenarios. 

\subsubsection{\label{sec:n(z)_err_consistency} Redshift Distribution Uncertainties}

\begin{figure}
\includegraphics[width=0.45\textwidth]{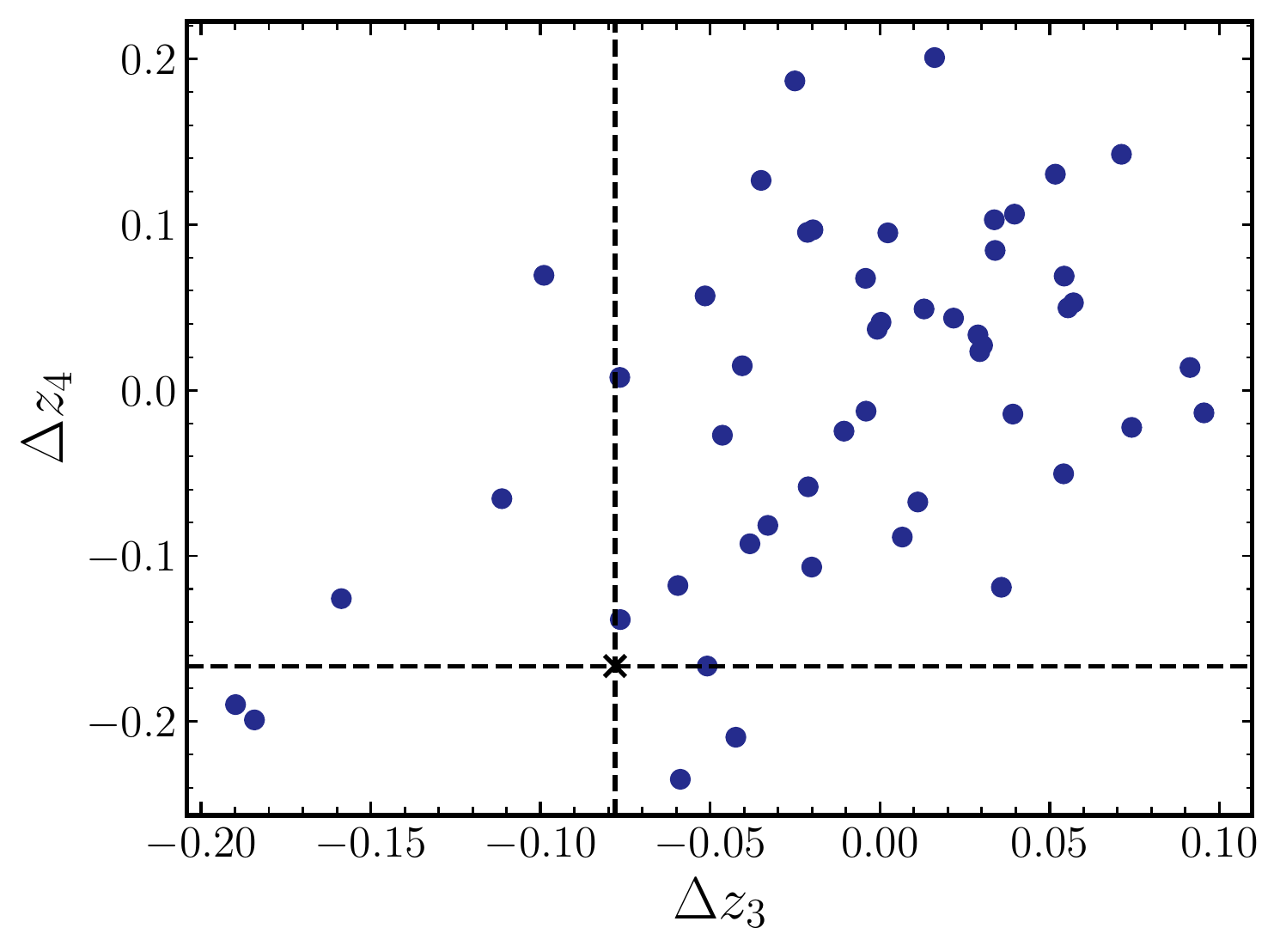}
\caption{\label{fig:deltaz_mocks} $\Delta z_3$ and $\Delta z_4$ values inferred from 50 noisy mock data vectors generated using the best-fit fiducial model, with the input $\Delta z_3$ and $\Delta z_4$ set to 0 (blue points). The fiducial inferred values of $\Delta z_3$ and $\Delta z_4$ (black cross), are larger than 90\% of the values inferred from these mocks, suggestive of detections of non-zero shifts, but with a significance at a less than $2\sigma$ level.}
\end{figure}

As described in Section~\ref{sec:nz_uncertainty}, there is a potential for biases in the photometric redshifts in our third and fourth redshift bins, especially because these bins are respectively only partially and not at all calibrated by the LRG clustering redshifts described in Section~\ref{sec:n(z)}. For this reason, in our fiducial analysis, we adopt a conservative, wide, flat prior for the shift of the mean redshift in these tomographic bins. Here, we explore the impact of using the Gaussian prior recommended by \cite{Rau2022} for $\Delta z_3$ and $\Delta z_4$. Table~\ref{tab:consistency_checks} and Figure~\ref{fig:consistency_checks} show a larger than $1\sigma$ shift in the $S_8$ constraint when using these Gaussian priors. As the uniform priors are the more conservative choice, we believe this to be evidence of systematic biases in these last two redshift bins. In fact, the conservative, flat priors used in the fiducial analysis give the constraints $\Delta z_3 = -0.076^{+0.056}_{-0.059}$ and $\Delta z_4 = -0.157^{+0.094}_{-0.111}$, which would be inconsistent with the Gaussian prior. To further test this, we use 50 noisy mock data vectors generated from the covariance matrix using the WMAP9 cosmology \cite{WMAP9} and the best-fit fiducial model for the astrophysical and observational systematics parameters, with the input $\Delta z_3$ and $\Delta z_4$ set to 0, and run the fiducial analysis on these mocks. We find that we only get a shift as large as the fiducial $\Delta z_3$ constraint 5 times, and a $\Delta z_4$ value as large as the fiducial constraint 5 times, as shown in Figure~\ref{fig:deltaz_mocks}. This indicates that these detections of shifts in the third and fourth redshift bin are significant at a $\sim$$2\sigma$ level. We leave the study and calibration of these biases to future work. We note that the conservative priors on $\Delta z_3$ and $\Delta z_4$ are a large part of the reason why our constraints on $S_8$ are not tighter compared to the Year 1 analyses \cite{Hikage2019}, despite the higher significance measurement of the power spectrum\footnote{The sky coverage of HSC Year 3 increases to  $416 \ \mathrm{deg}^2$ from $137 \ \mathrm{deg}^2$ for HSC Year 1. Based on this increase in area, we would expect the Y1 error of $\pm 0.0315$ to shrink by a factor of $\sim$$0.57$, giving an error of $\pm 0.018$. With the Gaussian informative prior on $\Delta z_3$ and $\Delta z_4$, the error on $S_8$ is $\pm 0.0245$, $\sim$$25\%$ smaller than the error for the fiducial analysis, $\pm 0.0325$. This explains a large percentage of the difference between our fiducial constraints and those expected from the increase in area. Further differences might be explained by a number of other conservative modeling choices made in this Y3 analysis, such as the scale cuts and use of the TATT model.}.

\subsubsection{\label{sec:psf_consistency} PSF Systematics Model}

As described in Section~\ref{sec:psf_model}, we adopt a model of the PSF systematics which accounts for correlations between the parameters. We explore the dependence of our $S_8$ constraint on this model choice by first checking the constraint when sampling the original set of PSF parameters $p = [\alpha^{(2)}, \beta^{(2)}, \alpha^{(4)}, \beta^{(4)}]$, such that the correlation between them is not accounted for. We also check our constraints when not modeling PSF systematics at all. Table~\ref{tab:consistency_checks} and Figure~\ref{fig:consistency_checks} show only small changes ($0.15\sigma$ and $0.22\sigma$ respectively)in the $S_8$ constraint from these different setups, indicating that constraints from cosmological parameters using this data set are not sensitive to the choice of PSF systematics model. However, we note that we still need higher order PSF moments, as shown by \cite{Zhang2022}, in order to self-consistently describe the shear two-point correlations and null tests. Moreover, our fiducial analysis results in significant detections of the PSF systematics parameters: $\alpha^{(2)} = 0.027^{+0.004}_{-0.003}$, $\beta^{(2)} = -0.394^{+0.033}_{-0.036}$, $\alpha^{(4)} = 0.176^{+0.014}_{-0.017}$, and $\beta^{(4)} = -0.194^{+0.075}_{-0.070}$. 

\subsubsection{\label{sec:neutrino_mass} Neutrino Mass}

In our fiducial analysis, we fix the sum of neutrino masses, $\sum m_{\nu} = 0.06 \mathrm{eV}$, as we do not expect our data to be able to constrain this parameter. The effect of a change in neutrino mass on large-scale strucutre observables, such as cosmic shear, would be absorbed by a change in $\sigma_8$, as a larger neutrino mass would lead to a suppression in the matter power spectrum amplitude at small scales over the range of redshifts that HSC can probe. One would only be able to constrain $\sum m_{\nu}$ by adding CMB constraints, which probe the amplitude of the matter power spectrum at higher redshifts. As such a joint constraint is outside the scope of this work, here we simply explore the effect of allowing neutrino mass to vary in our analysis, with a uniform prior from 0.06 eV to 0.6 eV. We note that this test, unlike the others described above and below, was performed after unblinding the analysis. We find very little change in our $S_8$ constraint, with a shift from the fiducial constraint of $0.22\sigma$.  

\subsection{\label{sec:results_internal_consistency} Internal Consistency}

We also check the robustness of our results to various splits of our data. This includes splitting our data by each field of the survey, using different photometric redshift codes for the source redshift distribution inference, removing each tomographic bin in turn, and adopting different sets of scale cuts. As with the tests of modeling and analysis choices, we require that the shifts in our $S_8$ constraint from each of these internal consistency tests be smaller than $1\sigma$ before unblinding. Any shifts larger than $1\sigma$ were investigated and understood prior to unblinding.

\begin{figure}
\includegraphics[width=0.45\textwidth]{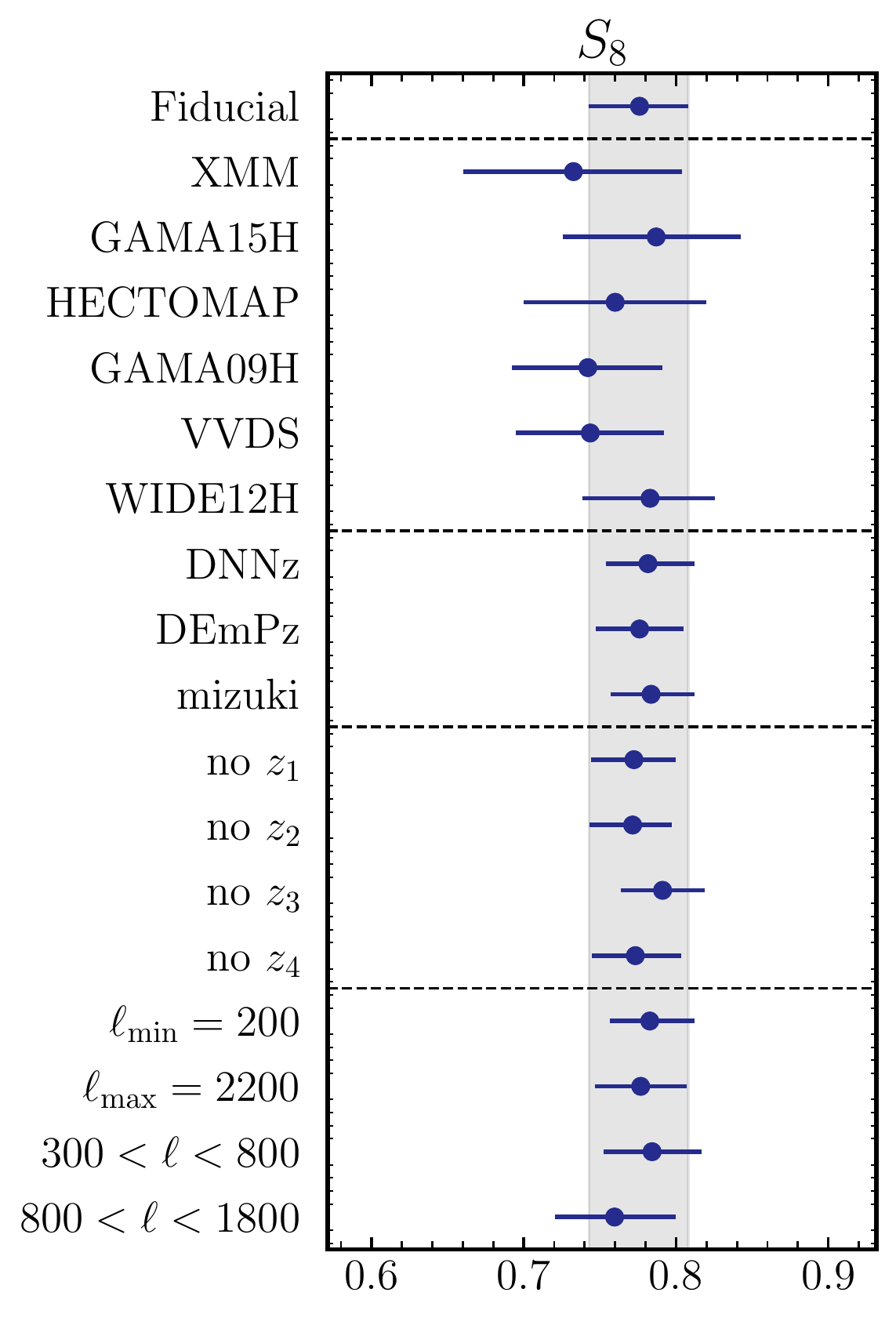}
\caption{\label{fig:internal_consistency_checks} Constraints on $S_8$ for the tests of internal consistency described in Section~\ref{sec:results_internal_consistency}.}
\end{figure}

The results of these tests are summarized in Table~\ref{tab:internal_consistency_checks} and Figure~\ref{fig:internal_consistency_checks}. Again, we use the \texttt{MultiNest} sampler for these checks, rather than \texttt{PolyChord}, but compare the resulting constraints to our fiducial constraints, which use \texttt{PolyChord}. We describe each of the tests in detail below. 
\begin{table}
\renewcommand*{\arraystretch}{1.8}
\caption{\label{tab:internal_consistency_checks}% 
$S_8$ constraints for the tests of internal consistency described in Section~\ref{sec:results_internal_consistency}. We report the  posterior mode value and 68\% confidence bounds of $S_8$ for each of these tests, along with the maximum posterior point of the chain (MAP) in parentheses. We report the shift in each of these tests from the fiducial constraint as a fraction of the fiducial 68\% confidence bounds.}
\begin{ruledtabular}
\begin{tabular}{p{3.5cm} p{3.5cm}p{1cm}}
\textrm{Consistency test}&
\textrm{$S_8$ posterior mode, (MAP)}&
\textrm{Shift ($\sigma_{\mathrm{fid}}$)}\\
\colrule
Fiducial & $0.776^{+0.032}_{-0.033} (0.792)$ & N/A\\
XMM & $0.733^{+0.071}_{-0.072} (0.710)$ & -1.32\\
GAMA15H & $0.787^{+0.055}_{-0.061} (0.837)$ & 0.34\\
HECTOMAP & $0.760 \pm 0.060 (0.777)$ & -0.49\\
GAMA09H & $0.742^{+0.049}_{-0.050} (0.782)$ & -1.05\\
VVDS & $0.744^{+0.048}_{-0.049} (0.724)$ & -0.98\\
WIDE12H & $0.783^{+0.043}_{-0.044} (0.785)$ & 0.22\\
\hline
\texttt{DNNz} & $0.782^{+0.030}_{-0.028} (0.822)$ & 0.18\\
\texttt{DEmPz} & $0.776 \pm 0.029 (0.798)$ & 0.0\\
\texttt{mizuki} & $0.784^{+0.028}_{-0.026} (0.802)$ & 0.25\\
\hline
No $z_1$ & $0.772^{+0.027}_{-0.028} (0.750)$ & -0.12\\
No $z_2$ & $0.771^{+0.026}_{-0.028} (0.782)$ & -0.15\\
No $z_3$ & $0.791^{+0.028}_{-0.027} (0.810)$ & 0.46\\
No $z_4$ & $0.773^{+0.030}_{-0.029} (0.770)$ & -0.09\\
\hline
$\ell_{\mathrm{min}} = 200$ & $0.783^{+0.030}_{-0.026} (0.819)$ & 0.22\\
$\ell_{\mathrm{max}} = 2200$ & $0.777 \pm 0.030 (0.785)$ & 0.03\\
$300 < \ell < 800$ & $0.784 \pm 0.032 (0.783)$ & 0.25\\
$800 < \ell < 1800$ & $0.760^{+0.040}_{-0.039} (0.765)$ & -0.49\\
\end{tabular}
\end{ruledtabular}
\end{table}

\subsubsection{\label{sec:fields_consistency} Individual Fields}

The HSC-Y3 data is split into six different fields on the sky, and as described in Section~\ref{sec:power_spectra}, we measure power spectra independently for each of these fields. Here, we assess the consistency in $S_8$ results from carrying out the fiducial analysis on each field individually. We note that these fields have a range of sizes - in order of increasing area, they are: XMM ($33.17 \ \mathrm{deg}^2$), GAMA15H ($40.87 \ \mathrm{deg}^2$), HECTOMAP ($43.09 \ \mathrm{deg}^2$), GAMA09H ($78.85 \ \mathrm{deg}^2$), VVDS ($96.18 \ \mathrm{deg}^2$), and WIDE12H ($121.32 \ \mathrm{deg}^2$). As shown in Table~\ref{tab:internal_consistency_checks} and Figure~\ref{fig:internal_consistency_checks}, we find that XMM shows a $\sim$$1.3\sigma$ shift in $S_8$, while GAMA09H and VVDS each show a shift of $\sim 1$$\sigma$. Since these fields are each quite small, and essentially independent, these shifts are not a cause for concern.

\subsubsection{\label{sec:n(z)_consistency} Source Redshift Distribution}

We compare the constraints on $S_8$ obtained from the fiducial source redshift distribution inference (see Section~\ref{sec:n(z)}) to the stacked photometric redshift PDFs from each of the three photo-$z$ codes described in Section~\ref{sec:photoz}. Table~\ref{tab:internal_consistency_checks} and Figure~\ref{fig:internal_consistency_checks} show that the resulting $S_8$ constraints are consistent with the fiducial model in every case, with shifts no larger than $0.25\sigma$. Furthermore, we find that the large shifts in the third and fourth bin redshift distributions (discussed in Section~\ref{sec:n(z)_err_consistency}) persist when using these different source redshift distributions, with $\Delta z_3$ and $\Delta z_4$ values that are consistent with the fiducial analysis.

\subsubsection{\label{sec:zbins_consistency} Redshift Bins}

The comparison of our $S_8$ constraints across different tomographic bins is an important internal consistency check (\cite{Efstathiou2018, Kohlinger2019}). For this test, we exclude one tomographic bin at a time and see whether the results are consistent with the fiducial result using all four bins. Our fiducial analysis adopts wide, flat priors on $\Delta z_3$ and $\Delta z_4$. The combination of these conservative priors and the removal of bins $z_1$ or $z_2$ would result in a major reduction in the constraining power of this test data set. To avoid this issue when conducting this test, we adopt informative priors on $\Delta z_3$ and $\Delta z_4$, taken from the fiducial posteriors of these parameters, $\Delta z_3 = -0.076 \pm 0.0575$ and $\Delta z_4 = -0.157 \pm 0.1025$. We find, as shown in Table~\ref{tab:internal_consistency_checks} and Figure~\ref{fig:internal_consistency_checks}, that the resulting $S_8$ constraints from removing each redshift bin are consistent with the fiducial constraint within $0.5\sigma$. However, we note that these cases are not entirely independent, as the priors used for $\Delta z_3$ and $\Delta z_4$ come from the fiducial posterior. 

\subsubsection{\label{sec:scales_consistency} Scale Cuts}

Finally, we check the internal consistency among different multipole bins. We first check that extending the upper and lower limits of our scale cuts to $\ell_{\mathrm{min}} = 200$ and $\ell_{\mathrm{max}} = 2200$ does not significantly change the value of $S_8$ (with shifts of $0.22\sigma$ and $0.03\sigma$ respectively). This suggests that our fiducial multipole range of $300 < \ell < 1800$ is a conservative choice. We then split the fiducial multipole range in half, first considering only large scales ($300 < \ell < 800$), and then considering only small scales ($800 < \ell < 1800$). While the $S_8$ constraint from large scales is consistent with the fiducial value (within $0.25\sigma$), there is a $\sim 0.5\sigma$ shift to a lower $S_8$ value when only considering small scales. However, since this shift is within $1\sigma$ of our fiducial constraint, this does not represent a significant bias. 

\subsection{\label{sec:comparison} Comparison to Other Constraints from the Literature}

\begin{figure}
\includegraphics[width=0.45\textwidth]{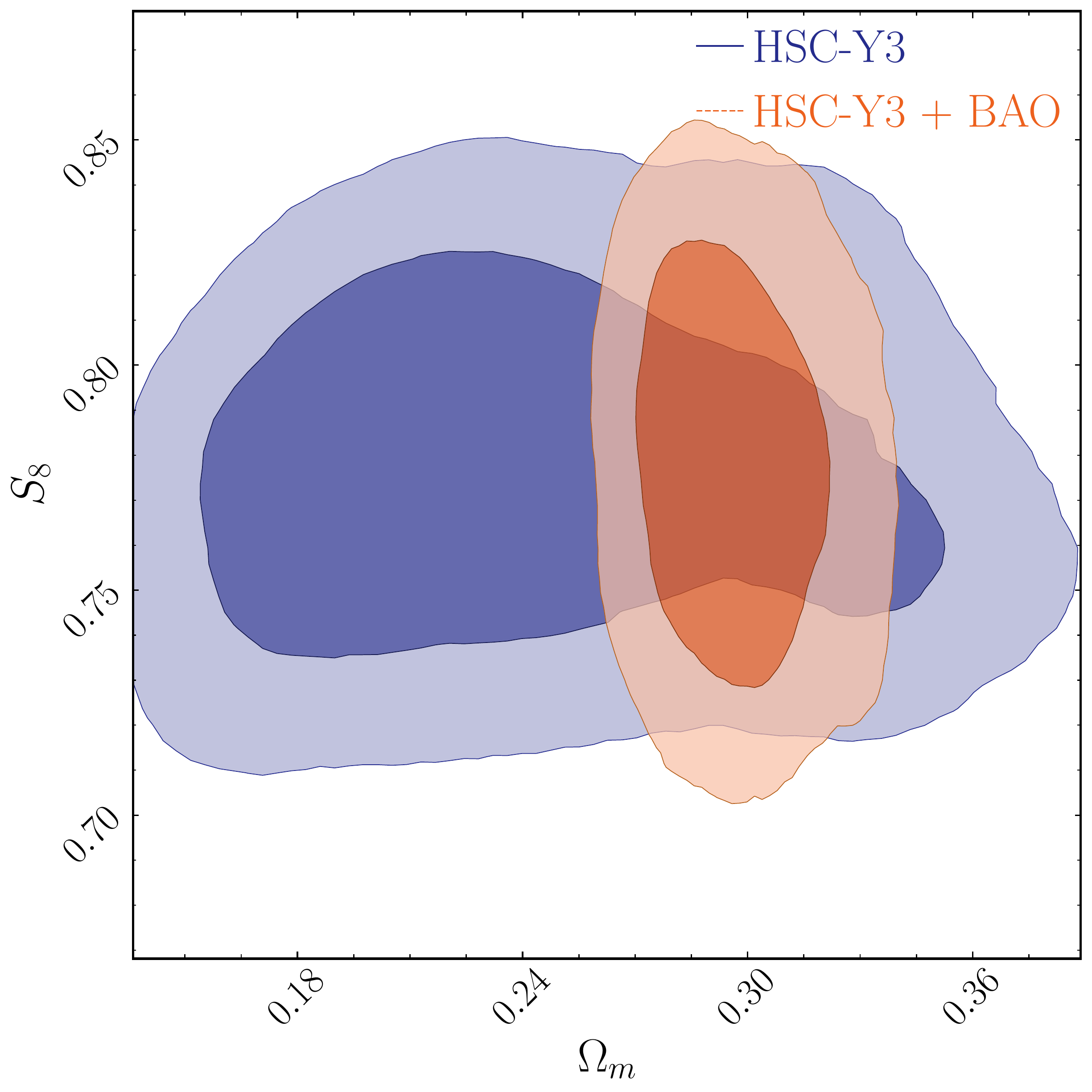}
\caption{\label{fig:results_bao_joint} Constraints on $\Omega_m$ and $S_8$ from this work (blue, solid), compared to the constraints from a joint analysis with the eBOSS DR16 BAO likelihoods (orange, dashed). Our fiducial $S_8$ constraint is unchanged by the addition of eBOSS data, while the joint analysis greatly improves our constraint on $\Omega_m$.}
\end{figure}

We first combine our fiducial constraints with those from Baryon Acoustic Oscillations (BAOs). We use the extended Baryon Oscillation Spectroscopic Survey (eBOSS) DR16 \cite{Alam2021} likelihood implemented in \texttt{CosmoSIS}. This analysis measures the BAO feature in the galaxy clustering power spectrum using a number of different galaxy samples, including the SDSS main galaxy sample (MGS, \cite{Ross2015}), BOSS DR12 galaxies \cite{Alam2017}, eBOSS Luminous Red Galaxies (LRGs, \cite{Bautista2021}), eBOSS Emission Line Galaxies (ELGs, \cite{deMattia2021}), eBOSS quasars (QSOs, \cite{Neveux2020}) and eBOSS Lyman-$\alpha$ Forest Samples (Ly$\alpha$, \cite{duMas2020}). We perform a joint likelihood analysis using our fiducial HSC-Y3 likelihood and the likelihoods for each of these BAO samples, assuming no correlations between the HSC and BAO likelihoods. The results of this analysis are shown in Figure~\ref{fig:results_bao_joint}. We find that our $S_8$ constraint is unchanged by the addition of BAO data, however the BAO data does greatly improve our constraint on $\Omega_m$ in the joint analysis.

We now compare our results to other constraints on $S_8$ and $\Omega_m$ from the literature, including:

\begin{itemize}
    \item \textit{Planck} 2018 cosmological constraints derived from primary CMB information (``TT, EE, TE+lowE''), without CMB lensing \cite{Planck2018Cosmology} \footnote{downloaded from the \textit{Planck} wiki: \url{https://pla.esac.esa.int/pla/aio/product-action?COSMOLOGY.FILE_ID=COM_CosmoParams_fullGrid_R3.01.zip}}. 
    \item Dark Energy Survey Year 3 (DES-Y3) cosmic shear constraints with the ``Maglim'' sample \cite{Amon2022, Secco2022} \footnote{downloaded from the DES website: \url{https://des.ncsa.illinois.edu/releases/y3a2/Y3key-products}}.
    \item The Kilo-Degree Survey 1000 (KiDS-1000) cosmic shear constraints \cite{Asgari2021} \footnote{downloaded from the KiDS website: \url{https://kids.strw.leidenuniv.nl/DR4/KiDS-1000_cosmicshear.php}}.
    \item The HSC Year 1 cosmic shear constraints from cosmic shear power spectra \cite{Hikage2019} \footnote{downloaded from the HSC Public Data Release 2 website: \url{https://hsc-release.mtk.nao.ac.jp/archive/filetree/s16a-shape-catalog/pdr1_hscwl/}}.
    \item The HSC Year 1 cosmic shear constraints from cosmic shear two point correlation functions \cite{Hamana2020} \footnote{downloaded from: \url{http://th.nao.ac.jp/MEMBER/hamanatk/HSC16aCSTPCFbugfix/index.html}}.
\end{itemize}
\begin{figure}
\includegraphics[width=0.45\textwidth]{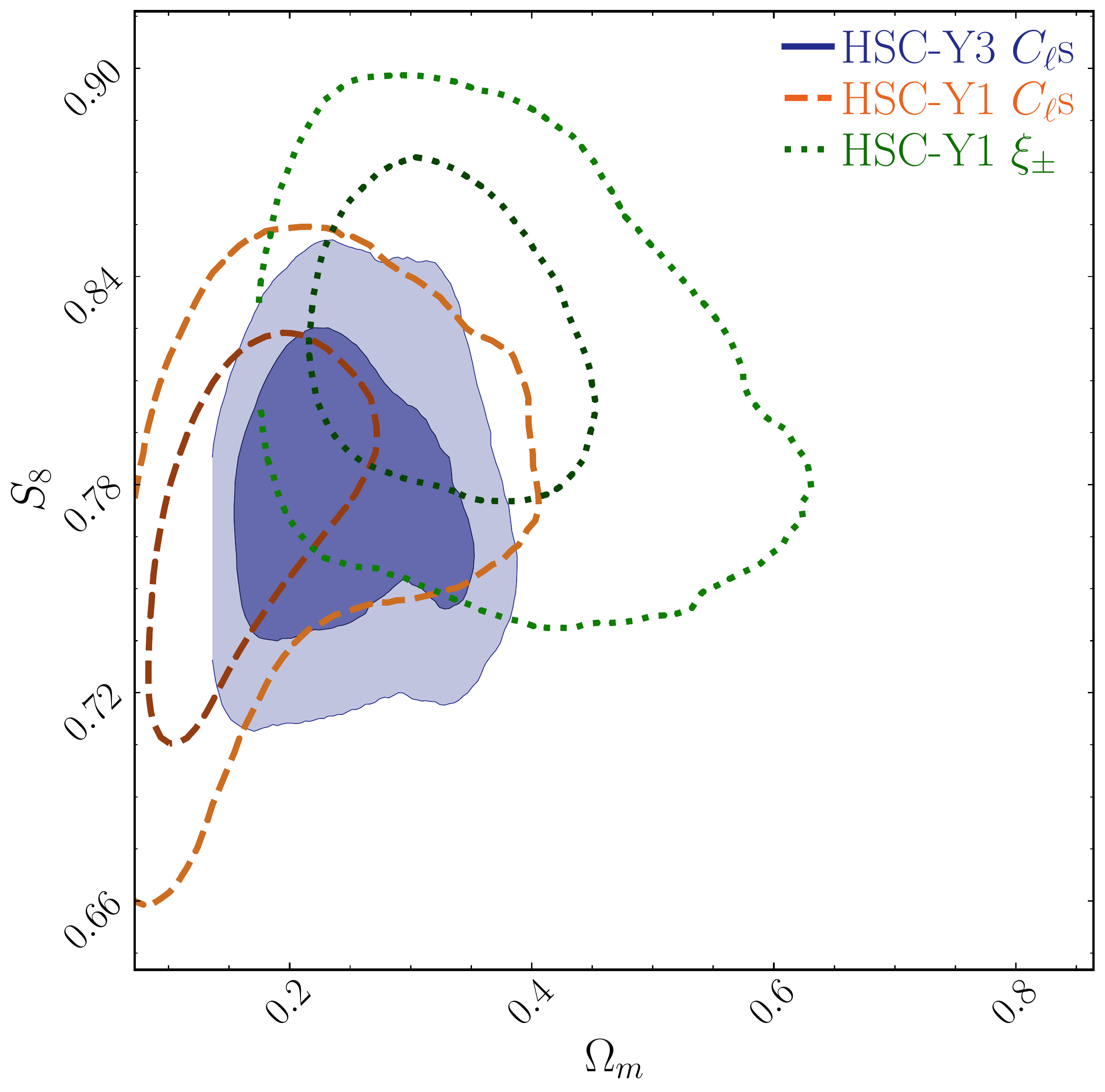}
\caption{\label{fig:results_comparison_year1} Constraints on $\Omega_m$ and $S_8$ from this work (blue, solid), compared to the constraints from the HSC Year 1 analyses based on $\Cl$s (orange, dashed) and 2PCFs (green, dotted). The  HSC-Y3 $\Omega_m$ contour is cut off at low values because the $A_s$ parameter posterior, which is poorly constrained by this analysis, hits the prior boundary (at the upper limit of $A_s = 10^{-8}$).}
\end{figure}

As shown in Figure~\ref{fig:results_comparison_year1}, we find that our results are generally in agreement with those from the HSC Year 1 analyses, although our central value of $S_8$ is lower than that from both Year 1 analyses. As discussed in Section~\ref{sec:photoz_err_discussion}, our uncertainties on $S_8$ are similar to those of the Year 1 analyses, because we have adopted a conservative flat prior on $\Delta z_3$ and $\Delta z_4$.
\begin{figure*}
\includegraphics[width=0.95\textwidth]{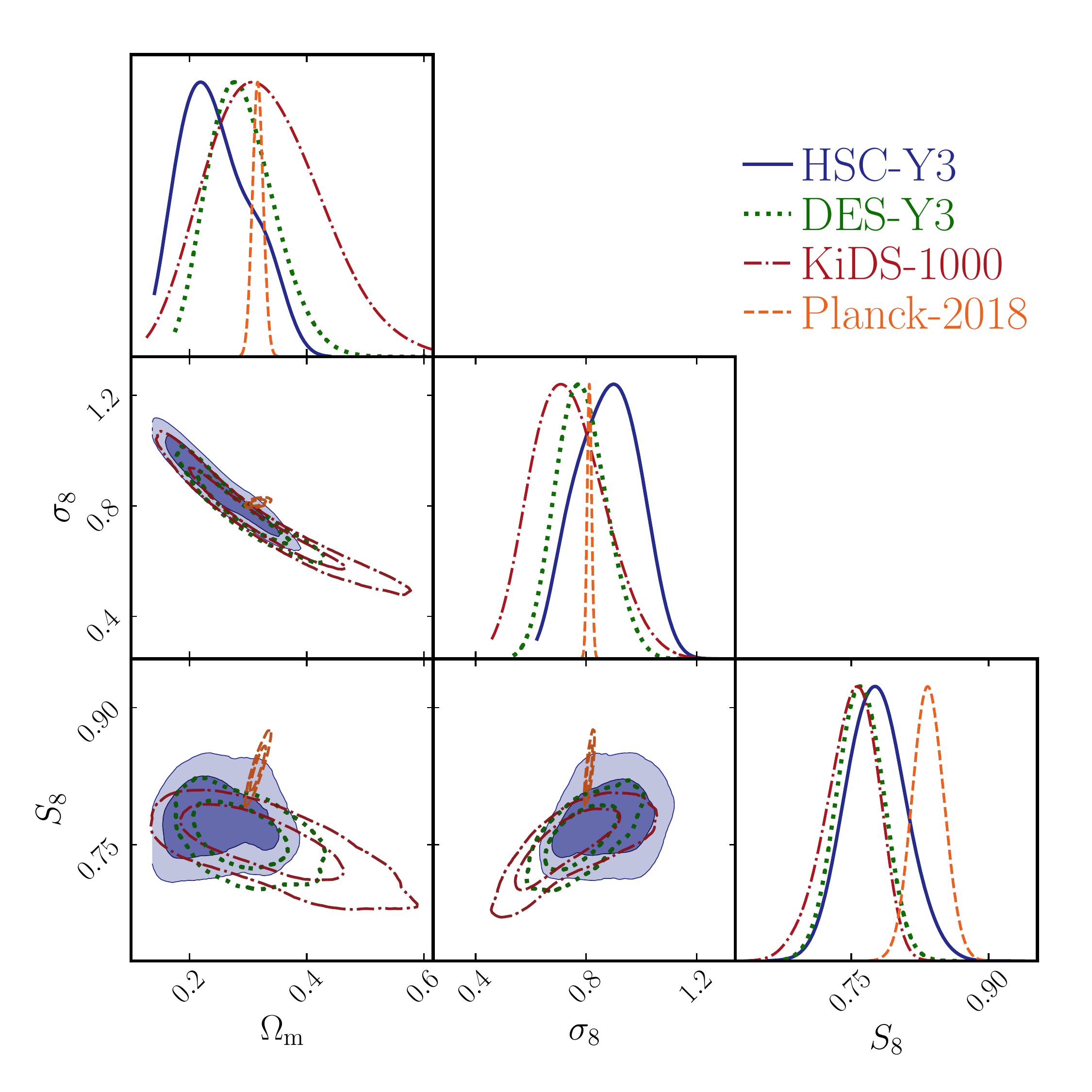}
\caption{\label{fig:results_comparison} Recent constraints on $\Omega_m$, $\sigma_8$ and $S_8$ from the external experiments listed in Section~\ref{sec:comparison}, including the constraints obtained in this work (HSC, blue, solid), from \textit{Planck} 2018 primary CMB data (orange, dashed), from DES-Y3 cosmic shear (green, dotted), and from KiDS-1000 cosmic shear (red, dashed-dotted). The 1D histograms are normalized to be the same height.}
\end{figure*}

A comparison of our results and those from \textit{Planck} 2018, DES-Y3, and KiDS-1000 is shown in Figure~\ref{fig:results_comparison}. We find that our results agree well with those from the recent cosmic shear experiments, DES-Y3 and KiDS-1000. However, our $S_8$ constraint appears to be in modest tension with the constraint from \textit{Planck} 2018. To investigate this further, we quantify the tension in $S_8$ using Method 3 of \cite{Charnock2017}. We first use importance sampling to generate chains of equal length from the original \textit{Planck} chain and our fiducial chain for this analysis. We then compute a new chain taking the difference between the posterior samples from \textit{Planck} and those from our analysis, $\Delta p$, and evaluate the probability enclosed within the posterior contour intersecting the point $\Delta p  = 0$. From this method, we find that our $S_8$ constraint has a 96\% probability of being in tension with the \textit{Planck} 2018 constraint, corresponding to a $2.02\sigma$ tension. This is a slightly smaller tension with \textit{Planck} than seen in DES-Y3 ($2.62\sigma$) and KiDS-1000 ($3.54\sigma$) analyses, but larger than the tensions seen in the HSC Year 1 $\Cl$ analysis ($1.71\sigma$) and HSC Year 1 $\xi_{\pm}$ analysis ($0.51\sigma$). The tensions for these analyses were computed using publicly available chains and the  tension metric described above.

\section{\label{sec:conclusion} Summary and Discussion}

We have presented results of the cosmic shear power spectrum analysis using the HSC third-year data. The data covers $\sim$$416 \ \mathrm{deg}^2$ of the sky, with exquisite depth and image quality, allowing us to measure precise shapes for galaxies in the redshift range $0.3 \leq z \leq 1.5$, with an effective number density of $\sim$$15 \ \mathrm{arcmin}^{-2}$ \cite{Li2022}. 

We have measured cosmic shear power spectra from the Y3 shape catalog using the Pseudo-$\Cl$ method to recover unbiased power spectra estimates. We obtain a power spectrum measurement with a high signal-to-noise ratio of 26.4 within the multipole range $300 \leq \ell \leq 1800$, and no significant detection of $B$ modes in this range. We measure the covariance matrix of these power spectra using the HSC mock shear catalogs \cite{Shirasaki2019}.

We fit the power spectra with a model which includes astrophysical effects, such as baryonic feedback and the intrinsic alignments of galaxies, as well as nuisance parameters to capture uncertainties in the source redshift distribution, PSF systematics, and shape measurement. We perform extensive model selection tests to ensure that model mis-specification would not significantly bias our cosmological parameter constraints. Our best-fit model fits the measured power spectra well, with a minimum $\chi^2$ of 55.38, and a corresponding $p$-value of 0.42. Throughout this process, we follow a careful and thorough blinding process to prevent confirmation bias from affecting our results. 

We constrain the parameter $S_8 \equiv \sigma_8(\Omega_m/0.3)^{0.5}$, and assuming a flat $\Lambda$CDM model, find the posterior mode $S_8 = 0.776^{+0.032}_{-0.033}$, with a maximum posterior at $S_8 = 0.792$. We conduct a number of tests of our modeling and analysis choices, as well as internal consistency checks, and show that this constraint is robust, with shifts generally not exceeding $0.5 \sigma$ (with one notable exception related to residual photometric redshift errors, discussed further below). Our constraints on $S_8$ agree extremely well with those from other HSC Year 3 cosmology analyses \cite{Li2023, Sugiyama2023, Miyatake2023}, and are also in agreement with constraints from other recent cosmic shear experiments \cite{Amon2022, Asgari2021}. However, these results are $2\sigma$ lower than the constraint from \textit{Planck} 2018 \cite{Planck2018Cosmology}, following a trend seen in a number of cosmic shear experiments \cite{Hikage2019, Amon2022, Doux2022, Asgari2021, Loureiro2022}. Further studies will be necessary to understand whether this trend is an indication that the $\Lambda$CDM model is not sufficient to describe both CMB and large scale structure data. 

Below, we discuss potential current limitations and future work that will be needed to better understand this apparent tension.

\subsection{\label{sec:photoz_err_discussion} Residual Photo-\texorpdfstring{$z$}{z} errors}

The constraining power of our analysis is similar to that of the HSC Year 1 analyses \cite{Hikage2019, Hamana2020}, despite the threefold increase in sky coverage. This is largely due to our choice to use a conservative, wide, flat prior on the shifts in the third and fourth redshift bins ($\Delta z_3$ and $\Delta z_4$). While \cite{Rau2022} suggests an informative, Gaussian prior on these two parameters (${\cal N}(0, 0.031)$ and ${\cal N}(0, 0.034)$ respectively), all HSC-Y3 cosmology analyses find evidence for biased redshifts in these bins \cite{Li2023, Sugiyama2023, Miyatake2023}. 

In particular, when using the Gaussian prior on these parameters, we see a large negative shift in $S_8$ when removing bin 4 from the analysis. Moreover, we find a $1.14\sigma$ shift in $S_8$ to higher values when using the Gaussian $\Delta z_3$ and $\Delta z_4$ priors as opposed to the flat, wide priors. In fact, our fiducial analysis detects large shifts in these redshift bins, $\Delta z_3 = -0.076^{+0.056}_{-0.059}$ and $\Delta z_4 = -0.157^{+0.094}_{-0.111}$, values which would not be allowed by the Gaussian priors. We test these detections using a set of noisy mock data vectors, generated based on our best-fit model, but with $\Delta z_3 = 0$ and $\Delta z_4 = 0$. Upon running our fiducial analysis on these mock data vectors, we find that our measured values of $\Delta z_3$ and $\Delta z_4$ are larger than those seen in 90\% of the mocks (Figure~\ref{fig:deltaz_mocks}), indicating that this is unlikely to be a spurious detection of a bias in the redshifts. 

In this analysis, we have taken the most conservative possible approach to these residual photometric redshift errors, by adopting a very wide, flat prior on these parameters. Improved calibration of the source redshift distribution at these high redshifts will allow the $S_8$ constraining power of the Y3 analysis to be improved beyond the Y1 analysis. This will be made possible with large spectroscopic galaxy catalogs from the ongoing Dark Energy Spectroscopic Instrument (DESI, \cite{DESI2016}) survey, as well as the upcoming Subaru Prime Focus Spectrograph (PFS, \cite{Takada2014, Tamura2022}).

\subsection{\label{sec:omegam_bao} Impact of \texorpdfstring{$\Omega_m$}{Omega m} prior}

As noted in Section~\ref{sec:priors}, our data and scale cuts are unable to constrain $\Omega_m$, as the ratio of $\Cl$s at two extremes of the $\Omega_m$-$\sigma_8$ degeneracy is nearly unity within our fiducial multipole range. However, our inference of $S_8$ should be unaffected by this. We explore the impact that a more informative $\Omega_m$ prior would have on our $S_8$ constraints. We use an $\Omega_m$ prior similar to recent constraints from Baryon Acoustic Oscillations, ${\cal N}(0.3, 0.01)$ \cite{Alam2017}. We find that our $S_8$ constraint is virtually unchanged by this more informative prior, with no shift in the posterior mode, and only a $\sim$$3\%$ increase in the posterior width (likely due to inconsistency with our true measured $\Omega_m$ value). 

Future work with the HSC-Y3 data may be able to improve constraints on $\Omega_m$ by extending the analysis to smaller angular scales ($\ell_{\mathrm{max}} \sim 4000$), where $\Cl$s are more able to differentiate between extreme regions of the $\Omega_m$-$\sigma_8$ degeneracy. However, this will come with the added challenge of modeling astrophysical effects at such small scales \cite{MacCrann2017, Huang2021, Yoon2021, Arico2023}. 

\subsection{\label{sec:astro_syst} Astrophysical Effects}

Our fiducial model includes the modeling of astrophysical effects, namely baryonic feedback and intrinsic alignments. Unlike the HSC Year 1 cosmology analyses \cite{Hikage2019, Hamana2020}, we obtain a value of $A_{\mathrm{bary}}$ that indicates deviation from the no baryonic feedback scenario, with $A_{\mathrm{bary}} = 2.43^{+0.46}_{-0.25}$. When using \texttt{HMCode2020} rather than \texttt{HMCode2016}, we find $T_{\mathrm{AGN}} = 7.68^{+0.27}_{-0.25}$. We have conducted a number of tests using both mock data and variations on our fiducial analysis with the HSC-Y3 data to show that our $S_8$ constraints are robust to our baryonic physics model. Moreover, we have shown that not modeling baryonic effects leads to a $0.5\sigma$ shift in $S_8$ to lower values. This might be indicative of support for the proposal that the apparent $S_8$ tension could be resolved by strong baryonic feedback \cite{Amon2022b}. Future cosmology analyses, especially ones aiming to use smaller scale information, will have to be careful in defining their scale cuts and baryonic feedback model, especially given this positive detection.

On the other hand, we do not see a clear detection of intrinsic alignments. In this analysis, we used the more conservative intrinsic alignment model, TATT, but this may not yet be necessary for our current level of constraining power. This will be an important choice for future analyses to explore. 

\subsection{\label{sec:future_data} Improvements with Future Data}

Our analysis, like many other preceding cosmic shear analyses \cite{Hikage2019, Amon2022, Doux2022, Asgari2021, Loureiro2022}, finds a value of $S_8$ that is $2\sigma$ lower than that inferred by the \textit{Planck} experiment in 2018 \cite{Planck2018Cosmology}. As weak lensing experiments continue to improve, we'll be able to see whether the central value of $S_8$ stays robust, and whether the significance of this apparent tension increases. In particular, we will soon be able to conduct this cosmic shear analysis with the final data release of HSC, covering over $1000 \ \mathrm{deg}^2$ of the sky, with the same extraordinary depth and image quality of these data. 

The depth and image quality of HSC are a preview of what we can expect from the upcoming Vera C. Rubin Observatory Legacy Survey of Space and Time (LSST, \cite{Ivezic2019}), which will cover $18,000 \ \mathrm{deg}^2$ of the sky, going one magnitude deeper than HSC. While the seeing will be similar to that of HSC, the additional number of visits contributing to each LSST co-added image will allow for far better control of the Point Spread Function. The addition of the $u$ band images will allow for improved photometric redshift estimation. At the same time, ongoing work to better study and develop modeling choices and analysis tools will improve our ability to obtain robust cosmological constraints from the data. We can also expect to see strong constraints from the weak lensing science from two upcoming space-based telescopes, \textit{Euclid} \cite{Euclid2011} and the \textit{Nancy Grace Roman Space Telescope} \cite{Roman2015}. Together, these data sets and weak lensing analyses will allow us to better understand the apparent tension in $S_8$ values inferred from weak lensing and from the CMB, and whether our current cosmological model is sufficient for describing the data. 

\begin{acknowledgments}
RD acknowledges support from the NSF Graduate Research Fellowship Program under Grant No.\ DGE-2039656. Any opinions, findings, and conclusions or recommendations expressed in this material are those of the authors and do not necessarily reflect the views of the National Science Foundation. XL, TZ, and RM are supported in part by the Department of Energy grant DE-SC0010118 and in part by a grant from the Simons Foundation (Simons Investigator in Astrophysics, Award ID 620789). AN was supported in part through NSF grants AST-1814971 and AST-2108126.
This work was supported in part by World Premier International Research Center Initiative (WPI Initiative), MEXT, Japan, and JSPS KAKENHI Grant Numbers JP18H04350, JP18H04358, JP19H00677, JP19K14767, JP20H00181, JP20J01600, JP20H01932, JP20H04723, JP20H05850, JP20H05855, JP20H05856, JP20H05861, JP21J00011, JP21H05456, JP21J10314, JP21H01081, JP21H05456, JP22H00130, JP22K03634, JP22J21612, JP22K03655 and JP22K21349 by Japan Science and Technology Agency (JST) CREST JPMHCR1414, by JSPS Core-to-Core Program Grant Numbers JPJSCCA20200002, by JST AIP Acceleration Research Grant Number JP20317829, Japan, and by Basic Research Grant (Super AI) of Institute for AI and Beyond of the University of Tokyo. SS was supported in part by International Graduate Program for Excellence in Earth-Space Science (IGPEES), WINGS Program, the University of Tokyo. TT was supported in part by FoPM, WINGS Program, the University of Tokyo.
AK was supported by National Science Foundation under Cooperative Agreement 1258333 managed by the Association of Universities for Research in Astronomy (AURA). Additional funding for Rubin Observatory comes from private donations, grants to universities, and in-kind support from LSSTC Institutional Members. The work of AAPM was supported by the U.S. Department of Energy under contract number DE-AC02-76SF00515. TS is supported by Grant-in-Aid for JSPS Fellows 20J01600 and JSPS KAKENHI Grant Number 20H05855.
DA acknowledges support from the Beecroft Trust, and from the Science and Technology Facilities Council through an Ernest Rutherford Fellowship, grant reference ST/P004474.

The Hyper Suprime-Cam (HSC) collaboration includes the astronomical communities of Japan and Taiwan, and Princeton University. The HSC instrumentation and software were developed by the National Astronomical Observatory of Japan (NAOJ), the Kavli Institute for the Physics and Mathematics of the Universe (Kavli IPMU), the University of Tokyo, the High Energy Accelerator Research Organization (KEK), the Academia Sinica Institute for Astronomy and Astrophysics in Taiwan (ASIAA), and Princeton University. Funding was contributed by the FIRST program from the Japanese Cabinet Office, the Ministry of Education, Culture, Sports, Science and Technology (MEXT), the Japan Society for the Promotion of Science (JSPS), Japan Science and Technology Agency (JST), the Toray Science Foundation, NAOJ, Kavli IPMU, KEK, ASIAA, and Princeton University.

This paper makes use of software developed for Vera C. Rubin Observatory. We thank the Rubin Observatory for making their code available as free software at http://pipelines.lsst.io/.

This paper is based on data collected at the Subaru Telescope and retrieved from the HSC data archive system, which is operated by the Subaru Telescope and Astronomy Data Center (ADC) at NAOJ. Data analysis was in part carried out with the cooperation of Center for Computational Astrophysics (CfCA), NAOJ. We are honored and grateful for the opportunity of observing the Universe from Maunakea, which has the cultural, historical and natural significance in Hawaii.

The Pan-STARRS1 Surveys (PS1) and the PS1 public science archive have been made possible through contributions by the Institute for Astronomy, the University of Hawaii, the Pan-STARRS Project Office, the Max Planck Society and its participating institutes, the Max Planck Institute for Astronomy, Heidelberg, and the Max Planck Institute for Extraterrestrial Physics, Garching, The Johns Hopkins University, Durham University, the University of Edinburgh, the Queen’s University Belfast, the Harvard-Smithsonian Center for Astrophysics, the Las Cumbres Observatory Global Telescope Network Incorporated, the National Central University of Taiwan, the Space Telescope Science Institute, the National Aeronautics and Space Administration under grant No. NNX08AR22G issued through the Planetary Science Division of the NASA Science Mission Directorate, the National Science Foundation grant No. AST-1238877, the University of Maryland, Eotvos Lorand University (ELTE), the Los Alamos National Laboratory, and the Gordon and Betty Moore Foundation.
\end{acknowledgments}

\appendix

\section{\label{app:bmodes} Large-Scale \texorpdfstring{$B$}{B} Modes}

\begin{figure*}[!htbp]
\includegraphics[width=0.95\textwidth]{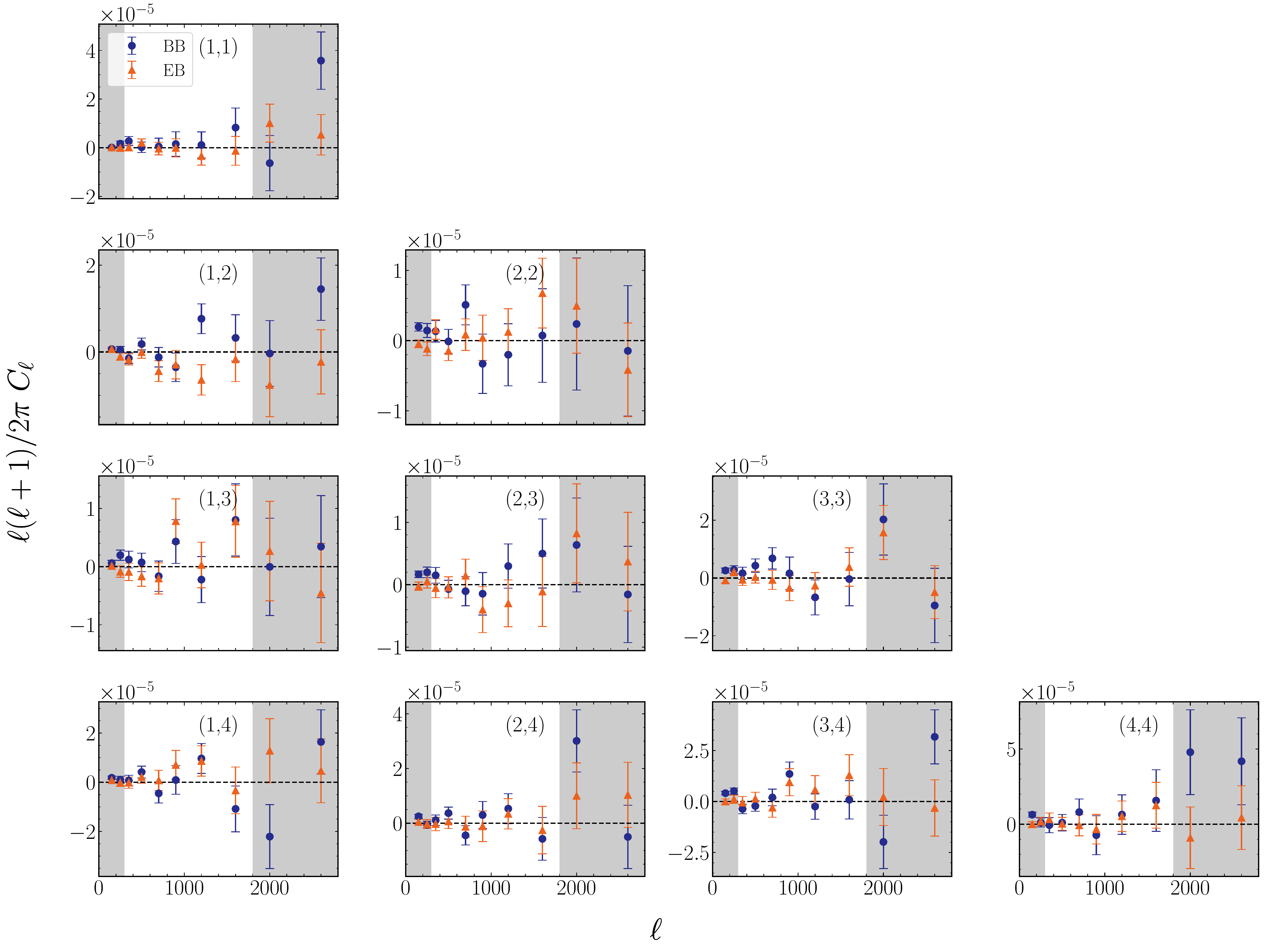}
\caption{\label{fig:bb_eb_spectra} Tomographic cosmic shear power spectra of $BB$ (blue circle) and $EB$ (orange triangle) modes, for the auto- and cross-correlations of the four tomographic redshift bins used in our analysis. Note the different y-axis scales from those in Figure~\ref{fig:power_spectra}. While no significant $B$ modes are detected for the range of scales used in our fiducial cosmological analysis, evidence of significant $B$ modes is seen for multipoles smaller than $\ell_{\mathrm{min}}=300$.}
\end{figure*}

As described in Section~\ref{sec:null_tests} and shown in Figure~\ref{fig:bb_eb_spectra}, we find evidence of significant $B$ modes at large scales, namely $\ell < 300$. Such large-scale $B$ modes were also observed in the HSC Year 1 analysis \cite{Hikage2019}. As the scale on which these $B$ modes appear corresponds approximately to the scale of the field of view of the HSC camera, they could be indicative of systematic effects due to variations with pointings on the sky. We do not include scales larger than $\ell_{\mathrm{min}} = 300$ in our cosmological analysis, due to the potential for contamination by systematic effects indicated by the presence of a $B$ mode signal.

\section{\label{app:cosmosis_validation} Validation of Likelihood Inference Setup}

We use a number of tests to ensure that our likelihood inference setup in \texttt{CosmoSIS} is unbiased. As described in Section~\ref{sec:sampler}, we simulate a data vector with our fiducial model, and analyze it with our fiducial setup. Given that the same model is used to generate and analyze the data, one might expect to retrieve the input parameters with no bias. However, the statistics we report throughout the paper, namely the 1D marginalized mode of the posterior, are subject to projection effects, i.e. they could be biased due to significant non-Gaussianities in the posterior. Indeed, when conducting this test with the fiducial model, we find that the projected mode of $\Omega_m$ is biased $0.77\sigma$ low from the input value, and $S_8$ is biased $0.48\sigma$ low compared to the input value.

To determine whether these biases are indicative of a bias in our likelihood inference setup, or whether they could be attributed to projection effects, we repeat this analysis using the ``maximum a posteriori'' (MAP) estimate from a minimization analysis, rather than the projected mode from nested sampling. This estimate should be robust to projection effects. We reanalyze the data vector simulated based on our fiducial model using the Powell minimizer (implemented in the \texttt{maxlike} sampler in \texttt{CosmoSIS}), and find a much smaller bias on our cosmological parameters: a $0.02\sigma$ bias on $\Omega_m$ and a $0.01\sigma$ bias on $S_8$ (see the first row of Table~\ref{tab:nonlinear_map_bias}). This indicates that the biases seen in the 1D marginalized mode likely arise from projection effects.

To further test for potential biases in our likelihood inference setup, we use the \texttt{star} sampler in \texttt{CosmoSIS}, which samples the parameter space uniformly across the prior, one dimension at a time, and computes the goodness of fit at each point. This removes some of the complications of the MAP estimation, which attempts to find the minimum likelihood point simultaneously across all dimensions of the parameter space. When taking 2000 samples of each parameter, while keeping the other parameters fixed to their input values, we find that the sampler is able to recover the point closest to the input value as the minimum $\chi^2$ point. This indicates that the likelihood is unbiased.

\section{\label{app:other_params} Posterior Distributions of the Fiducial Model Parameters}

\begin{figure*}[!htbp]
\includegraphics[width=0.95\textwidth]{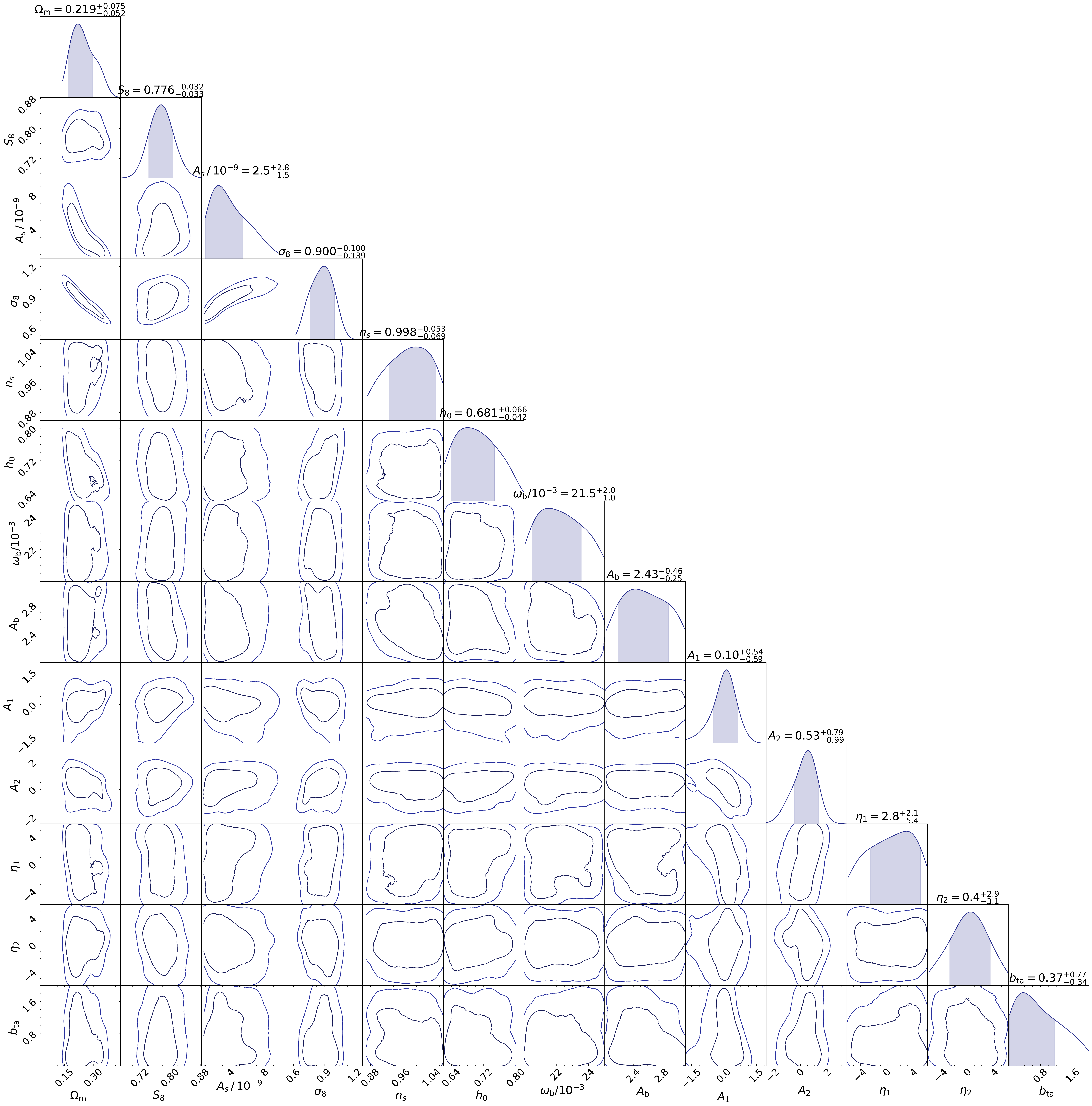}
\caption{\label{fig:all_param_constraints} Fiducial posteriors of all cosmological ($\Omega_m$, $S_8$, $A_s$, $\sigma_8$, $n_s$, $h_0$, $\Omega_b h^2$) and astrophysical ($A_{\mathrm{bary}}$, $A_1^{\mathrm{IA}}$, $A_2^{\mathrm{IA}}$, $\alpha_1^{\mathrm{IA}}$, $\alpha_2^{\mathrm{IA}}$, $b_{\mathrm{TA}}^{\mathrm{IA}}$) parameters in our fiducial model. Here, $S_8$ and $\sigma_8$ are derived parameters.}
\end{figure*}

Figure~\ref{fig:all_param_constraints} shows the marginalized one-dimensional and two-dimensional posteriors of cosmological and astrophysical (intrinsic alignment and baryonic physics) parameters. The fiducial constraints for all the parameters sampled in this analysis are shown in Table~\ref{tab:fid_parameters_values}.

\begin{table}%[b]
\renewcommand*{\arraystretch}{1.8}
\caption{\label{tab:fid_parameters_values}%
Fiducial constraints of the cosmological, astrophysical and observational systematics parameters sampled in our analysis. We report the marginalized mode of each parameter and the asymmetric $\pm 34\%$ confidence intervals, along with the maximum posterior point (MAP) of the nested sampling chain in parentheses.}
\begin{ruledtabular}
\begin{tabular}{ll} 
Parameter & Constraint \\ \hline
\multicolumn{2}{l}{\bf{Cosmological parameters}}\\
$\Omega_\mathrm{m}$                 & $0.219^{+0.075}_{-0.052} (0.226)$\\
$A_\mathrm{s} \,(\times 10^{-9})$   & $2.5^{+2.8}_{-1.5} (5.98)$\\
$h_0$                               & $0.681^{+0.066}_{-0.042} (0.652)$\\
$\omega_\mathrm{b} \equiv \Omega_b h^2$                 & $0.0215^{+0.0020}_{-0.0001} (0.0245)$\\
$n_\mathrm{s}$                      & $0.998^{+0.053}_{-0.069} (0.920)$\\
\hline
\multicolumn{2}{l}{\bf{Baryonic feedback parameters}}\\
$A_\mathrm{b}$                      & $2.43^{+0.46}_{-0.25} (2.97)$ \\
\multicolumn{2}{l}{\bf{Intrinsic alignment parameters}}\\
$A_1$                               & $0.10^{+0.54}_{-0.59} (0.07)$ \\
$\eta_1$                            & $2.8^{+2.1}_{-5.4} (-2.4)$ \\
$A_2$                               & $0.53^{+0.79}_{-0.99} (0.78)$ \\
$\eta_2$                            & $0.4^{+2.9}_{-3.1} (-2.6)$ \\
$b_\mathrm{ta}$                     & $0.37^{+0.77}_{-0.34} (0.38)$ \\
\hline
\multicolumn{2}{l}{\bf{PSF systematics} (Section~\ref{sec:psf_model})}\\
$\tilde{\alpha}_2$                          & $-0.03^{+0.98}_{-0.97} (0.04)$\\
$\tilde{\beta}_2$                           & $-0.1^{+1.0}_{-1.1} (-0.3)$\\
$\tilde{\alpha}_4$                          & $0.0^{+1.0}_{-1.1} (0.9)$\\
$\tilde{\beta}_4$                           & $-0.02^{+0.97}_{-1.03} (-0.07)$\\
\multicolumn{2}{l}{\bf{Redshift distribution uncertainties} (Section~\ref{sec:nz_uncertainty})}\\
$\Delta z^{(1)}$                    & $0.006^{+0.024}_{-0.021} (0.010)$ \\
$\Delta z^{(2)}$                    & $-0.005 \pm 0.023 (0.005)$ \\
$\Delta z^{(3)}$                    & $-0.075^{+0.056}_{-0.059} (-0.046)$ \\
$\Delta z^{(4)}$                    & $-0.157^{+0.094}_{-0.111} (-0.144)$ \\
\multicolumn{2}{l}{\bf{Shear calibration biases} (Section~\ref{sec:shear_calib})}\\
$\Delta m^{(1)}$                    & $-0.002^{+0.012}_{-0.010} (-0.007)$ \\
$\Delta m^{(2)}$                    & $-0.005^{+0.010}_{-0.012} (0.009)$ \\
$\Delta m^{(3)}$                    & $-0.001^{+0.010}_{-0.011} (0.004)$ \\
$\Delta m^{(4)}$                    & $-0.0006^{+0.0098}_{-0.0124} (-0.003)$ \\
\end{tabular}
\end{ruledtabular}
\end{table}

\bibliography{apssamp}

\end{document}